\documentclass[dvipsnames, 12pt]{article}
\usepackage[margin=30mm]{geometry}
\usepackage{amsmath}\allowdisplaybreaks
\usepackage{amsfonts}
\usepackage{amssymb}
\usepackage{verbatim}
\usepackage{setspace}
\usepackage{rotating}
\usepackage{booktabs}
\usepackage{longtable}
\usepackage{tabularx}
\usepackage[x11names]{xcolor}
\usepackage{subcaption}
    \usepackage[flushleft]{threeparttable}

\usepackage{etoolbox}\AtBeginEnvironment{thebibliography}{\linespread{1}\selectfont}
\newcommand{\zerodisplayskips}{%
  \setlength{\abovedisplayskip}{-5pt}%
  \setlength{\belowdisplayskip}{5pt}%
  \setlength{\abovedisplayshortskip}{-5pt}%
  \setlength{\belowdisplayshortskip}{5pt}}
\appto{\normalsize}{\zerodisplayskips}
\appto{\small}{\zerodisplayskips}
\appto{\footnotesize}{\zerodisplayskips}
\usepackage{tikz}
\usetikzlibrary{fit,positioning}
\usepackage{amsthm}
\usepackage{bm}
\usepackage{dsfont}
\usepackage{lscape}
\usepackage{geometry}
\usepackage{float}
\usepackage{verbatim}
\usepackage{subcaption}
\usepackage[toc,page,header]{appendix}
\usepackage{minitoc}
\usepackage{algorithm}
\usepackage{algpseudocode}

\usepackage{dsfont}
\usepackage{tablefootnote}
\usepackage{multirow}
\usepackage{mathtools}

\usepackage{hyperref}
\hypersetup{
    colorlinks=true,
    linkcolor=cyan,
    filecolor=cyan,      
    urlcolor=cyan,
    citecolor=cyan,
    breaklinks=true}

\usepackage[square,numbers]{natbib}
\usepackage{authblk}
\usepackage{caption}
\usepackage{subcaption}

\usepackage{framed}
 
\usepackage{listings}

\begin{document}
\doparttoc 
\faketableofcontents 

\title{Artificially Intelligent Opinion Polling}

\author{Roberto Cerina \\\texttt{r.cerina@uva.nl}\\Institute for Logic, Language and Computation\\University of Amsterdam \and Raymond Duch\\\texttt{raymond.duch@nuffield.ox.ac.uk}\\Nuffield College\\University of Oxford}

\date{\today}
\maketitle
\thispagestyle{empty}

\begin{abstract}
\noindent We seek to democratise public-opinion research by providing practitioners with a general methodology to make representative inference from cheap, high-frequency, highly unrepresentative samples. 
To this end, we provide two major contributions:  1) we introduce a general sample-selection process which we name \emph{online selection}, and show it is a special-case of selection on the dependent variable. We improve MrP for severely biased samples by introducing a bias-correction term in the style of King \& Zeng to the logistic-regression framework.  We show this bias-corrected model outperforms traditional MrP under online selection, and achieves performance similar to random-sampling in a vast array of scenarios; 2) we present a protocol to use Large Language Models (LLMs) to extract structured, survey-like data from social-media. We provide a prompt-style that can be easily adapted to a variety of survey designs. We show that LLMs agree with human raters with respect to the demographic, socio-economic and political characteristics of these online users. The end-to-end implementation takes unrepresentative, unsrtuctured social media data as inputs, and produces timely high-quality area-level estimates as outputs. This is \emph{Artificially Intelligent Opinion Polling}. We show that our AI polling estimates of the $2020$ election are highly accurate, on-par with estimates produced by state-level polling aggregators such as \texttt{FiveThirtyEight}, or from MrP models fit to extremely expensive high-quality samples.\end{abstract}

\newpage
\pagenumbering{arabic} 
\section{Introduction}\label{intro}

\noindent The use of increasingly unrepresentative samples contributes to systematic bias in the sub-national predictions of public opinion polling \cite{kennedy2018evaluation,sturgis2016report}.  Multilevel Regression and Post-Stratification (MrP) \cite{gelman1997poststratification,park2004bayesian} is a statistical technique that enables estimation of sub-national opinion from unrepresentative samples. It does so by adjusting estimates for a variety of non-response biases \cite{gelman2019nonresponse}.  There are two distinct examples of MrP that have been used successfully to perform model-based pre-election opinion polling with unrepresentative samples.  A foundational implementation exploits an extremely large unrepresentative sample of Xbox gamers \cite{wang2015forecasting} to generate state-level forecasts of the 2012 U.S. presidential election.  MrP performed well at the sub-national level with an extremely large but unrepresentative low quality opt-in sample.  A second example of successful MrP implementation employs smaller – although still large by conventional standards – higher quality samples consisting of self-selected online panelists \cite{lauderdale2020model,lauderdale2019MRP,hanretty2019MRP}.  These efforts have also successfully predicted sub-national election results.  In both of these implementations characteristics of the data likely play an important role in the MrP success – either data quality and/or sample-size.  But should the effectiveness of MrP polling models lie with Big Data, or the curation of a large, diverse and high quality online panel \cite{twyman2008getting}?	We believe the real promise of MrP modeling is a world in which the data collection protocol can be increasingly ignored and the reduction in bias can be assured by improvements in statistical modeling.  We propose a number of methodological steps in this direction.\\

\noindent The goal of this paper is to democratise public-opinion research by providing practitioners with a general methodology to make representative inference from cheap, high-frequency, highly unrepresentative samples. 
To this end, we provide two major contributions:  First, we introduce a general sample-selection process that we name \emph{online selection}, and show it is a special-case of selection on the dependent variable. This method improves MrP for severely biased samples by introducing a bias-correction term in the style of King \& Zeng \cite{king2001logistic} to the logistic-regression framework.  This bias-corrected model outperforms traditional MrP under online selection and achieves performance similar to random-sampling in a vast array of scenarios. Second, we present a protocol to use Large Language Models (LLMs) \cite{vaswani2017attention,devlin2018bert} to extract structured, survey-like data from social-media. The prompt-style we implement can be easily adapted to a variety of survey designs. We show that LLMs agree with human raters with respect to the demographic, socio-economic and political characteristics of these online users.\\

\noindent We illustrate the potential of our approach with data from the $2020$ US election. In the run-up of the election we collect a large corpus of Tweets. We use Amazon Mechanical Turk (AMT) workers and LLMs to extract user-level information, such as demographics and voting intention. We then apply the bias-corrected MrP to the extracted survey-like object. We note this procedure is in principle fully automated: a series of \texttt{R} scripts were used to download the data from the Twitter streaming API via the \texttt{rtweet} \cite{kearney2019rtweet} package; the \texttt{openai} \cite{openai2023} package is used to access the OpenAI API and extract the survey-object from the corpus; bias-corrected MrP is implemented in Stan \cite{carpenter2017stan} via \texttt{rtstan} \cite{rstan2023}. The end-to-end implementation takes unrepresentative, unsrtuctured social media data as inputs, and produces timely high-quality state-level estimates of the vote as outputs. This is \emph{Artificially Intelligent Opinion Polling}. We show that our AI polling estimates of the $2020$ election are highly accurate; on-par with estimates produced by state-level polling aggregators such as \texttt{FiveThirtyEight}, or from MrP models fit to extremely expensive high-quality samples such as the American National Election Study (ANES).\\

\noindent The paper proceeds as follows: Section \ref{statistical} outlines the statistical context under which we operate, including a description of the population of interest, the sampling mechanism, and the Hierarchical Bayesian structured MrP model we use to make representative inference; Section \ref{simulation_study} presents the results of a simulation study to ascertain the properties of the chosen inferential framework; Section \ref{data} presents our social media feature extraction approach, including our LLM prompt style and a description of various auxiliary datasets to enable a full comparison of the AI polling performance; Section \ref{results} summarises the results from the application of AI polling to the $2020$ US Presidential Election, and tentatively explores the connection between the quality of the underlying state-level estimates and the  quality of the social media data annotations. We present a high-level discussion of the limitations of this study, the significance of the findings, and areas of future research in Section \ref{discussion}.



\section{Statistical Context}\label{statistical}
A population of interest consists of $N$ individuals, indexed by $i \in \{1,\dots,N\}$. The population is stratified according to $M$ mutually exclusive \emph{cells}, $\mathcal{K} = \{k_1,\dots, k_M\}$. Each individual belongs to one of the cells: 

$\forall i, \exists! \mbox{ } m \in \{1, ..., M\}: i \in k_m$. 

\noindent The number of individuals belonging to each cell is:

\begin{equation}
    w_{m} = \sum^N_{i} \mathds{1}(i \in k_m).
\end{equation} 

\noindent Cells are defined over a set of attributes $X$, such that $X_m = \{x_{m1},\dots,x_{mP}\}$. To keep this exposition general, we leave the definition of $X$ and its components vague for now. This information is stored in a \emph{stratification frame}, which is known to the researcher prior to the study. An extract from the stratification frame used in this paper, derived primarily from the American Community Survey (ACS) $5$ years aggregation ($2014$ - $2019$) is presented in Table \ref{stratification_frame}.\\

\begin{table}[ht]
\caption{Stratification Frame Extract}
\scalebox{0.645}{
\centering
\begin{tabular}{r|ccccccccc|c}
  \hline
  \hline
 $m$ & \emph{state} & \emph{gender} & \emph{ethnicity} & \emph{age} & \emph{college.degree} & \emph{household.income} & \emph{vote.2016} & \emph{state$\_$R.vote.2016} & $\cdots$  & $w$ \\ 
  \hline
1 & Alabama & M & Black & $55-64$ & 0 & $50,000-75,000$ & R & 1.16 & $\cdots$ & 198.43 \\ 
  2 & Alabama & M & Black & $55-64$ & 0 & $50,000-75,000$ & D & 1.16 & $\cdots$ & 109.81 \\ 
  3 & Alabama & M & Black & $55-64$ & 0 & $50,000-75,000$ & other & 1.16 & $\cdots$ & 11.37 \\ 
  4 & Alabama & M & Black & $55-64$ & 0 & $50,000-75,000$ & stay home & 1.16 & $\cdots$ & 65.75 \\ 
  5 & Alabama & F & Black & $55-64$ & 0 & $50,000-75,000$ & R & 1.16 & $\cdots$ & 211.25 \\ 

  $\vdots$ & $\vdots$ & $\vdots$ & $\vdots$ & $\vdots$ & $\vdots$ & $\vdots$ & $\vdots$ & $\vdots$ & $\vdots$ & $\vdots$ \\ 
  
  117840 & Wyoming & F & Asian & $65+$ & 1 & $50,000-75,000$ & stay home & 1.60 & $\cdots$ & 0.06 \\ 
  117841 & Wyoming & M & Other & $65+$ & 1 & $25,000-50,000$ & R & 1.60 & $\cdots $ & 0.87 \\ 
  117842 & Wyoming & M & Other & $65+$ & 1 & $25,000-50,000$ & D & 1.60 &   $\cdots$ & 0.28 \\ 
  117843 & Wyoming & M & Other & $65+$ & 1 & $25,000-50,000$ & other & 1.60 & $\cdots$ & 0.14 \\ 
  117844 & Wyoming & M & Other & $65+$ & 1 & $25,000-50,000$ & stay home & 1.60 & $\cdots$ & 0.13 \\ 
   \hline
   \hline
\end{tabular}}
\begin{tablenotes}
\item[]\footnotesize Note: An extract from the stratification frame. The variable \emph{state\_R.vote.2016} is a standardised measure of the \% of votes the Republican party obtained in a given state. More state-level covariates are available in the frame, but omitted here for clarity. $w$ is not an integer since the frame has been extended according to the MrsP procedure \cite{leemann2017extending} to include $2016$ vote choice as an individual-level predictor.
\end{tablenotes}
\label{stratification_frame}
\end{table}

\noindent Each member of the population is further subject to a discrete choice. Individual $i$ considers set $\mathcal{C} = \{c_1,...,c_J\}$. Their choice is recorded in a random variable $y$, such that $y_i = c_j$ indicates the event: \emph{individual $i$ has opted for the $j^{th}$ choice}.\\ 

\noindent In our application to the $2020$ US election, we operationalise the choice-set facing voters as $\mathcal{C} = \{\mbox{Republican}, \mbox{Democrat}, \mbox{Libertarian}, \mbox{Green}, \mbox{Stay Home} \}$. This separates us from Lauderdale et al. \cite{lauderdale2020model}, who prefer a two-tiered choice-set, such that voters first choose whether to turn-out or stay-home, and further express a preference conditional on their choice of turnout. This allows for training turnout and vote-choice models on separate samples. Lauderdale et al. suggest this is an advantage as turnout can be better measured and verified a posteriori of an election, whilst self-reported turnout from surveys can be biased \cite{jackman2019does}. In this paper, we are interested in testing the viability of artificially intelligent opinion polling as a holistic methodology for surveying opinion, and hence focus on models entirely trained on AI polls. \\

\noindent We can aggregate the choices made by each member of the population to reveal the cell-level choice-probability:

\begin{equation}\label{cell_level_probability}
\pi_{mj} = \frac{1}{w_m} \sum_{i \in g_m}  \mathds{1}(y_i = c_j).
\end{equation}

\noindent Cell-level choice-probabilities can be further aggregated to reveal the marginal distribution of choice over any combination of the $P$ dimensions which define the cells. This is also known as a \emph{stratified} measure of preferences. \\

\noindent To illustrate, let $l\in \{1,\dots,L\}$ represent a set of categories to which a cell can belong to. Let $\mathcal{O} = \{o_1,...,o_L\}$ indicate the set of cells belonging to each category. Finally let $x_1$ be a categorical variable, such that:

$m \in o_l \iff x_{m1} = l$.

\noindent For a less abstract conceptualisation, take variable `state' in our stratification frame. This is a categorical variable taking any one of $L = 51$ values across the $M = 117,844$ cell in the frame: \emph{state} $\in$ $\{$Alabama, Alaska, $\dots$, Wisconsin, Wyoming$\}$. In our frame, the first $2,392$ cells belong to Alabama, the next $2,152$ to Alaska, and so on: $\mathcal{O} =$ $\{$ $o_{\mbox{\tiny AL}} = [1; \dots; 2,392],$ $o_{\mbox{\tiny AK}} = [2,393; \dots; 4,544],$ $\dots,$ $o_{\mbox{\tiny WI}} = [113,613; \dots; 116,008],$ $o_{\mbox{\tiny WY}} = [113,614; \dots; 117,844]$ $\}$.\\

\noindent The marginal probability of choosing option $j$ over $x_1$ is then: 

\begin{equation}\label{post_stratified_estimate}
 \theta_{lj} = \frac{\sum_{m \in o_l} \pi_{mj} \times w_m}{\sum_{m \in o_l} w_m}. 
\end{equation}

\noindent In our application, $\theta_{lj}$ represents the vote share of party $j$ in state $l$. $\theta$ is the parameter we wish to estimate. 

\subsection{Sampling}\label{sampling}

\noindent Unfortunately, $\bm{\pi}_j$ is not known to the researcher prior to the study, and must be estimated to obtain $\theta$. To generate plausible estimate of $\pi$ we sample from the above population and observe a set of choices.\\

\noindent No specific sampling scheme is assumed at this stage. A sampling event is indicated for each individual by random variable $\varrho_i \in\{0,1\}$. The set of sampled individuals is $\mathcal{S} = \{s_1,\dots,s_n\}$. A complementary set $\mathcal{S}^{\prime} = \{s^{\prime}_{1},\dots,s^{\prime}_{N-n}\}$ represents non-sampled individuals, such that $i \in \mathcal{S} \iff \varrho = 1$, and $i \in \mathcal{S}' \iff \varrho = 0$. Let $\iota \in \{1,...,n\}$ index the sample observations, then:

$\forall \iota, \exists! \mbox{ } i : s_{\iota} = i$.

\noindent Our goal is to estimate the posterior distribution $p( \bm{\pi}_j \mid \bm{y},\bm{\varrho} = 1)$.\\

\subsection{Bayesian Inference}\label{bayes}

We seek to estimate $p( \bm{\pi}_j \mid \bm{y},\bm{\varrho} = 1)$ via Hierarchical Bayesian modeling \cite{gelman2013bayesian}. We specify a DGP to be learned by our model. 
The model should return plausible estimates of the posterior that are independent of sample selection. The validity of our estimate of the population distribution of $\bm{\pi}_j $ depends crucially on the \emph{ignorability} assumption \cite{van2018flexible}. To see this, let us reformulate the problem in the context of missing data.\\

\noindent Recall that the cell-level population probability depends on the individual-level choices (Equation \ref{cell_level_probability}). We observe a subset of those choices $y_i, \forall i \in \mathcal{S}$, which we call $\bm{y}^{obs}$; we do not observe $y_i, \forall i \in \mathcal{S}^{\prime}$, denoted by $\bm{y}^{mis}$. In order to produce a valid posterior distribution for  $\bm{\pi}_j $, we need to have a complete set of $\bm{y} = (\bm{y}^{obs},\bm{y}^{mis})$. Hence we must generate plausible values of $\bm{y}^{mis}$ from the posterior distribution $p(\bm{y}^{mis}\mid \bm{y}^{obs} , \bm{\varrho})$.\\

\noindent For the estimated posterior to be valid, we must be able to ignore the sampling selection model:

\begin{align*}
&p(\bm{y}^{mis}\mid \bm{y}^{obs} , \bm{\varrho}) = p(\bm{y}^{mis}\mid \bm{y}^{obs}), \\
\therefore \mbox{ } & p(\bm{y} \mid  \bm{y}^{obs}, \bm{\varrho} = 0 ) =   p(\bm{y} \mid  \bm{y}^{obs},\bm{\varrho} = 1 ).
\end{align*}

\noindent If ignorability is violated, values generated from the estimated posterior predictive distribution of $\bm{y}$ may be unrepresentative. This could lead to bias in our estimate of $\bm{\pi}_j$ as we aggregate up to the cell-level. The phenomena which is generally considered responsible for violations of the ignorability assumption is \emph{non-response bias} \cite{groves2011survey}. We will look at one example from this type of violation in Section \ref{selection_bias}. Note that because we ultimately aggregate $\bm{y}$ up to the cell-level, we benefit from the law of large numbers \cite{wooldridge2015introductory}. Hence prediction errors at the individual level do not translate linearly to the cell-level. \\


\noindent We propose a Bayesian Hierarchical model to estimate the posterior distribution of choice probability: 

\begin{equation*}
p(\bm{\pi}_j \mid \bm{y}) \propto p(\bm{y} \mid \bm{\pi}_j)p(\bm{\pi}_j),
\end{equation*}

\noindent where $p(\bm{\pi}_j \mid \bm{y}) $ is the likelihood of the sample observation given $\bm{\pi}_j$, and $p(\bm{\pi}_j)$ is the prior distribution.

\subsubsection{Likelihood Declaration}

\noindent Let $y_\iota$ be a nominal random variable indicating a discrete choice. Instead here we model $y_\iota = c_1$, $y_\iota = c_2$ $\dots$ as a sequence of independent binary variables. This allows us to estimate $\bm{\pi}_j$ separately for each $j$, assuming conditionally independent Bernoulli likelihoods. Note: typically, the likelihood for this type of data would be modeled using a multinomial distribution. We find this approach can be computationally wasteful and does not seem to produce any gain over the proposed model at the desired level of stratification (see Section \ref{simulation_study}). \\


\noindent The probability of choosing any alternative within the choice set is conditionally independent of any other alternative:

\begin{align*}
\mbox{Pr}(y_\iota = c_1, \dots, y_\iota = c_J \mid X_\iota) = \mbox{Pr}(y_\iota = c_1\mid X_\iota)\mbox{Pr}(y_\iota = c_2\mid X_\iota)  \dots  \mbox{Pr}(y_\iota = c_J\mid X_\iota). 
\end{align*}

\noindent The event $y_\iota = c_j$ can be re-coded as a binary variable $q_{\iota j} = \mathds{1}(y_\iota = c_j)$. We can then assume that $q_{\iota j}$ follows a Bernoulli distribution, governed by a probability parameter $\pi_{\iota j} = \mbox{Pr}(q_{\iota j} = 1)$. Throughout the paper we use $\pi$ to indicate the probability of choice at different levels, where these levels are indicated by the indexing - so $\pi_{\iota j}$ is the choice probability for an individual in our sample, $\pi_{i j}$ for an individual in the population, $\pi_{m j}$ for a cell in the population, $\pi_{j}$ is the population prevalence of choice $c_j$, etc. . To satisfy our conditional independence assumption, we model a latent parameter $\bm{\mu}_j$ as a linear function $f$ of $\Omega_j$ and $X$ . $\Omega_j$ is a vector of choice-specific parameters which map the attribute-set $X$ onto the latent parameter $\bm{\mu}_j$.

\begin{align*}
    q_{\iota j} &\sim \mbox{Bernoulli}(\pi_{\iota j}) ;\\
    \pi_{\iota j} &= \frac{\mbox{exp}(\mu_{\iota j})}{1+\mbox{exp}(\mu_{\iota j})};\\
    \mu_{\iota j} & = f(\Omega_j,X_{\iota}).
\end{align*}

\subsubsection{Prior Declaration}
We embrace a parsimonious approach to prior specification that leverages prior knowledge to produce a structured posterior \cite{gao2021improving}. Our prior specification is further informed by the Penalised Complexity \cite{simpson2017penalising} paradigm.\\

\noindent Our priors facilitate compliance with the ignorability assumption. Bayesian hierarchical modeling can be a powerful tool to adjust for non-response bias \cite{gelman2019nonresponse}. It allows for weighted pooling of information from likelihood and priors. The effect of pooling information on the model parameters is referred to as \emph{shrinkage}. Structuring priors produces shrinkage of parameters towards a desirable functional form of latent variable $\bm{\mu}_j$. Structured priors allow us to leverage prior knowledge to regularise our linear predictor \cite{gao2021improving,hanretty2018comparing}. While shrinkage is generally desirable, excessive pooling can lead to under-coverage and low-correlation in MrP estimates. To relax the partial pooling of coefficients up to an optimal standard, a robust set of unstructured fixed-effect predictors at the desired levels of analysis (e.g. areal and/or temporal units) is necessary \cite{buttice2013does,lax2013should}. This level-specific predictor adds further structure to the latent-variable. MrP models can be extremely sensitive to proper specification of this predictor and efforts to automate its selection \cite{broniecki2022improved} merit further scrutiny.\\


\noindent Table \ref{X_recode} defines $X$ and $\Omega_j$ for our application. Given this set of covariates and parameters, we can express the latent propensity $\bm{\mu}_j$ as a linear function: 

\begin{align}
    \mu_{\iota j} = & \alpha_j +\gamma^{\Lambda}_{l[\iota]j} + \gamma^{\Delta}_{d[\iota]j} +\gamma^A_{a[\iota]j} + \gamma^G_{g[\iota]j} + \gamma^R_{r[\iota]j} + \gamma^H_{h[\iota]j}  + \gamma^E_{e[\iota]j} + \gamma^V_{v[\iota]j} + \sum^{12}_b \beta_{bj} x^{\star}_{\iota b};
\end{align}
where $X^\star = \{x_1^\star = x_9,\dots,x_{12}^\star = x_{20}\}$ are the subset continuous level-specific predictors, detailed in Table \ref{X_recode}.\\

\begin{table}[ht]
\caption{Model Predictors and Parameters}
\label{X_recode}
\scalebox{0.7}{
\begin{tabular}{cllcc|cc}
\hline
\hline

 \multicolumn{5}{c|}{$X$}  &  \multicolumn{2}{c}{$\Omega_j$}  \\

 \emph{predictor} & \emph{level} & \emph{description} & \emph{index} & \emph{domain} & \emph{parameter} & \emph{prior correlation structure} \\[0.15cm]
 \hline
\hline
$\mathbf{1}$ & global & / & / & / & $\alpha_j$ &
\begin{tabular}{@{}l@{}} iid \end{tabular} \\[0.15cm]
\hline

$x_1$ & state & state\_id & $l$ & \{1,\dots,51\} & $\gamma^{\Lambda}_{lj}$ & \begin{tabular}{@{}l@{}} spatial (BYM2) \end{tabular} \\[0.15cm]
\hline

$x_2$ & day & day\_id & $ $d$ $ & \{30,\dots,0\} & $\gamma^{\Delta}_{dj}$ & \begin{tabular}{@{}l@{}} random walk \end{tabular} 
\\[0.15cm]
\hline 

$x_3$ & \multirow{5}{*}{individual} & age\_id & $ $a$ $ & \{1,\dots,6\} & $\gamma^{A}_{aj}$ & random-walk \\[0.15cm]

$x_4$ & &income\_id & $ $h$ $ & \{1,\dots,5\} & $\gamma^{H}_{hj}$ & random-walk \\[0.15cm]

$x_5$ & &sex\_id & $ $g$ $ & \{1,2\} & $\gamma^{G}_{gj}$ & unstructured + shared variance \\[0.15cm]

$x_6$ & &race\_id & $ $r$ $ & \{1,\dots,5\} & $\gamma^{R}_{rj}$ &  unstructured + shared variance  \\[0.15cm]

$x_7$ & &edu\_id & $ $e$ $ & \{1,2\} & $\gamma^{E}_{ej}$ &  unstructured + shared variance \\[0.15cm]

$x_8$ & &vote16\_id & $ $v$ $ & \{1,\dots,4\} & $\gamma^{V}_{vj}$ &  unstructured + shared variance  \\[0.15cm]
\hline

$x_9$ & \multirow{9}{*}{state}  &  $2016$ $j$ share & $  \multirow{9}{*}{/} $ &  \multirow{9}{*}{$\mathbb{R}$} & $\beta_{1j}$ & \multirow{9}{*}{iid}  \\[0.15cm]

$x_{10}$ &  & $2012$ $j$ share & & & $\beta_{2j}$ & \\[0.15cm]

$x_{11}$ &  &  \% white  & & &  $\beta_{3j}$ &  \\[0.15cm]

$x_{12}$ &  &  \% evangelical  & & &  $\beta_{4j}$ &  \\[0.15cm]

$x_{13}$ & &  \% college degree  & & & $\beta_{5j}$ & \\[0.15cm]

$x_{14}$ & &  region = `midwest'  & & &  $\beta_{6j}$&\\[0.15cm]

$x_{15}$ &  & region = `northeast'  & & & $\beta_{7j}$ &  \\[0.15cm]

$x_{16}$ &  & region = `south' & & &  $\beta_{8j}$ &\\[0.15cm]

$x_{17}$ &  & region = `west'  & & &  $\beta_{9j}$ & \\[0.15cm]
\hline 

$x_{18}$ & \multirow{2}{*}{day}  & economic index  & $  \multirow{2}{*}{/} $ &  \multirow{2}{*}{$\mathbb{R}$} & $\beta_{10j}$ &   \multirow{2}{*}{iid} \\[0.15cm]

$x_{19}$ & & incumbent approval  & & &  $\beta_{11j}$ &  \\[0.15cm]
\hline 

$x_{20}$ & state - day & cumulative COVID-19 deaths  & $  \multirow{1}{*}{/} $ &  \multirow{1}{*}{$\mathbb{R}$} & $\beta_{12j}$ & iid\\[0.15cm]
 \hline
 \hline
\end{tabular}
}
\begin{tablenotes}
\item[]\footnotesize Note: Predictors and parameters specific to our application. `iid' refers to fully independent parameters, or `fixed' effects \cite{gelman2013bayesian}. `unstructured + shared variance' priors refers to classic random-intercepts. Random-walk and spatial correlation structures are explained in detail below.
\end{tablenotes}
\end{table}
\noindent \textbf{Global intercept.}
Our global intercept parameter $\alpha_j$ is assigned a weakly informative prior. No correlation structure amongst choices is used to inform the baseline rate of choice:

\begin{equation}
    \alpha_j \sim N(0,10).
\end{equation}

\noindent \textbf{Unstructured effects.} A number of classic random intercepts are used to describe the effects of nominal variables with no specific structure. Allowing slight abuse of notation, let $U$ denote a given categorical predictor and $u$ represent the levels within that predictor. The random intercept prior is then:


\begin{align}
    \gamma^U_{uj} \sim&  N(0,\sigma_j^U), && \mbox{ } \forall \mbox{ } U \in \{G,R,E,V\};
\end{align}

\noindent While sex and educational attainment are binary variables, they are modeled using the unstructured random intercept prior. For the sake of interpretation, we marginally prefer the soft sum-to-zero constraint obtained via sharing $\sigma$ to the traditional corner-constraint. We further prefer this approach to generalise data-cleaning functions. In practice we do not expect these estimates to be significantly different from the fixed-effect estimates, as shrinkage is minimal under weakly-informative priors when the number of levels is $< 3$ \cite{park2004bayesian}.\\

\noindent Priors for the standard deviation parameters are assigned according to the recommendations of the Stan team \cite{gelman2020default}, and are weakly-informative on the log-odds scale: 

\begin{align}
\sigma_j^U \sim N^{+}(0,1), && \forall \mbox{ } U \in \{G,R,E,V\}.
\end{align}

\noindent \textbf{Spatial structure.} Previous studies \cite{gao2021improving,hanretty2018comparing} have suggested that explicit modeling of geographic proximity can improve estimates of the distribution of preferences across states. An account of the distribution of spatial preferences is presented by the Besag-York-Mollié (BYM) \citep{besag1991bayesian} family of models. We  focus on the BYM2 formulation \cite{riebler2016intuitive}:

\begin{align}
\gamma^{\Lambda}_{lj} =& \mbox{ }\sigma^{\Lambda}_j \left( \phi_{lj}\sqrt{(1-\xi_j)} + \psi_{lj}\sqrt{(\xi_j/\epsilon)}  \right); \\
\phi_{lj} \sim &  N(0,1) ;\\
\psi_{lj}  \mid \psi_{{l}^\prime j}  \sim & N\left(\frac{\sum_{{l} ^{\prime} \neq l} \psi_{{l} ^\prime j} }{\nu_{l}},\frac{1}{\sqrt{\nu_{l}}} \right) ;\\
\xi_j \sim & \mbox{Beta}\left(\frac{1}{2},\frac{1}{2}\right);\\
\sigma_j^{\Lambda} \sim & N^{+}(0,1);
\end{align}

\noindent where the total spatial effect $\gamma^{\Lambda}_{lj}$ is the convolution  of unstructured random intercepts $\phi_{lj}$ and intrinsic-conditionally-autoregressive (ICAR) effects $\psi_{lj}$. The autoregressive element allows estimation of $\psi_{lj}$ to be conditional on the average neighbourhood effect, where $\psi_{{l^{\prime}} j}$ represents the effect of a neighbour. The neighbourhood structure is dictated by an adjacency matrix, typically derived from a map. The ICAR prior standard deviation decreases as the number of neighbours $\nu_l$ increases. $\xi_j \in (0,1)$ is a mixing parameter which imposes an identifiability constraint. This is necessary to optimise posterior exploration and sensibly assign variance amongst competing explanations. Spatial and unstructured effects share a standard deviation parameter $\sigma_j^\lambda$. For this assumption to be sensible, $\psi_{lj}$ and $\phi_{lj}$ must be on the same scale. This is typically not the case, as the scale of $\psi_{lj}$ is defined by the local neighbourhood, whilst $\phi_{lj}$ is scaled across all areas. To ensure the shared-variance assumption holds, we calculate a scaling factor $\epsilon$ from the adjacency matrix, and use it to re-scale our spatial effects appropriately. Notation for islands is omitted in the above, but note these are a special case of the model, for which $\xi_j = 0$ \cite{donegan2022geostan,donegan2022flexible}. The islands still contribute to partial-pooling for the unstructured effects, but are ignored for the spatial component.\\ 

\noindent \textbf{Random-walk structure.} Gao et al. \cite{gao2021improving} have shown improved accuracy of MrP estimates by including structures to account for correlations amongst effects of neighbouring levels in ordinal variables. We focus on the \emph{random walk} structure:

\begin{align}
\gamma^U_{uj}  \mid \gamma^U_{u-1\mbox{ }j} \dots  \gamma^U_{1j} \sim & N( \gamma^U_{u-1\mbox{ }j}, \sigma_j^{U}),&& \forall \mbox{ } u > 1,\mbox{ }  U \in  \{\Delta,A,H\}; \\
\sigma_j^{U} \sim & N^{+}(0,1).
\end{align}

\noindent Again a sum-to-zero constraint $\sum_u \gamma^U_{uj} = 0$ is used to ensure identifiability.\\

\noindent \textbf{Independent linear Predictors.} The final set of priors to specify is for the fixed-effect regression coefficients $\beta$. These are independet weakly informative priors on the log-odds scale: 
\begin{align}
    \beta_{bj} \sim& N(0,1),&& \forall \mbox{ } b \mbox{ } \in \{1,\dots,12\}.
\end{align}
Traditional MrP approaches make use of fixed-effects at the area-level. We have an interest in testing the ability of AI polls to capture not merely the cross-states distribution of vote-choICe, but also temporal trends. As such we introduce a day-level, and a state-by-day level, set of fixed-effects. The number of days-to-election levels $(d \in \{1,\dots,30\})$ is large-enough, and there is enough variance across these, that we can expect a time-varying fixed-effect predictor to enrich our estimates of temporal dynamics. 
 
\subsubsection{Implementation in Stan}
We fit our models using the probabilistic programming language Stan \cite{carpenter2017stan, stan2018rstan}. Stan performs Bayesian inference via the the `No-U-Turn' sampler \cite{hoffman2014no}, a version of the Monte Carlo Markov Chain (MCMC) algorithm known as Hamiltonian Monte Carlo (HMC) \cite{gelman2013bayesian}. At convergence, sampling from the full conditional distribution of each parameter will be equivalent to sampling from the joint posterior $p(\Omega_j \mid \bm{q}_j )$.\\

\noindent Prior to fitting the model, we perform a number of operations designed to encourage efficient posterior exploration. We standardise our continuous correlates $X^\star$. A non-centered parametrisation is implemented for all our random-effects \cite{papaspiliopoulos2007general}. The ICAR prior is efficiently  specified in Stan via the following improper (non-generative) prior \cite{morris2019bayesian}: 

\begin{align*}
\mbox{log } p(\bm{\psi}_{j} ) \propto& \mbox{exp} \left\{ -\frac{1}{2} \sum_{l^{\prime} \neq l}(\psi_{lj} - \psi_{l^\prime j} )^2 \right\};
\end{align*}

\noindent a sum-to-zero constraint $\sum_l \psi_{lj} = 0$ is implemented to ensure identifiability. The Stan code for this model is provided in Listings \ref{lst:stan_dist} to \ref{lst:stan_dist3} in the Appendix\footnote{The model presented in the Listings includes an offset parameter, which will be explained in Section \ref{selection_bias}. Given the models are otherwise identical, the no-offset model is omitted. Note further that the model makes use of the spatial functions developed by Connor Donegan \cite{donegan2022flexible}. Functions developed by Mitzi Morris \cite{morris2019bayesian} were used for calculating the adjacency matrix.}.\\

\noindent We fit the same model separately for each choice $j$. For each choice-model, we generate $8$ chains of samples worth $500$ iterations each. From each chain we discard a warmup of $250$ iterations, and apply thinning-factor of $4$ to minimise auto-correlation. We keep a total of $504$ high-quality posterior samples for each parameter.\footnote{We are not particularly concerned about strict convergence. The goal of this implementation is for the posterior samples of the cell-level choice-probabilities, aggregated up to the desired level of analysis (see Equation \ref{post_stratified_estimate}), to be stable across runs. This aggregate, which converges faster than any given model parameter (law of large numbers), is the primary output of our estimation procedure. We do not seek to make inference for any given parameter. Lauderdale et al. \cite{lauderdale2020model} describe reaching stability across runs by producing shorter chains ($36$ chains of $25$ iterations, $25$-iteration warmup). We find our relatively small posterior samples are enough to ensure stability of the $5^{th}$, $50^{th}$ and $95^{th}$ percentiles of our state- and day- level estimates.}

\subsubsection{Posterior Prediction}

\noindent Let $\chi$ index the posterior samples obtained via the MCMC procedure. Allowing for the slightly abusive notation introduced earlier to indicate all random effects, posterior samples for the latent propensity $\bm{\mu}_j$ are derived: 

\begin{align}\label{post_pred}
    \{\mu\}^{\chi}_{mj} = & \{\alpha\}^{\chi}_j + \sum_U \{\gamma^{U}\}^{\chi}_{u[m]j} + \sum_b \{\beta\}^{\chi}_{bj}  x^{\star}_{m b}, && \forall \mbox{ } \chi \in \{1,\dots,504\}.
\end{align}

\noindent We obtain posterior samples of the choice-probability $\bm{\pi}_j$ via the inverse-logit link\footnote{In the specific application to the $2020$ US Presidential election, we perform a normalisation step to calculate the distribution of preferences across individuals who turn-out. Let $j = 5$ denote the `stay home' option, and $j \in \{1,\dots,4\}$ represent choosing the Republican, Democrat, Libertarian or Green-party candidates. We normalise the choice probabilities to account for turnout as follows: $\pi^{\star}_{mj} = \frac{\pi_{mj}}{\sum^{4}_j \pi_{mj}}, \mbox{ } \forall j \in \{1,\dots,4\}$.}:


\begin{align*}
\{\pi\}^{\chi}_{mj} = \frac{\mbox{exp}\left(\mbox{ }\{\mu\}^{\chi}_{mj}\mbox{ }\right)}{1 + \mbox{exp}\left(\mbox{ }\{\mu\}^{\chi}_{mj}\mbox{ }\right)}.
\end{align*}


\noindent Finally, calling back to Equation \ref{post_stratified_estimate}, we can obtain posterior simulations for the desired marginal probability of choosing option $j$ over any of the categorical predictor $U$ of interest: 

\begin{equation}\label{posterior_post_stratified_estimate}
 \{\theta\}^{\chi}_{uj} = \frac{\sum_{m \in o_u} \{\pi\}^{\chi}_{mj} \times w_m}{\sum_{m \in o_u} w_m}. 
\end{equation}

\subsection{Online Selection}\label{selection_bias}

By assuming a sampling design that conforms to the ignorability assumption, it is valid to use the posterior distribution to input unobservable choices.  Unfortunately, the choices of observable individuals may be unrepresentative due to \emph{selection effects}. Factors which are not included in the analysis can determine the individual's probability to select into sample. If these factors are correlated with preferences, our estimates will be biased. In this section we present a methodology to address a simple form of selection bias in the context of MrP.\\

\noindent Selection bias resulting from selecting on the dependent variable is addressed in a seminal paper by King \& Zeng \cite{king2001logistic}. They develop a bias-correction method in the context of rare-events, though their results generalise to various instances of exogenous selection on the dependent variable leading to unbalanced choice counts in the sample \cite{cerina2023explaining}.\\ 

\noindent In the context of pre-election opinion polling, we are primarily concerned with unbalanced samples of respondents from social media and online panels. We assume a simple selection mechanism that we label \emph{online selection}. Online selection affects the number of individuals eligible for selection per cell $w_m$ when only a proportion of the eligible individuals will `survive' the selection. Notice that these cell-counts can be broken down into the sum of the number of individuals who make a specific choice: $w_m = \sum_j \pi_{mj} w_m = \sum_j \sum_{i \in g_{m}} \mathds{1}(y_i = c_j)$. Let survival proportion per cell-choice combination be $(1 - \Upsilon_{mj})$, where $\Upsilon_{mj}$ is an online selection penalty. The individual selection probability under this mechanism can then be expressed as:  

\begin{align}
& \mbox{Pr}(\varrho_i = 1 \mid i \in g_m, y_i = c_j) =  \frac{w^\star_{mj}}{\sum_j \sum_{m} w^\star_{mj}}\\
& w^\star_{mj} = (1 - \Upsilon_{mj}) \sum_{i \in g_{m}} \mathds{1}(y_i = c_j);\\
&\Upsilon_{mj} \sim \mbox{Beta}\left(\mu_j^\Upsilon,\sigma_j^\Upsilon\right);\hspace{30pt}
\mu_j^\Upsilon \in (0,1);\hspace{30pt}
\sigma_j^\Upsilon  \in \left(0,\mu^\Upsilon(1-\mu^\Upsilon) \right);
\end{align}

\noindent where $\mu^{\Upsilon}_j$ is a choice-specific inclination to opt-out of a given medium; $\sigma^{\Upsilon}_j$ is the cross-cells deviation from the central tendency, which is bounded at $\mu^\Upsilon(1-\mu^\Upsilon)$ to ensure the Beta distribution is proper; and $\Upsilon_{mj}$ is the resulting cell-heterogeneous choice-specific selection effect.\\ 

\noindent It is easy to see this selection at work in self-selection on social-media. The social media site Gab \footnote{\url{https://gab.com}} has notoriously attracted right-leaning voters, and within this group is has attracted a specific subset of cells \cite{jasser2023welcome}. Twitter and Facebook have historically been liberal-leaning platforms \cite{mellon2017twitter}, though they attract different segments of this population. These considerations would translate directly to a platform-specific survival-probability $\Upsilon_{mj}$, dominated by a choice-specific tendency $\mu^{\Upsilon}_j$. 
Sampling at random from the population of these social-media communities is then akin to noisy sampling on the dependent variable, in the manner described above. Note that this violates the ignorability assumption: posterior predictive samples generated by models trained under this sampling protocol will be different from those obtained by training a model on the unobserved data. In other words: 

\begin{equation}
    p(\bm{y} \mid \bm{y}^{obs}, \bm{\varrho} = 0) \neq p(\bm{y} \mid \bm{y}^{obs}, \bm{\varrho} = 1) .
\end{equation}

\noindent How can we then estimate a valid posterior distribution $p(\bm{\pi}_j \mid \bm{y}) $ from samples selected as above? In our DGP we have proposed a structured logistic-regression as a plausible model for the DGP of our data. Even assuming we can fully control for factors which influence both selection and choice probabilities, the proposed model will still be biased. The consequences of this type of selection are principally manifest in a biased intercept $\alpha_j$. We propose to apply King \& Zeng's prior correction \cite{king2001logistic} to account for choice-specific online selection. \\ 

\noindent Let sample size $n = n^0_j + n^1_j$, where $n^1_j = \sum^n_\iota \mathds{1}(y_\iota = c_j)$, the total number of individuals who choose option $j$ in our sample (\emph{cases}), and $n^0_j = \sum^N_\iota \mathds{1}(y_\iota \neq c_j)$, the number of those who do not (\emph{controls}). These quantities have known population counterparts: $N = N^0_j + N^1_j$, where $N^1_j = \sum^N_i \mathds{1}(y_i = c_j)$ and $N^0 = \sum^N_i \mathds{1}(y_i \neq c_j)$. We can define the conditional probabilities of sampling cases and controls under our selection mechanism as follows: 

\begin{align}
\Pr(\varrho_{\iota} =  1 \mid y_\iota = c_j) =  & \frac{n^1_j}{N^1_j} ; \\
\Pr(\varrho_{\iota} = 1 \mid y_\iota \neq c_j) =  & \frac{n^0_j}{N^0_j}.
\end{align}

\noindent King \& Zeng show that the log-odds of selection can be used as an offset in a logistic regression model to correct bias associated with the intercept: 

\begin{align}
    \mu_{\iota j} = & \log\left(\frac{n^1_j/N^1_j}{n^0_j/N^0_j}\right) + \tilde{\alpha}_j + \sum_U \gamma^{U}_{u[\iota]j} + \sum_b \beta_{bj}  x^{\star}_{\iota b};\\
     \tilde{\alpha}_j = &\alpha_j - \log\left(\frac{n^1_j/N^1_j}{n^0_j/N^0_j} \right).
\end{align}

\noindent 
Posterior samples can be generated for a representative sampling protocol by omitting the offset from the prediction equation:

\begin{align}
    \tilde{\mu}_{m j} = & \tilde{\alpha}_j + \sum_U \gamma^{U}_{u[m]j} + \sum_b \beta_{bj}  x^{\star}_{m b};\\
    \tilde{\pi}_{mj} = &\frac{\exp(\tilde{\mu}_{m j})}{1 + \exp( \tilde{\mu}_{m j} )}.
\end{align}

\subsubsection{Exogenous Prevalence}
The prior correction relies on knowledge of $N^0_j$ and $N^1_j$ - or more succinctly, knowledge of the prevalence $\pi_j = \frac{N^1_j}{N^1_j + N^0_j}$. This is unknown and unobserved in our setup, as described in Section \ref{sampling}. We must therefore find a way to estimate this quantity. \\

\noindent MrP is often used to obtain small-area estimates in the context of pre-election opinion polling. This methodology is powerful because it allows us to address a variety of non-response biases by controlling for relevant covariates in the regression equations \cite{gelman2019nonresponse}. These adjustments are vital in the context of small-area estimation \cite{kennedy2018evaluation}, where obtaining representative samples appears more challenging than at the national level. Despite occasional misses, there is strong evidence that national polling has remained accurate over time and across countries \cite{jennings2018election}. It follows that national polling aggregators are a relatively objective source of population-prevalence during an election campaign. \\

\noindent We propose to leverage prevalence estimates obtained via polling averages to inform our bias-correction term. The methodology of aggregating polls at the national and sub-national level during an election campaign is well established \cite{jackman2005pooling,linzer2013dynamic,heidemanns2020updated}. It would be feasible to leverage national polling data and an appropriate aggregation model to produce a fully-Bayesian estimation of prevalence, simultaneously with the other model parameters. This would however impose greater computational costs on the estimation procedure. Moreover, given the bias induced by the online sampling protocol, it is preferable to take prevalence as a known constant. \\

\noindent In our application to the $2020$ US presidential election we use the average of the \texttt{FiveThirtyEight} national-level predictions \cite{f382020forecast}. Note that we are producing an election-day estimate, so we assume the entire time-series of forecasts up-to election-day is known. The prevalence per party is then the simple average of these series across the days of the campaign. 

\section{Simulation Study}\label{simulation_study}
What are the implications of introducing the King \& Zeng bias-correction for estimates of the stratified (Equation \ref{post_stratified_estimate}) and cell-level (Equation \ref{cell_level_probability}) choice probabilities? 
To explore the properties of the bias-correction mechanism, we perform a simulation study. Table \ref{simulation_scenarios} presents the various models and sampling protocols we compare in the simulation study. Our primary interest is to compare our proposed modeling strategy \texttt{(S.8)} against its uncorrected version \texttt{(S.4)}, and the `best-case-scenario' of random-sampling \texttt{(S.0)}. We further use the simulation study to evaluate the gains afforded by the use of structured priors under the proposed DGP, as well as any disadvantage arising from using a sequential Bernoulli likelihood as opposed to a Multinomial likelihood. We evaluate each of the scenarios according to the following metrics:

\begin{align}
\mbox{Bias:} &&\mathcal{B} =&  \frac{1}{n}\sum_i f_i - \hat{f}_i;\\
\mbox{Root Mean Squared Error:}&&\mathcal{RMSE} =&  \sqrt{\frac{1}{n}\sum_i  (f_i - \hat{f}_i)^2}; \\
\mbox{Pearson Correlation:}&& \rho =&  \frac {\sum _{i=1}^{n}(\hat{f}_{i}-{\bar {\hat{f}}})(f_{i}-{\bar {f}})}{{\sqrt {\sum _{i=1}^{n}(\hat{f}_{i}-{\bar {\hat{f}}})^{2}}}{\sqrt {\sum _{i=1}^{n}(f_{i}-{\bar {f}})^{2}}}};\\
\mbox{Coverage } (90\%): && \Gamma =& \frac{1}{n}\sum_i  \mathds{1}( \hat{f}_i^{5\%}<f<\hat{f}_i^{95\%}).
\end{align}

\begin{table}[ht]
\caption{Summary of the modeling and sampling scenarios.}
\label{simulation_scenarios}
\begin{center}
\scalebox{0.65}{
\begin{tabular}{c|cccc|cccc|c}
\hline
\hline
\emph{Scenario\_ID} & \emph{Sampling} & \emph{Likelihood} & \emph{Structured Priors} & \emph{Bias-correction} & \emph{Bias} & \emph{RMSE} & \begin{tabular}{@{}l@{}} 
 \emph{Pearson} \\
 \emph{Correlation}
 \end{tabular} &
 \begin{tabular}{@{}l@{}} 
 \emph{Coverage}\\
 \emph{(Distance}\\
 \emph{from 90\%)}
\end{tabular} & \emph{Colour} \\
\hline 
\hline
\texttt{(S.0)} & random & Bernoulli & \texttt{TRUE} & \texttt{FALSE} & / & /& /& / & \textcolor{DodgerBlue1}{\rule{0.5cm}{1mm}}\\ 
\texttt{(S.1)} & random  & Bernoulli  & \texttt{FALSE} &\texttt{FALSE} & 0 & 0.004 & -0.014 & -0.005& \textcolor{Blue4}{\rule{0.5cm}{1mm}}\\  
\hline
\texttt{(S.2)} & random  & Multinomial & \texttt{TRUE} &\texttt{FALSE} & 0 & 0 & -0.002 & -0.012& \textcolor{Aquamarine1}{\rule{0.5cm}{1mm}}\\ 
\texttt{(S.3)} & random  & Multinomial & \texttt{FALSE} &\texttt{FALSE} & 0 & 0.005 & -0.016 &-0.014 & \textcolor{Aquamarine4}{\rule{0.5cm}{1mm}}\\  
\hline
\texttt{(S.4)} & selected  & Bernoulli  & \texttt{TRUE} &\texttt{FALSE} & 0.093 & 0.084 & -0.088 & -0.445& \textcolor{OrangeRed1}{\rule{0.5cm}{1mm}}\\ 
\texttt{(S.5)} & selected & Bernoulli  & \texttt{FALSE} &\texttt{FALSE} & 0.094 & 0.087 & -0.104 &-0.435 & \textcolor{Red4}{\rule{0.5cm}{1mm}}\\  
\hline
\texttt{(S.6)} & selected & Multinomial & \texttt{TRUE} &\texttt{FALSE} & 0.092 & 0.082 & -0.087 & -0.447& \textcolor{LightSalmon1}{\rule{0.5cm}{1mm}} \\
\texttt{(S.7)} & selected & Multinomial & \texttt{FALSE} &\texttt{FALSE} & 0.093 & 0.085 & -0.104& -0.439& \textcolor{LightSalmon3}{\rule{0.5cm}{1mm}}\\  
\hline
\texttt{(S.8)} & selected & Bernoulli  & \texttt{TRUE} &\texttt{TRUE} & 0.015 & 0.028 & -0.079 & -0.150 & \textcolor{DarkOliveGreen1}{\rule{0.5cm}{1mm}}\\  
\texttt{(S.9)} & selected & Bernoulli  & \texttt{FALSE}  & \texttt{TRUE} & 0.016 & 0.033 & -0.097& -0.149 & \textcolor{DarkOliveGreen4}{\rule{0.5cm}{1mm}}\\ 
\hline
\hline
\end{tabular}
}
\end{center}
\begin{tablenotes}
\item[]\footnotesize Note: Summary of the modeling and sampling scenarios evaluated in the simulation study. The scoring metrics presented here are averages differences relative to \texttt{(S.0)}, the best-case scenario.
\end{tablenotes}
\end{table}


\noindent We take inspiration from Leemann \& Wasserfallen \cite{leemann2017extending} to calibrate the simulation. We simulate a population of size $N = 1,000,000$, and we focus on $J = 3$ options. We explore samples of size $n \in [100,\dots 10,000]$. The simulated data broadly follows the DGP described in Section \ref{bayes}. Like \cite{leemann2017extending} we use arbitrary cutoffs to discretise a set of correlated individual-level covariates $X$. Individuals are assigned to areas according to a Dirichlet-Multinomial process to explore both even and uneven population distributions across areas. Spatially correlated effects $\bm{\psi}_j$ are sampled from a Spatial Autoregressive (SAR) DGP. 
We exaggerate the degree of spatial auto-correlation by simulating $1,000$ SAR parameters and selecting the combination that gives the highest Moran I in each simulation round. This is then mixed with $\bm{\phi}_j$ according to a mixing parameter $\xi_j$, as per the BYM2 model. The area-level effect $\beta_j$ is assigned a uniform prior to explore a range of contextual variables' effect sizes. We simulate a total of $150$ populations, generating performance scores for $450$ choices in total ($J = 3$ for every simulation). The simulation study's DGP follows: 

\begin{small}
\begin{align*}
&\pi_{ij} = \mbox{Softmax}(\alpha_j + \gamma^1_{u^1[i]j} + \gamma^2_{u^2[i]j} +\gamma^3_{u^3[i]j} + \gamma^{\Lambda}_{l[i]j} + \beta_j z_{l[i]});\\
&\alpha_j \sim  N(0,1);\hspace{45pt}\gamma^k_{u^kj}  \sim N(0,1);\hspace{45pt}
\beta_j \sim  \mbox{Unif}\left(-1,1\right);\hspace{45pt}z_l \sim  N(0,1);\\
&
\bm{\gamma}^{\Lambda}_{j} \overset{\text{BYM2}}{\sim}  \left(
\bm{\nu},\bm{\phi}_j,\bm{\psi}_j,\xi_j\right);\hspace{42pt}
\bm{\phi}_j \sim N(0,1);\hspace{42pt}
\bm{\psi}_j \sim \mbox{SAR}  ;\hspace{42pt}
\xi_j \sim  \mbox{Unif}(0,1);\\
&u^k_{i} = \begin{cases}
1 & x_{ik}< -1\\
2 & -1 \geq x_{ik} < 0\\
3 & 0 \geq x_{ik} < 1\\
4 & x_{ik}>1
\end{cases};\hspace{37.5pt}\begin{pmatrix}x_{i1}
\\
x_{i2}\\
x_{i3}
\end{pmatrix} \sim  N
\begin{bmatrix}
\begin{pmatrix}
0\\
0\\
0
\end{pmatrix},
\begin{pmatrix}
1 & \rho^x & \rho^x\\
\rho^x& 1 & \rho^x\\
\rho^x & \rho^x & 1
\end{pmatrix}
\end{bmatrix};
\hspace{15pt}\rho^x \sim \mbox{Unif}(0,1);\\
& l_i \sim  \mbox{Dirichlet-Multinomial}\left(n = 1,\alpha_1 =1 ,\dots,\alpha_{51} = 1 \right).\\
\end{align*}
\end{small}

\noindent Figure \ref{sims.comparison.theta} presents a performance comparison of every model and sampling strategy combination (on the x-axis) against \texttt{(S.0)} (on the y-axis). Figure \ref{sims.comparison.pi} presents the same comparison for the estimation of cell-level probabilities $\bm{\pi}_j$. Our simulation study suggests the following: \emph{i. \textbf{likelihood}}: for the estimation of the stratified preferences $\bm{\theta}_j$ there are no substantial differences between models using Bernoulli and Multinomial likelihoods. On the other hand, at the cell-level, the best Multinomial estimates of $\bm{\pi}_j$ under random sampling have lower average RMSE ($-0.017$), somewhat greater correlation ($+0.021$) and significantly greater coverage ($+0.143$); \emph{ii. \textbf{structure}}: across metrics, at the stratified and cell levels, structured models systematically outperform unstructured ones. However, the gains from structured priors for the DGP under consideration appear extraordinarily minor. The metric most affected by structure appears to be correlation, where structured models provide an increase in correlation around $+0.01$; \emph{iii.  \textbf{bias-correction}}: bias-corrected models afford unequivocal advantages under online-selection. Looking at estimates of the stratified preferences $\bm{\theta}_j$, the best bias-corrected model under online selection \texttt{(S.8)}
alleviates the absolute bias of the best performing uncorrected alternative \texttt{(S.6)}
by $-0.077$; the RMSE falls by $-0.054$; the correlation increases $+0.008$ and the coverage shoots up $+0.297$. In estimates of $\bm{\pi}_j$ at the cell level, we see similar improvements in bias ($-0.074$), RMSE ($-0.044$), correlation ($+0.009$) and coverage ($+0.16$).\\

\begin{figure}
    \caption{Effect of sample size $n$ on estimation performance for $\bm{\theta}_j$.}
  \label{sims.properties.theta.n}
    \includegraphics[width = \textwidth]{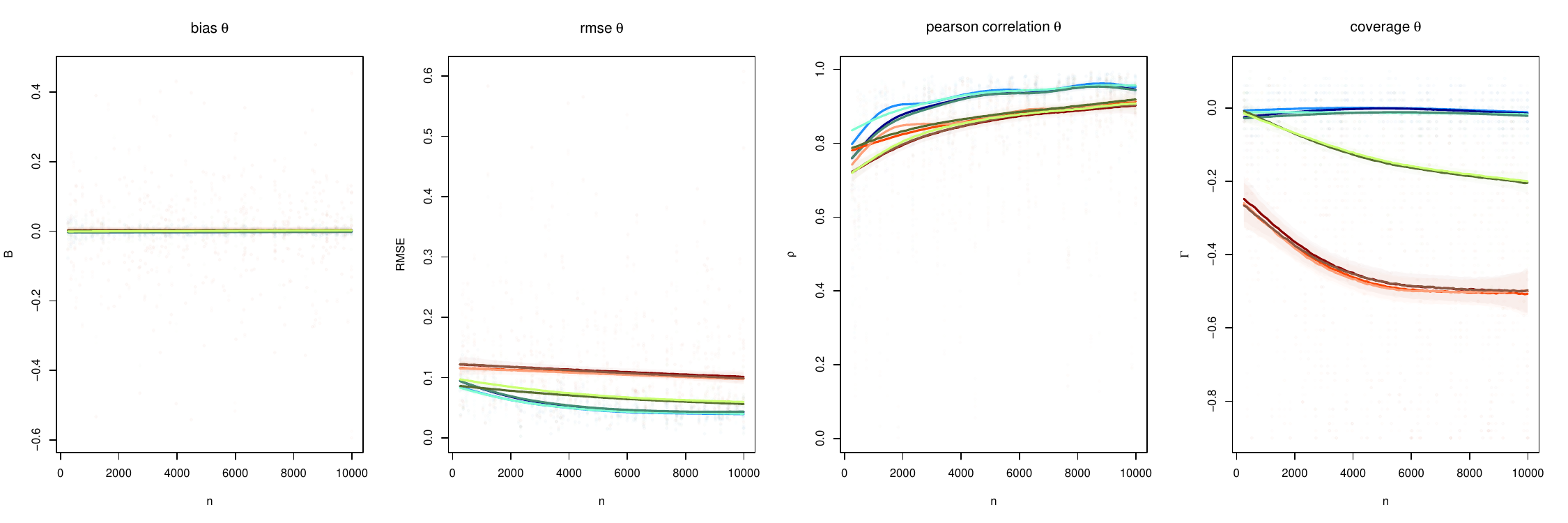}
\begin{tablenotes}
\item[]\footnotesize Note: See Table \ref{simulation_scenarios} or Figure \ref{sims.comparison.theta} for colour coding. 
\end{tablenotes}
\end{figure}
\begin{figure}
    \caption{Effect of population prevalence $\pi$ on estimation performance for $\bm{\theta}_j$.} \label{sims.properties.theta.prevalence.pi}
    \includegraphics[width = \textwidth]{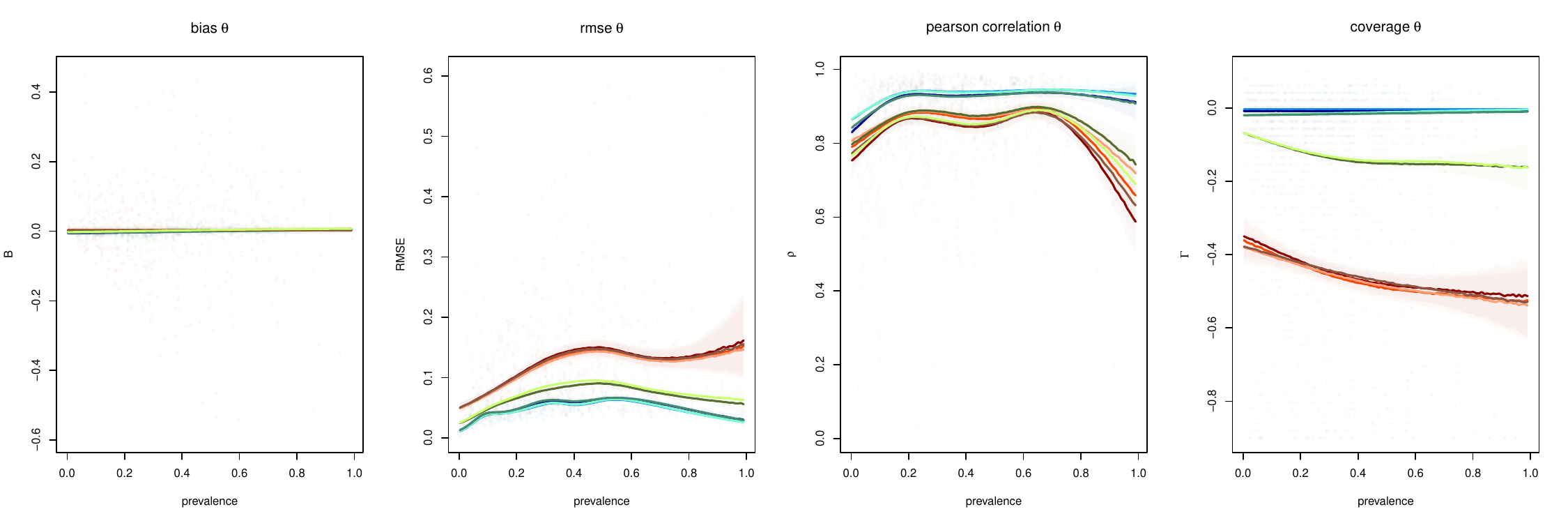}
\begin{tablenotes}
\item[]\footnotesize Note: See Table \ref{simulation_scenarios} or Figure \ref{sims.comparison.theta} for colour coding.
    \end{tablenotes}
\end{figure}
\begin{figure}
        \caption{Effect of online selection penalty $\mu^\Upsilon_j$ on estimation performance for $\bm{\theta}_j$.} \label{sims.properties.theta.penalty}
    \includegraphics[width = \textwidth]{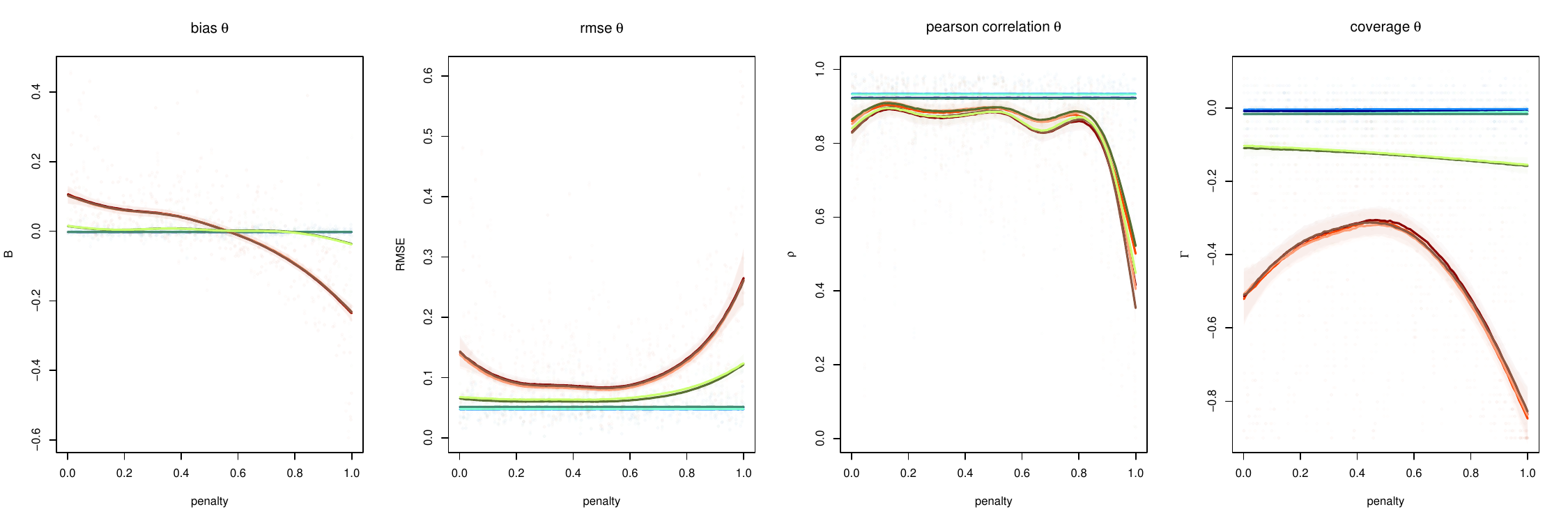}
    \begin{tablenotes}
    \item[]\footnotesize Note: See Table \ref{simulation_scenarios} or Figure \ref{sims.comparison.theta} for colour coding.
    \end{tablenotes}
\end{figure}
\begin{figure}
        \caption{Effect of sample bias $(\hat{\pi}_j - \pi_j)$ on estimation performance for $\bm{\theta}_j$.} \label{sims.properties.theta.sample.prevalence.bias}
    \includegraphics[width = \textwidth]{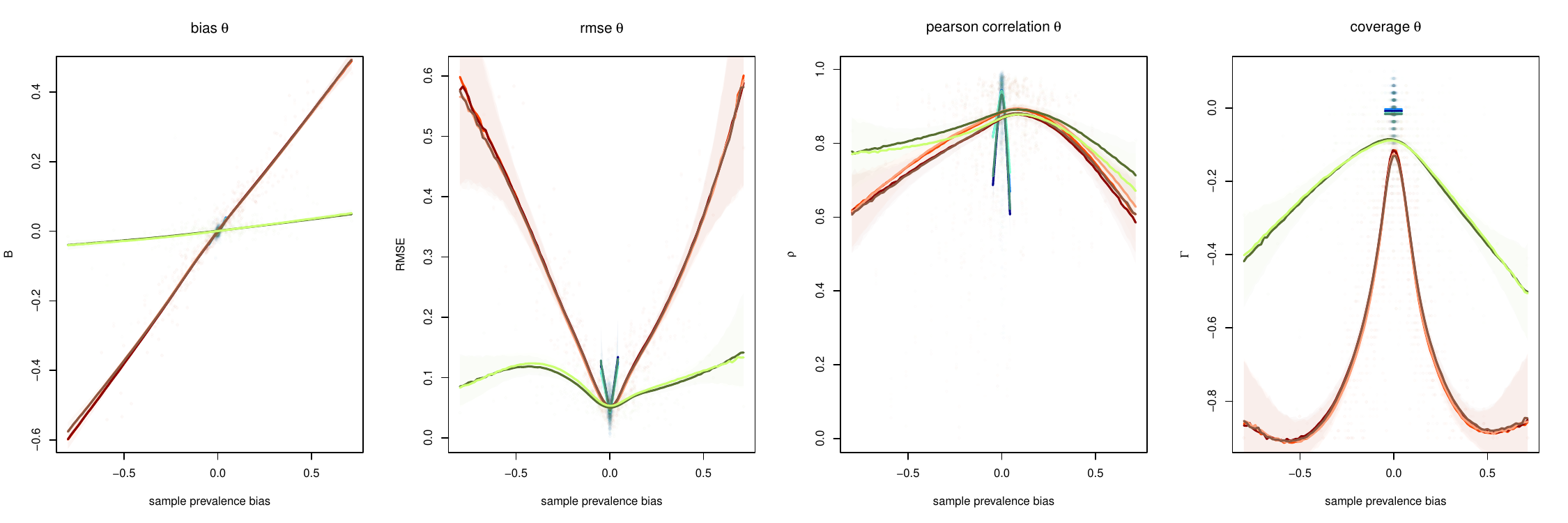}
    \begin{tablenotes}
    \item[]\footnotesize Note: See Table \ref{simulation_scenarios} or Figure \ref{sims.comparison.theta} for colour coding.
    \end{tablenotes}
\end{figure}

\noindent We further explore how the performance of each estimation strategy responds to: i. changes in sample size $n$; ii. changes in population-prevalence $\pi_j$; iii. changes in the central tendency of the online selection penalty $\mu^\Upsilon_j$; iv. changes in the size of the severity of sample-prevalence-bias (the difference between population-prevalence $\pi_j$ and sample-prevalence $\bar{\pi}_j$). Figures \ref{sims.properties.theta.n} to \ref{sims.properties.theta.sample.prevalence.bias} present the distribution of each of the scoring metrics for each of the four stimuli, as it pertains to the estimation of $\bm{\theta}_j$. Similar plots relevant to the estimation of the cell-level probabilities $\bm{\pi}_j$ are available in Figures  \ref{sims.properties.pi.n} to  \ref{sims.properties.pi.sample.prevalence.bias}. A detailed description of these figures is available in the Appendix. \\

\noindent We can summarise the findings of our analysis as follows: i.(Figures \ref{sims.properties.theta.n} and \ref{sims.properties.pi.n}): Returns of additional samples reach a quasi-plateau around $n=8,000$ for random-selection, whilst no clear plateau is reached for non-random samples. There is evidence for decreasing returns at the limit of the studies samples sizes ($n \approx 10,000$). Still, more $n$ appears to be better for reducing RMSE and increasing correlation in estimates from non-random samples. Large sample-sizes tend to degrade the coverage of models trained on non-random samples. Bias-corrected models appear more robust at any sample size, and reach similar levels of RMSE as random samples at large sample sizes; ii.(Figures \ref{sims.properties.theta.prevalence.pi} and \ref{sims.properties.pi.prevalence.pi}): RMSE is minimised at lower prevalence levels, whilst correlation plateaus at $\pi_j \approx 0.2$, and degrades beyond  $\pi_j \approx 0.7$. Bias-corrected models appear robust to any prevalence levels, and behave similarly to random-samples throughout; iii.(Figures \ref{sims.properties.theta.penalty} and \ref{sims.properties.pi.penalty}): increases in the central selection penalty $\mu^{\Upsilon}_j$ substantially degrade performance for non-random samples. If a party is relatively under-selected ($\mu^{\Upsilon}_j < \bar{\mu^{\Upsilon}}$) we induce positive-bias to structured MrP estimates, whilst over-selection brings about a more severe negative-bias. These translate into degradation of RMSE, correlation and coverage. Bias-correction makes these models extremely robust to central-selection pressure, meaningfully deviating from random-sampling performance for bias, RMSE and correlation only beyond $\mu^{\Upsilon}_j \approx 0.7$. This is a massive level of selection which suggests less than $30\%$ of individuals for a given party are eligible for selection into the subject pool, on average. Under-coverage remains an issue, though it is massively alleviated by bias-correction relative to uncorrected models; iv.(Figures \ref{sims.properties.theta.sample.prevalence.bias} and \ref{sims.properties.pi.sample.prevalence.bias}): sample prevalence bias is a major driver or poor performance. Uncorrected models suffer dramatically from sample prevalence bias in all metrics. Bias-correction generates robust estimates which appear to perform at similar levels as random samples for moderate levels of bias, and are relatively robust at any level of bias. \\

\noindent We conducted a simulation study to assess the performance of a bias-corrected model under online selection against a series of scenarios outlined in Table \ref{simulation_scenarios}. We examined performance across four metrics: bias, RMSE, pearson correlation and coverage of the $90\%$ prediction interval. Our simulation study ultimately shows that, under severely unrepresentative samples, uncorrected MrP is liable to fail on every metric. Our findings further suggest that performance of our bias-corrected model under online selection is comparable to that of an uncorrected model under random sampling for a broad range of plausible scenarios. 
Based on the simulation study, we produce the following recommendations: 1.) When interested in cell-level estimates of preferences, use a Multinomial likelihood. Otherwise, a series of Bernoulli models is preferable given computational considerations. 2.) Structured priors are preferable to unstructured ones, though the gains from these are very minimal compared to other modeling choices. 3.) When the true population prevalence $\pi_j$ is available, there is no reason to perform uncorrected MrP to estimate cell- or stratified-level quantities - we always recommend to implement the King \& Zeng prior correction. 4.) With respect to sample size, we confirm that more is better, though we caution that after $n > 10,000$ the gains in terms of RMSE and correlation are minimal, and there is an associated loss of coverage for selected samples. 


\section{Structuring Digital Traces}\label{data}
Having built a solid theoretical foundation for making close-to-representative inference under online selection, we now turn to data collection. \emph{digital trace} data are a cheap and plentiful alternative to random-digit-dial surveys. We define digital trace data as unobtrusively measured information belonging to a real-life person that can be found online. A rich corpus of digital traces can be obtained from social media companies such as Twitter or \texttt{reddit} via publicly available APIs \cite{barrie2021academictwitter,baumgartner2020pushshift}. We focus in this paper on data obtained via the Twitter streaming API, though the methodology proposed here is general to any social medium.\\

\noindent Digital trace data from social media contains signals about public opinion and voting preferences \cite{cerina2020measuring}, partisanship \cite{barbera2015birds,hemphill2016polar}, demographics \cite{wang2019demographic}, geographics \cite{stock2018mining} and other individual-level information. Social media data is unrepresentative \cite{mellon2017twitter,tufekci2014big}, and selection onto different platforms is often directly dependent on political preferences \cite{jasser2023welcome}. We can think of social media data as an imperfect online panel \cite{diaz2016online} constructed of largely unstructured data. Our challenge is to structure this data into a survey object, amenable to analysis via the inferential framework outlined above.\\ 

\noindent In this section we present a procedure to extract survey-like information from social-media data via artificial or human intelligence. We also present the alternative samples we use in this paper to validate and compare our feature-extraction and estimation strategies. Table \ref{sample_selection} presents the $6$ samples we use in our analysis.  

\begin{landscape}
\begin{table}[htp]
\flushleft
\scalebox{0.65}{
\begin{tabular}{l|llllc|cccc}
\hline
\hline
 \emph{Sample} & \emph{Subject-pool} & \begin{tabular}{@{}l@{}} \emph{Selection} \\ \emph{Mechanism} \end{tabular}  & \begin{tabular}{@{}l@{}} \emph{Interview} \\ \emph{Mode} \end{tabular} & \begin{tabular}{@{}l@{}} \emph{Interview} \\ \emph{Language} \end{tabular} & \begin{tabular}{@{}l@{}} \emph{Dates} \\ \emph{Range} \end{tabular} &$n$ & \begin{tabular}{@{}l@{}}\emph{Response} \\ \emph{rate (\%)}  \end{tabular} & \begin{tabular}{@{}c@{}} \emph{Cost}\\
 \emph{(estimated} \\
 \emph{per respondent)} \end{tabular}\\
 
 \hline
 \hline
 
 ANES & \begin{tabular}{@{}l@{}}citizens \\ age$>18$ \\ non-institutional \end{tabular} & 
 \begin{tabular}{@{}l@{}} random sampling \\ households \\ physical letters + email \end{tabular} & 
 \begin{tabular}{@{}l@{}}
 online survey \\
 telephone - CATI \\
 video - CAVI
 \end{tabular} 
 & 
  \begin{tabular}{@{}l@{}}
 english\\
 spanish
  \end{tabular} & 
  $[$\texttt{18/08}\mbox{ , }\texttt{3/11}$]$ 
  & $8,280$ & $36.7 $ & $c>\$70$ \\

 \hline 
 
 ABC \& WaPo & 
 \begin{tabular}{@{}l@{}}citizens \\ age$>18$ \\ non-institutional \\ registered voters \\ battle-ground states\tablefootnote{The ABC / Washington Post samples comprise of independent \emph{random digit dial} polls conducted in Arizona Florida, Michigan, Minnesota, North Carolina, Pennsylvania and Wisconsin.} \end{tabular}
&\begin{tabular}{@{}l@{}} random sampling\\ telephone numbers \end{tabular}  & telephone - CATI &\begin{tabular}{@{}l@{}}
 english\\
 spanish
  \end{tabular} & 
    $[$\texttt{8/09}\mbox{ , }\texttt{29/10}$]$ 
  & $8,933$ & $\sim 3$ \tablefootnote{The response rate of the ABC \& WaPo polls, identified with $id:\#1216$, is not known. The pollster's sample dispositions tables (\url{https://docs.google.com/spreadsheets/d/1wGoDaDPeEiy2xsDLOi4MSG1uLyGtyRvOjDv56OLz0GY/edit\#gid=183034761}) provide response rates for other pre-election polls, such as $\#$1215 and $\#$1217. The average response rate ($RR1$ from the accepted standards \cite{definitions2023final}) appears to be stable across polls.} & $ \$5 \leq c \leq \$25$ \tablefootnote{Costs were not reported by ABC \& WaPo. The estimate is based on the authors' experience with costs associated with CATI interviews, as well as the pollster's sampling design.}\\

\hline 
 
AMT &  workers & \begin{tabular}{@{}l@{}}self-selection\\ AMT marketplace\end{tabular} & online survey & english & 
\begin{tabular}{@{}l@{}} $[$\texttt{4/10}\mbox{ , }\texttt{2/11}$]$ \end{tabular}
& $2,492$ & / & $ \$ 1.5 $ \tablefootnote{\label{amt_sampling_design}We collect responses from AMT workers and for Twitter users using the same survey instrument. Our budget is $ \sim \$ 10,000$. Each worker was paid $ \$1.25 $  to complete a task, and from each task we obtained $2$ surveys: a) the worker's self-reported preferences and demographics; b) the worker's labels for a given Twitter user. We retain a single response per worker to avoid ballot-stuffing bias. We throw away a number of responses due to data-quality issues evidenced by attention- and bot- checks. Ultimately we collected unique and high-quality responses from $2,492$ unique AMT workers, and for $4,293$ Twitter users. The cost per response converges to around $\$1.5$}  \\

\hline
 Twitter - Human & \begin{tabular}{@{}l@{}} Twitter users \\
 mention \emph{Biden} or \emph{Trump}\\
 posted $\geq 10$ tweets \\
 tweet language is english
 \end{tabular}&  
 \begin{tabular}{@{}l@{}} 
 random sampling \\
 human labeling
 \end{tabular} & \begin{tabular}{@{}l@{}} 
 self-reported description \\
 posted tweets \\
 Twitter meta-data \end{tabular} & english &
\begin{tabular}{@{}l@{}} $[$\texttt{24/07}\mbox{ , }\texttt{2/11}$]$ \end{tabular} & $4,293$ & / & $ \$ 1.5 $ \footref{amt_sampling_design}  \\
\hline

 Twitter - GPT + Large Prompt & \begin{tabular}{@{}l@{}} Twitter users \\
 mention \emph{Biden} or \emph{Trump}\\
 posted $\geq 10$ tweets \\
tweet language is english
 \end{tabular}&  
 \begin{tabular}{@{}l@{}} 
 random sampling \\
 \texttt{gpt-3.5-turbo} labeling
 \end{tabular} & \begin{tabular}{@{}l@{}} 
 self-reported description \\
 posted tweets \\
 Twitter meta-data 
 \end{tabular} & english & 
 $[$\texttt{24/07}\mbox{ , }\texttt{2/11}$]$
 & $4,453$ & / & $ c< \$ 0.01 $   \\
\hline

 Twitter - GPT + Small Prompt &\begin{tabular}{@{}l@{}} Twitter users \\
 mention \emph{Biden} or \emph{Trump}\\
 posted $\geq 5$ tweets \\
tweet language is english
 \end{tabular}&  
 \begin{tabular}{@{}l@{}} 
 random sampling \\
 \texttt{gpt-3.5-turbo} labeling
 \end{tabular} & \begin{tabular}{@{}l@{}} 
 self-reported description \\
 posted tweets \\
 Twitter meta-data 
 \end{tabular} & english & 
  $[$\texttt{03/10}\mbox{ , }\texttt{3/11}$]$
 & $29,276$ &/ & $ c< \$ 0.005 $  \\
 \hline
 \hline
\end{tabular}
}
\caption{Description of the Study Samples.}
\label{sample_selection}
\end{table}
\end{landscape}

\subsection{Twitter Corpus}
Data collection from Twitter was performed with the \texttt{rtweet} package \cite{kearney2019rtweet}. We stream tweets from July $24^{th}$ $2020$ to election-day November $3^{rd}$ $2020$. The earliest tweet in our pool is dated \texttt{2020-07-16} at \texttt{01:42:46 UTC}, whilst the latest was created on \texttt{2020-11-03} at \texttt{00:12:55 UTC}. Our query returns a sample of tweets containing the words `\emph{Biden}' or `\emph{Trump}'. We make use of a congressional districts map to specify a point-radius search around the centroid of each district, to ensure good geographic coverage. We screen-out non-english tweets to limit the amount of non-US-residents in our pool. This curated corpus contains $492,539$ unique users, responsible for $3,019,184$ tweets. 

\subsection{Feature Extraction via LLMs}
We use LLMs to generate surveys from social media data. 
The LLM of choice is \texttt{gpt-3.5-turbo} from OpenAI. In constructing our online digital trace sample, we rely on \texttt{gpt-3.5-turbo} for two feature extraction tasks that are very much a strength of LLMs. First, with limited contextual information, \texttt{gpt-3.5-turbo} generates socio-demographic profiles of individuals.  This is a simple categorization exercise that requires no demanding causal reasoning or directed exploration where GPT has performed less well \cite{binz2023using}. Second, there is considerable evidence that \texttt{gpt-3.5-turbo} performs as well, if not better, than humans in characterizing the partisanship of Twitter content \cite{tornberg2023chatgpt,ornstein2022train,gilardi2023chatgpt}.   \\

\noindent An important distinction must be made with the work of Argyle et al. \cite{argyle2023out}. In Argyle, the generative quality of the LLM model is of interest - the model is asked to generate new survey-like samples after conditioning. The authors argue that with proper conditioning the machine can generate close-to-random samples of populations of interest. We propose a very different approach: We advocate using LLMs to annotate existing high-frequency, unrepresentative, cheap and unstructured samples. Relying on the machine to generate a reasonably representative synthetic sample requires an understanding of the DGP of LLMs which, under the current paradigm, is unknowable: there is a total lack of explainability \cite{gunning2019xai}, despite recent advances \cite{bills2023language}. Its not clear that relying on conditioning of the kind outlined in Argyle is sufficient to ensure representative inference. The issue of time-relevance is paramount in pre-election opinion polling: the samples need to contain valuable information about the daily changes in preferences across population categories. While it is possible to correct for selection biases to a degree by conditioning the LLM with the aid of exogenous representative samples (e.g. the ANES), it is not possible to have access to equally relevant and representative samples in the run-up to an election. The point of opinion-polling is to collect such samples. Our approach is able to capture changes over time by providing regular, timely, high-frequency, unobtrusively observed expressions of preferences. \\

\noindent Figures \ref{location.prompt} and \ref{demo.prompt} present the prompts provided to \texttt{gpt-3.5-turbo}, along with a sample of answers. The LLM was asked to classify users according to the survey-like categories presented in Figure \ref{demos}. We set \texttt{temperature = 0} to ensure relatively stable classification. For each Twitter user we provide the LLM with two prompts: the first to extract the location of the user, and the second to extract the socio-demographic and political characteristics. The reason for these prompts to be separate has to do with the nature of Twitter self-reported location data and the importance of a clean location signal for MrP. The self-reported location of the user tends to be reasonably low in noise, so it makes sense to extract this in a separate prompt. Including it in the socio-demographic prompt could increase the level of noise in the response. Moreover the formatting of the prompt for location relies on the knowledge of the LLM regarding what a `State' in the US is, as well as what a `Territory' is. By using this inherent knowledge without having to specify $50$ states + DC as separate options in the prompt, as we would have to in the $2^{nd}$ prompt's format, we save on valuable tokens. \\

\begin{figure}
    \begin{framed}
    \textbf{Prompt:}\\
    \texttt{paste(}\\
    `A person writes their location in their bio as follows:$\ll$',\texttt{users\$location[i]},`$\gg$.\\
    Which state in the US do they live in ?\\
    For this answer consider Washington DC and other Territories of the US as states.\\
    Write out just the full name of the state, and if not from US, write ``Not from US"'.
    \texttt{)}
    \end{framed}
    \begin{framed}
        \textbf{Answer:}\\
 $[1]$ `Pennsylvania'\\                                                                                                                                                                      
 $[2]$ `Not enough information is provided to determine which state the person lives in. They simply stated their location as ``USA''.'\\                                                    
 $[3]$ `Minnesota' \\                                                                                                                                                                        
 $[4]$ `Texas' \\                                                                                                                                                                       
 $[5]$ `Indiana' $\cdots$

    \end{framed}
    \caption{Example of the location-field prompt, followed by a sample of $5$ answers. The object \texttt{user\$location[i]} is passed to the prompt for the $i^{th}$ user in the sample.}
    \label{location.prompt}
\end{figure}

\noindent For the second prompt, notice the line of code randomising a set of categories (\texttt{demos \_string}) across which we want the LLM to classify users. We randomise to account for the auto-regressive nature of LLMs \cite{lecun2023large}. The category - and the LLM's classification of the user within that category - that is placed earlier in the prompt has an effect downstream to the future classifications. We presume, for example, that if the LLM generates output classifying a given user as `highly educated', it may be more likely to classify the same user as a Democrat, regardless of the content of the Tweets. This is in virtue of the probability of the word `Democrat' following the words `highly educated' being more likely than the word `Republican', again independently of the content of the tweets \cite{wolfram2023chatgpt}. By randomising the categories across which we wish to classify users in the prompt-order, we avoid systematic bias due to auto-regressive behaviour in the full annotated sample.\\

\begin{figure}
    \begin{framed}
    \textbf{Prompt:}\\
    \texttt{paste(}\\
    `A person has in their Twitter bio the following information:\\
    $\ll$',\texttt{users\$bio[i]},`$\gg$ ;\\
    Further, they have written the following tweets:\\
    $\ll$',\texttt{users\$text[i]},`$\gg$.\\
    I will now show you a number of categories to which this user may belong.\\
    The categories are preceded by a header (e.g. ``AGE:'' or ``SEX:'' etc.) and an identifier (e.g. ``A1'', ``A2'' or ``E2'' etc.).
    Please select, for each header, the most likely category to which this user belongs to.\\
    In your answer present, for each header, the selected identifier.',
    \texttt{paste0(sample(demos\_string,replace = FALSE),collapse = `$\backslash$n')\\
    )}.
    \end{framed}
    \begin{framed}
        \textbf{Answer:}\\
        \begin{footnotesize}
        HIGHEST EDUCATIONAL QUALIFICATION: Q2) completed high-school but did not go to college\\
        THIS INDIVIDUAL IS REGISTERED AS: R2) a Democrat\\
        ETHNICITY: E1) White\\
        2016 US PRESIDENTIAL ELECTION VOTE: L3) voted for Hillary Clinton, the Democrat candidate\\
        SEX: S1) Male\\
        AGE: A5) 45 to 54 years old\\
        MARITAL STATUS: M1) Married\\
        2020 US PRESIDENTIAL ELECTION VOTE: V3) voted for Joe Biden, the Democrat candidate\\
        2018 MIDTERM ELECTION VOTE: T3) voted for the Democratic Party\\
        HOUSEHOLD INCOME BRACKET: H5) more than $100000$ USD per year $\cdots$
        \end{footnotesize}
    \end{framed}
    \caption{Example of the socio-demographic and political characteristics prompt, followed by a sample answer. The object \texttt{users\$bio[i]} is passed to the prompt for the $i^{th}$ user in the sample, and contains the location information of the user, their self-reported description and their screen-name. The object \texttt{users\$text[user]} is also looped-over, and provides the LLM with a list of recent tweets produced by the user. Finally, the \texttt{R} code line \texttt{paste0(sample(demos\_string,replace = FALSE),collapse = `$\backslash$n')} samples attributes in random order from the object \texttt{demos\_string}, which contains all the combinations of headers and identifiers. Every user is classified into exclusively one option for every category.}
    \label{demo.prompt}
\end{figure}

\noindent We ask the LLMs to annotate two sets of samples. \emph{Sample (1)} was generated dynamically during the election campaign. Users from our evolving Twitter corpus were selected daily starting on October $1^{st}$ $2020$ and ending November $1^{st}$. On a given day during the campaign a subset of users was sampled at random from the corpus, conditional on having tweeted at least $10$ times up to that point. A total of $4,590$ users from this sample were fed to the LLM. \emph{Sample (2)} consists of every user from our corpus who: a) tweeted for a total of $5$  times of more during the campaign; b) produced at least $1$ tweet in the last month of the campaign. This second sample, which spans $30,154$ unique users, serves to examine the effect of decreasing the amount of context available to the LLM in favour of increasing sample size. 

\subsection{Feature Extraction via Amazon Mechanical Turks}
To validate our LLM classification we use Amazon Mechanical Turk workers to annotate  \emph{Sample (1)}. We use a simple survey instrument coded in \texttt{html} on the AMT platform to present the workers with user information. Much like the LLM, workers have access to the last $10$ tweets generated by the user, the user's name, location and description. Beyond this information, workers also see the dates on which the tweets were posted, which device the tweets were generated on, the users' profile and background pictures, and whether the account was verified. We implement attention-checks to control data-quality, as well as to screen-out bots and VPN workers falsifying their location \cite{kennedy2020shape}. A total of $2,492$ unique workers participated in the human intelligence task.


\subsection{Benchmarking}

To put the performance of bias-corrected structured MrP in context, we fit the model to a series of reference datasets. We use the $2020$ American National Election Study (ANES) time-series sample as the `optimal' scenario - namely a close-to-random, high-quality sample. Poor performance of the model on this dataset would indicate serious issues with the modeling framework. To measure the gains obtained via modeling, we further compare results against the raw, unweighted and unmodeled ANES data. In a separate scenario, we generate a new training set by augmenting the ANES with samples from swing-state rdd polls produced by ABC and The Washington Post. This makes the sample less representative, though it is common to over-sample swing-states in classic MrP applications \cite{lauderdale2019MRP,lauderdale2019constructing}, and hence it is of interest to measure performance under this combination. To compare performance against unrepresentative samples recruited via online surveys, as opposed to social-media data, we use a survey of Amazon Mechanical Turks. Survey responses were collected from Turks who contributed to the human-intelligence annotation task. Finally, 
we compare performance against the uniform swing model \cite{jackman2014predictive}. We use the true national swing from $2016$ to $2020$ to fit this model.

\section{Results}\label{results}

\begin{table}[!htb]
\caption{Summary of results for Democrats and Republicans.}
\label{results_summary_table}
\scalebox{0.575}{
\begin{tabular}{r|r||cccc|cc|ccc}
\hline
\hline
& & \multicolumn{4}{c|}{\emph{Non-probability Samples}}
&\multicolumn{2}{c|}{\emph{Quasi-probability Samples}} & \multicolumn{3}{c}{\emph{Other Benchmarks}}\\
\hline
& & 
\begin{tabular}{@{}c@{}} 
Twitter\\
\texttt{gpt-3.5-turbo}\\
(10 tweets)
 \end{tabular}
 & \begin{tabular}{@{}c@{}} 
Twitter\\
\texttt{gpt-3.5-turbo}\\
(5 tweets)
 \end{tabular} & \begin{tabular}{@{}c@{}} 
Twitter\\
Human
 \end{tabular}  &\begin{tabular}{@{}c@{}} 
Amazon\\
Mechanical\\
Turks
 \end{tabular} & ANES & 
 \begin{tabular}{@{}c@{}} 
 ANES +\\
 ABC \& WaPo
 \end{tabular}
 &  \begin{tabular}{@{}c@{}} 
 ANES \\
 `Raw'
 \end{tabular} & \begin{tabular}{@{}c@{}} 
Uniform \\
Swing
 \end{tabular} & \texttt{FiveThirtyEight}\\
 \hline
 \hline
\multirow{3}{*}{Bias} & \texttt{D}& 0.02 & 0.04 & 0.04 & \underline{\textbf{0}} & -0.01 & 0.01 & 0 & -0.01 & 0.03\\
&\texttt{R} & -0.03 & -0.05 & -0.05 & \underline{\textbf{0}} & 0 & -0.01 & -0.09 & 0 & -0.02 \\
&\texttt{R-D}& -0.05 & -0.09 & -0.09 & \underline{\textbf{0}} & 0 & -0.02 & -0.08 & 0.01 & -0.04 \\
  \hline
 \multirow{3}{*}{RMSE} & \texttt{D} &\textbf{ 0.03} & 0.05 & 0.05 & 0.04 & 0.02 & 0.01 & 0.07 & 0.02 & 0.03 \\
&\texttt{R}& \textbf{0.03}& 0.06 & 0.06 & 0.05 & 0.02 & 0.02 & 0.12 & 0.02 & 0.02 \\
&\texttt{R-D}& \textbf{0.06}& 0.1 & 0.11 & 0.09 & 0.03 & 0.03 & 0.17 & 0.03 & 0.05 \\
  \hline
\multirow{3}{*}{ \begin{tabular}{@{}r@{}} 
 Pearson \\
 Correlation
 \end{tabular}} &\texttt{D}& \underline{\textbf{0.99}} &  \underline{\textbf{0.99}} &0.97 & 0.93 & 0.99 & 0.99 & 0.87& 0.99 & 0.99\\
&\texttt{R}& \underline{\textbf{0.99}}& \underline{\textbf{0.99}} &0.97  & 0.93 & 0.99 & 0.99 & 0.75 & 0.98 & 0.99 \\
&\texttt{R-D}& \underline{\textbf{0.99}} & \underline{\textbf{0.99}} & 0.98 & 0.93 &0.99  & 0.99 & 0.83 & 0.99 & 0.99 \\
   \hline
 \multirow{3}{*}{\begin{tabular}{@{}r@{}} 
  Coverage\\
  $90\%$
 \end{tabular}} &\texttt{D}& 0.63 & 0.22 & 0.76 & \underline{\textbf{0.84}}& 0.96& 0.98 & / & / & 1 \\
&\texttt{R} & 0.71& 0.08 & 0.38 & \underline{\textbf{0.86}} & 0.98 & 0.94 & / & / & 1 \\
&\texttt{R-D} & 0.65 & 0.14 & 0.47  & \underline{\textbf{0.84}} & 1 & 0.98 & 0.83 & 0.99 & 0.88 \\
 \hline
 \multirow{2}{*}{ \begin{tabular}{@{}r@{}} 
 $R^2$ Non-Unif.\\
 Swing
 \end{tabular}} &\texttt{D}& \underline{\textbf{0.22}} & 0.18 & 0.13 & 0.01& 0.61 & 0.72&/ & 0 & 0.62 \\
&\texttt{R}& \underline{\textbf{0.45}} & 0.22 & 0.08 & 0.30 & 0.56 & 0.61 &  / & 0 & 0.61 \\
  \hline
  \hline
\end{tabular}
}
\begin{tablenotes}
\item[]\footnotesize Note: Summary of the main results for the two major parties.  For each metric we show in underlined bold the best performing non-representative sample.
\end{tablenotes}
\end{table}

Table \ref{results_summary_table} summarizes the accuracy of state-level election day vote-share predictions for six different sampling strategies and three alternative modeling benchmarks arranged over the columns. The rows of Table \ref{results_summary_table} are the different evaluation metrics.  Figure \ref{area_predictions} presents the election-day state-level predictions for each choice under consideration in the $2020$ US election (rows) by each model/data combination (columns). Figure  \ref{area_predictions_delta} summarises the ability of forecasts to capture the change since the last election. Figures \ref{area_predictions_538} and \ref{area_predictions_delta_538} present the same comparisons, though here Libertarians and Greens are aggregated into an `other' category to enable comparison with state-of-the-arts area-level polling aggregator \texttt{FiveThirtyEight}. Turnout is also omitted from this comparison as \texttt{FiveThirtyEight}'s calculation could not be sensibly reconciled with our approach. Figure \ref{nat_temporal_predictions} displays the national predictions over the course of the campaign. Below we highlight key results regarding state-level predictions and predicted campaign trends.

\subsection{Election-day State-level Predictions}

\noindent  \emph{(1). Bias-corrected structured MrP achieves state-of-the-arts performance on high-quality quasi-random samples}. Firstly, a sanity check against disaggregation: estimates of the Republican-Democrat (R-D) margin from the raw ANES data are significantly worse, with an RMSE close to $0.14$ points higher. This is caused by an anti-Republican bias of $0.09$ points. Training our model with the same ANES data sees a reduction of this bias to $0$. Secondly, a check on overall performance: again looking at the R-D margin, we witness the following performance across metrics for the bias-corrected model trained on ANES data: \{bias = 0, RMSE = 0.03, correlation = 0.99, coverage = 1\}. Comparing this against \texttt{FiveThirtyEight}'s forecasting model \{bias = -0.04, RMSE = 0.05, correlation = 0.99, coverage = 0.88\}, we can be confident our approach can compete with state-of-the-arts polling aggregators under optimal sampling. It's worth noting that a uniform-swing model using the true national swing also outperforms the \texttt{FiveThirtyEight} forecast. This is in line with existing literature about the predictive power of the uniform swing \cite{jackman2014predictive}, though we note it performs exceedingly well here largely due to the extremely high correlation between the $2016$ and $2020$ election.\\ 

\noindent \emph{(2). Bias-corrected structured MrP achieves satisfactory performance on all selected samples}. Estimates of the Democratic vote for selected samples are generally satisfactory, performing within the following ranges \{bias = [0,0.04], RMSE = [0.03,0.04], correlation = [0.93,0.99], coverage = [0.22,0.84]\}; Republican vote estimates present a similar performance range \{bias = [-0.05,0], RMSE = [0.03,0.06], correlation = [0.93,0.99], coverage = [0.08,0.86]\}; Turnout estimation performance is also of interest \{bias = [-0.07,0.02], RMSE = [0.02,0.08], correlation = [0.11,0.73], coverage = [0.45,0.88]\}. Third-party predictions generally have low bias and RMSE, high coverage, but also lower correlation compared to the two main parties. We observe the following patterns in results: 

a. Twitter data under-sampled Republicans, and over-sampled Democrats to a degree that was not entirely accounted for by the King \& Zeng bias-correction. This was true whether the data was annotated by humans or machine. Amazon Mechanical Turks on the other hand displayed no residual systematic bias; 

b. Performance on turnout for selected samples was variable: with the exception of the model trained on human-labeled Twitter users, turnout predictions were generally satisfactory, showing a maximum RMSE of $0.04$. The best performing model was that trained on the  small-context AI-labeled Twitter data, which achieved a correlation of $0.73$ - superior to a turnout model trained on ANES data. Human annotators tended to under-estimate the propensity for Twitter users to show-up on election day, generating a severe bias in the turnout estimates of $-0.07$; 

c. As predicted by the simulation study, coverage worsened as sample size increased on selected samples. The worst coverage was associated with the small-context AI-labeled data, which had the largest sample size ($n>30,000)$. The best coverage was associated with the Amazon Mechanical Turk survey, which had the smallest sample size ($n<3,000$); 

d. Performance on the R-D margin on selected samples tended to be worse due to compounding biases - \{bias = [-0.09,0], RMSE = [0.06,0.11], correlation = [0.93,0.99], coverage = [0.14,0.84]\} - though the best performing models were still in line with state-of-the-art MrP applications \cite{lauderdale2020model}.\\

\noindent \emph{(3). AI-annotated social-media surveys with large context (10 tweets) outperform other selected samples and achieve state-of-the-arts performance}. The performance of the AI-annotated polls with large-context on the R-D margin \{bias = -0.05, RMSE = 0.06, correlation = 0.99, coverage = 0.65 \} is extremely close to the \texttt{FiveThirtyEight} forecasting model performance, generally displaying similar levels of bias, RMSE and correlation per choice-model, and losing out primarily in coverage.  The small-context AI-annotated polls performed on the whole slightly worse, with compounding biases leading to a relatively large R-D RMSE of $0.1$. This provides some evidence as to the optimal prompting style for generating social-media polls using LLMs, in an MrP application of this sort. The human-labeled social-media polls performed worst out of the selected samples, with a R-D RMSE of $0.11$. The survey of Amazon Mechanical Turks has variable performance: a relative large RMSE of $0.09$ is paired with a completely unbiased prediction on the R-D margin. Given the unbiased nature of the estimates and relatively high coverage, it's possible that under larger sample sizes, the sample of Turk workers might have rivaled the performance of models trained on large-context AI-annotated social-media data. \\

\noindent \emph{(4). Bias-corrected models trained on AI-annotated social-media surveys can explain a substantial portion of the non-uniform swing for each party}. Figures \ref{area_predictions_delta} and \ref{area_predictions_delta_538} showcase the ability of the models to capture non-uniform swings from the $2016$ election. To get a sense of this, we square the correlation coefficients to gain an interpretation in terms of the proportion of the cross-states variance in swings explained by our models. The $R^2$ for the swing in Democratic vote across top-performing models are \{ gpt\_10\_tweets = $22.1\%$, ANES + WaPo = $72.3\%$,  \texttt{FiveThirtyEight} = $62.4\%$ \}; for the Republicans we have  \{ gpt\_10\_tweets = $42.3\%$, ANES + WaPo = $60.8\%$,  \texttt{FiveThirtyEight} = $60.8\%$ \}; values for third-parties are omitted as the correlations are generally close to $1$ for most models; for turnout we have  \{ gpt\_10\_tweets = $9.6\%$, gpt\_5\_tweets = $23.0\%$, ANES = $17.6\%$, ANES + WaPo = $7.3\%$ \}. Selected samples generally explain a lower portion of the non-uniform swing from $2016$ than random samples. These proportions are nonetheless substantial, and broadly justify a non-uniform modeling approach. Amongst selected samples, large-context AI-annotated social-media polls provide the best predictions for the Republican and Democratic swings, whilst the low-context AI annotations provide the better explanations for third-party and turnout swings.\\

\noindent The national-level election-day prediction of the vote from the best performing AI poll is in line with that produced by traditional polling aggregators, despite the AI being trained on extremely unrepresentative samples.

\subsection{National Campaign Trends}

\begin{landscape}

\begin{figure}
    \caption{Predictive performance at the national level over the course of the last $30$ days of the campaign.}
    \label{nat_temporal_predictions}
    \includegraphics[width = 750pt]{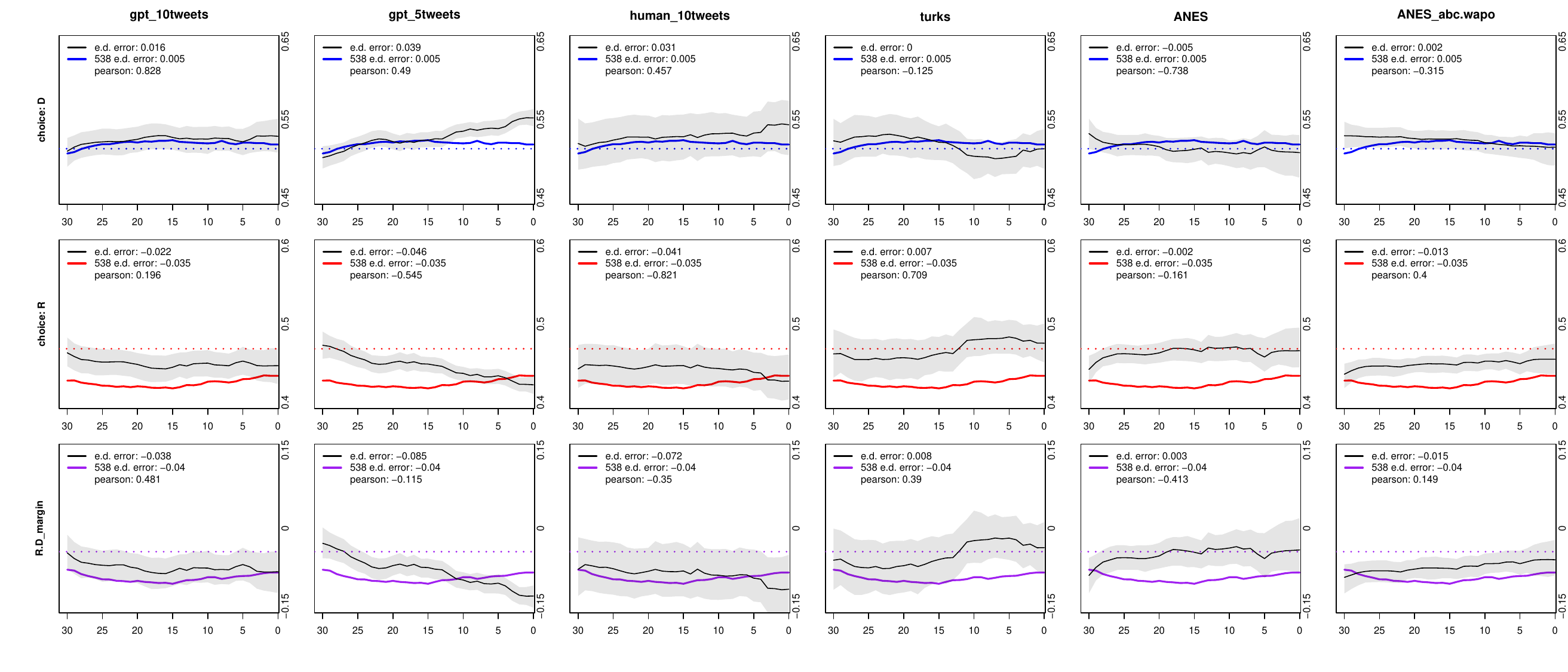}
    \begin{tablenotes}
    \item[]\footnotesize Note: Predictive performance at the national level over the course of the last $30$ days of the campaign. The finely dotted line in each plot represents the true election result. The model predictions are always in black, enclosed in a $90\%$ prediction interval. The pearson correlation coefficient between our point estimates and the \texttt{FiveThirtyEight} polling averages are presented in the legends.
    \end{tablenotes}
\end{figure}

\end{landscape}

\noindent \emph{Bias-corrected models trained on AI-annotated social-media surveys provide reasonable estimates of the national vote and campaign-trends}. If we are to use AI polls to monitor changes in support over the course of the election-campaign, these should show similar national trends in the last $30$ days of the election campaign as the \texttt{FiveThirtyEight} polling averages over the same period. Figure \ref{nat_temporal_predictions} presents this comparison.\\


\noindent Unlike election-day state-level vote shares, daily polling averages are uncertain quantities. For ease of analysis, here we take them as `truth', though there is generally substantial disagreements day-to-day swings in the vote across pollsters. The temporal variance in the last $30$ days of the campaign is extremely small by definition, hence metrics such as bias and RMSE are not very meaningful for cross-model comparisons. The Pearson correlation coefficient seems like a better metric, capturing the extent to which two pollsters can `order' the daily preferences. But here again it is easier to `order' quantities which are extremely different - such as the election-day state-level vote-shares, as opposed to quantities which are extremely similar - such as daily national estimates of voter preferences. Republicans win around $5\%$ of the vote in Washington D.C., and around $70\%$ of the vote in Wyoming - whilst their national vote varies from $41.8\%$ to $43.4\%$ in the last $30$ days in the campaign. These differences - though potentially very consequential, are very challenging to capture. Literature further suggests that, whilst campaign events matter, minor daily swings in voting preferences show a high degree of stochasticity \cite{wlezien2002timeline,linzer2013dynamic}. For all these reasons, we should not expect correlation levels comparable to the state-level election-day estimates, when it comes to comparing daily national swings estimates.\\

\noindent We are most interested in the ability of the model to replicate trends observable in traditional polls. The metric which captures this ability is the correlation coefficient. Large-prompt AI-annotated social-media polls perform well on this metric for the Democratic vote, with a correlation of $0.83$ with the \texttt{FiveThirtyEight}  average, whilst the pattern around the Republican vote is more muddled - with a relatively low correlation of $0.2$.

\pagebreak 

\newgeometry{left = 1.25cm,bottom = 0.5cm,top = 0.5cm}

\restoregeometry
\pagebreak 

\subsection{Human-Machine Disagreement}
We seek to understand the strong performance of the Artificially Intelligent Polls. The better performance of the \texttt{gpt-3.5-turbo} model, prompted with $10$ tweets, compared to the performance of the human-annotated sample is especially meaningful. The two samples deal with the same Twitter users. This would then suggest the annotations from the LLM were more useful to the MrP framework than those produced by humans. This does not necessarily mean the AI labels are more accurate. It is plausible that \texttt{gpt-3.5-turbo} annotates the sample of Twitter users to be more congruent with the election result. \texttt{gpt-3.5-turbo} was trained with data up-to September $2021$ \footnote{\url{https://platform.openai.com/docs/models/gpt-3-5}}, hence it is in some sense aware of the election result. It could therefore be applying some implicit raking \cite{fienberg1970iterative}, and generating individual-level labels which are more consistent with the marginal distribution of the $2020$ vote. Here we will show the degree, and patterns, of agreement/disagreement between humans and LLMs.

\subsubsection{Kippendorff's $\alpha$}

\begin{figure}[htp]
 \caption{Bootstrap distribution of Krippendorff's $\alpha$ for the annotated variables used in the model.}
        \label{kip_human_gpt10}        \includegraphics[width=\textwidth]{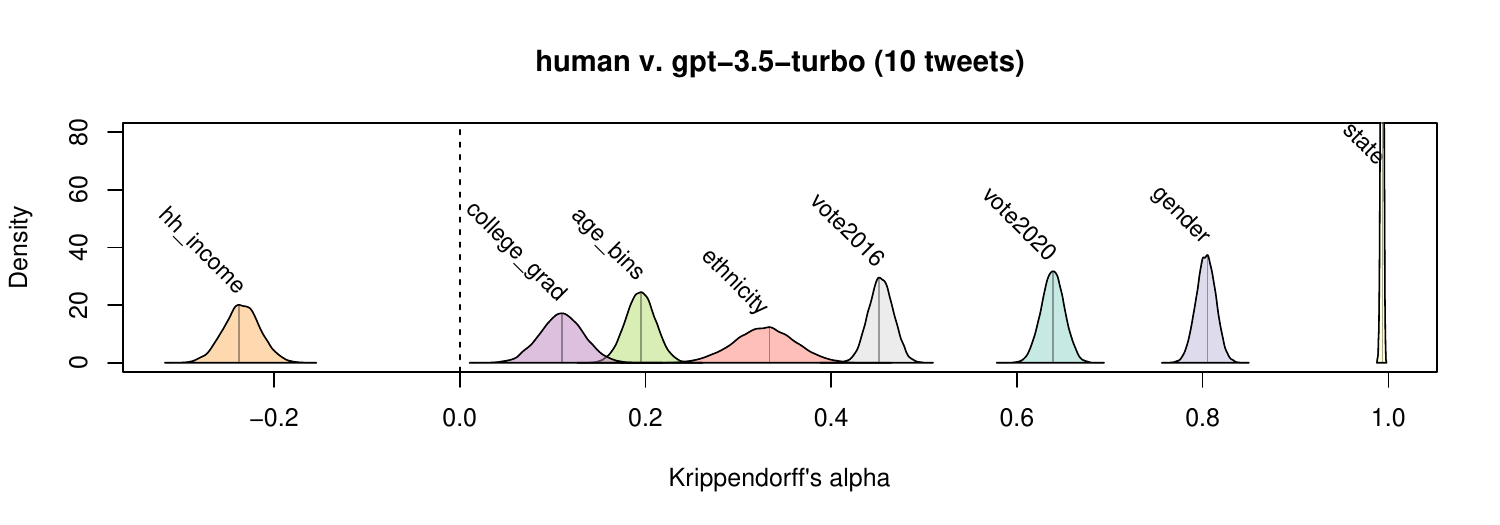}
\begin{tablenotes}
\item[]\footnotesize Note: Bootstrap distribution of Krippendorff's $\alpha$ for the annotated variables used in the model. The annotators being compared here are humans  and \texttt{gpt-3.5-turbo} with $10$-tweets-prompt.
\end{tablenotes}
     \end{figure}

\noindent Figure \ref{kip_human_gpt10} presents the bootstrap distribution of Krippendorff's $\alpha$ \cite{krippendorff2011computing} for annotation comparisons between humans and \texttt{gpt-3.5-turbo} with $10$-tweets-prompt.  Figures \ref{kip_human_gpt5} and \ref{kip_gpt10_gpt5} present similar comparisons for human versus \texttt{gpt-3.5-turbo} with $5$-tweets-prompt; and \texttt{gpt-3.5-turbo} with $10$-tweets-prompt versus \texttt{gpt-3.5-turbo} with $5$-tweets-prompt. Comparisons are provided for each relevant variable which was used in modeling (Table \ref{X_recode}). Krippendorff's $\alpha$ is a measure of inter-rater reliability relative to a `chance' level of agreement. It is calculated as $1 - \frac{\mbox{observed disagreement}}{\mbox{expected disagreement}}$. We use the package $\texttt{icr}$ \cite{staudt2023package} for efficient computation.\\

\noindent Krippendorff's $\alpha$ is the state-of-the-arts metric of global agreement across raters. Figure \ref{kip_human_gpt10} (and \ref{kip_human_gpt5}) shows statistically significant agreement between humans and the LLM across variables, with the exception of the variable `household income', despite performing the adjusted computation to account for its ordinal nature. The magnitude of the degree of agreement is highly variable across variables, with only the `$2020$ vote', `gender' and `state' touching Krippendorff's minimal arbitrary threshold of $\alpha = \frac{2}{3}$ for \emph{`the data under consideration [to be] at least similarly interpretable by other scholars (as represented by different coders)'} \cite{krippendorff2004reliability}. We should note here that this threshold is minimally relevant for our purposes: whilst Krippendorff's $\alpha$ tells us that there are substantial differences in how humans and LLMs annotate Twitter users' profiles, there is evidence that LLMs may provide better annotations than crowd-workers \cite{gilardi2023chatgpt}, and especially so in the realm of political data \cite{tornberg2023chatgpt}. Moreover, the context of  Krippendorff's quote is that of reliability data, whereby strict thresholds may make sense depending the consequences of disagreement. Finally, we note that the level of analysis of interest for us is the `aggregate' level - namely the cell-level or the stratified-level - and there are many sets of annotations that would be consistent with optimal performance at those levels.

\subsubsection{(Dis)Agreement Network}

\begin{table}[ht]
\caption{Rater-agreement matrix $A$ for the variable `2020 vote'.}
\label{agreement_matrix}
\centering
\begin{tabular}{r|r||rrrrr}
  \hline
    \hline
  & &\multicolumn{5}{l}{\texttt{gpt-3.5-turbo}} \\ 
   \hline
\parbox[t]{2mm}{\multirow{5}{*}{\rotatebox[origin=c]{90}{\texttt{human}}}} & & D & G & L & R & stay home \\ 
  \hline
    \hline
 & D & 1906 &   3 &   4 &  50 & 234 \\ 
 & G &   3 &   5 &   0 &   0 &   0 \\ 
 & L &   0 &   0 &   7 &   0 &   2 \\ 
 & R &  40 &   0 &   6 & 864 &  59 \\ 
 & stay home & 185 &   2 &   1 &  95 &  71 \\ 
   \hline
   \hline
\end{tabular}
\end{table}

\noindent We complement Krippendorf's $\alpha$ with a fine-grained analysis of the agreement per variable. We do not have access to an underlying `truth' for the characteristics of Twitter users, hence this analysis cannot confirm or disprove the absolute accuracy of the AI labels. We can however identify patterns of disagreement between AIs and humans. From these patterns we can assess the directions of bias affecting each annotated sample of Twitter users. This in turn can inform us on the reasons for the relative success of AI polls compared to relevant alternatives.  \\

\noindent We borrow from the network-science literature and treat contingency-tables of annotations for each rater-pair as generations from a bipartite network \cite{young2021reconstruction}. Table \ref{agreement_matrix} shows an example of the agreement matrix for the dependent variable `vote\_2020'. To each of these matrices we fit a Bayesian Poisson mixture-model of the following form:

\begin{align*}
&A_{ij} \sim  \mbox{Poisson}(\mu_{ij});\\
&\mbox{log}(\mu_{ij}) = \beta^0+ \beta^1_i + \beta^2_j + \mbox{log}(1 + rB_{ij});\\
& B_{ij} \sim \mbox{Bernoulli}(\pi_{ij}), \hspace{80pt} \pi_{ij} \sim  \mbox{Beta}\left(\frac{1}{2},\frac{1}{2}\right), \hspace{74pt} r \sim  \mbox{Exp}(0.01);\\
& \beta^0 \sim   N(0,10),  \hspace{45pt}  \beta^1_i \sim   N(0,\sigma_1), \hspace{45pt} \beta^2_j \sim  N(0,\sigma_2), \hspace{45pt} \bm{\sigma} \sim  \mbox{Unif}(0,5);
\end{align*}

\noindent where $A$ is the agreement matrix; $i, j \in \{1,\dots,L\}$ index the category of a given variable the two raters are annotating (e.g. for the variable `2020 vote', each rater can choose any one of $L = 5$ categories); $\beta^0$ is the global abundance term, representing the overall sampling effort; $\beta^1$ and $\beta^2$ model the relative propensity of each rater to classify someone in each category; $r$ is the sampling premium if the two raters are `linked' in the latent network; and $B$ is the incidence matrix, representing the \emph{latent (dis)agreement network} of annotations across raters. \\

\noindent This model is a `twist' on the classic saturated log-linear model for contingency tables, where the `twist' is that the interaction term has a latent-variable parametrisation which allows the estimation of a network structure. $B_{i,j} = 1$ represents a link between the annotations produced by the two raters, meaning when rater $1$ chooses annotation $i$, rater $2$ will preferentially choose annotation $j$, net of any inherent tendency for any rater to choose any other option. $rB_{ij}$ is responsible for the counts in the agreement matrix not explained by rating propensity. Upon estimating the model, we can generate $S$ plausible network patterns $B^\star$ from the posterior predictive distribution of $B$. The monte carlo mean reveals the posterior predictive probability of a link between annotation levels across raters:

\begin{equation*}
    \bar{B^\star}_{ij} = \frac{1}{S}\sum^S_s {B^\star}_{sij}.
\end{equation*}

\begin{figure}[!htp]
    \caption{Perfect Agreement}
    \label{perfect}
    \includegraphics[width =1.0\textwidth]{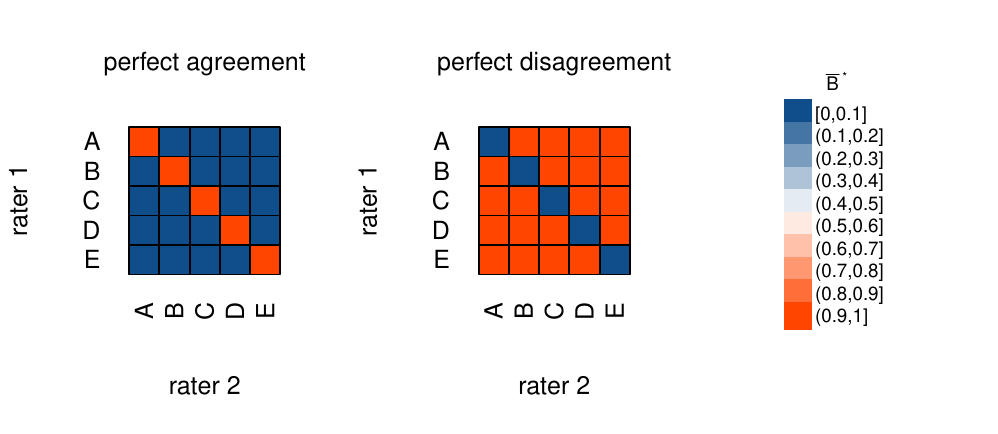}
    
    \begin{tablenotes} 
    \item[]\footnotesize 
    Note: From left to right, an example of `perfect agreement' between two raters across 5 annotations, followed by an example of `perfect disagreement' under the latent network model.
    \end{tablenotes}
\end{figure}

\begin{figure}[htp]
       \caption{Posterior predictive incidence across annotations:  humans versus \texttt{gpt-3.5-turbo} model prompted with $10$ tweets} 
        \label{net_human_gpt10}
\centering
\captionsetup[subfigure]{justification=centering}
     \begin{subfigure}[t]{0.3\textwidth}
         \centering
         \includegraphics[width=\textwidth]{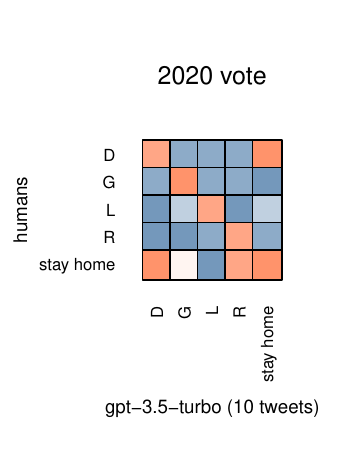}
         \caption{\scriptsize{\\Human $\rightarrow$ \emph{stay home} \\LLM $\rightarrow$ Democrats \emph{stay home}}}
         \label{net_human_gpt10_vote2020}
     \end{subfigure}
     \hfill
     \begin{subfigure}[t]{0.31\textwidth}
         \centering
         \includegraphics[width=\textwidth]{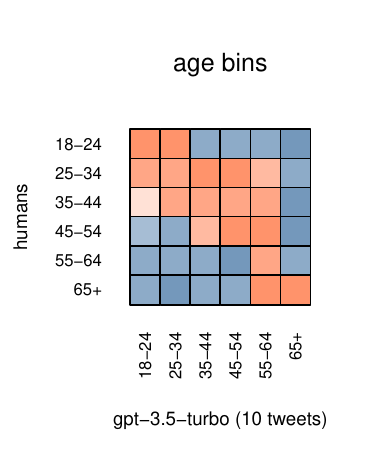}
         \caption{\\ \scriptsize{\emph{middle-aged} confusion\\
         LLM $\rightarrow$ some \emph{old-age}}}
         \label{net_human_gpt10_age.bins}
     \end{subfigure}
     \hfill
     \begin{subfigure}[t]{0.335\textwidth}
         \centering
         \includegraphics[trim={0 0.5cm 0 0},clip,width=\textwidth]{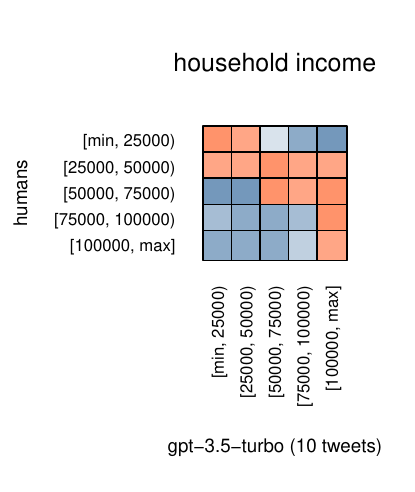}
         \caption{\\ \scriptsize{LLM $\rightarrow$ \emph{higher-income}}}
         \label{net_human_gpt10_hh.income}
     \end{subfigure}

     \hfill
     \begin{subfigure}[t]{0.34\textwidth}
         \centering
         \includegraphics[width=\textwidth]{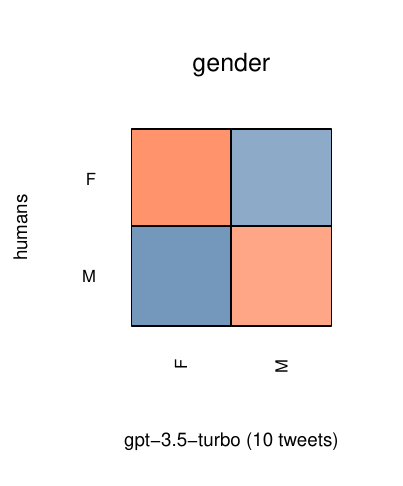}
         \caption{\\ \scriptsize{Quasi-perfect agreement}}
         \label{net_human_gpt10_gender}
     \end{subfigure}
     \hfill
     \begin{subfigure}[t]{0.33\textwidth}
         \centering
         \includegraphics[width=\textwidth]{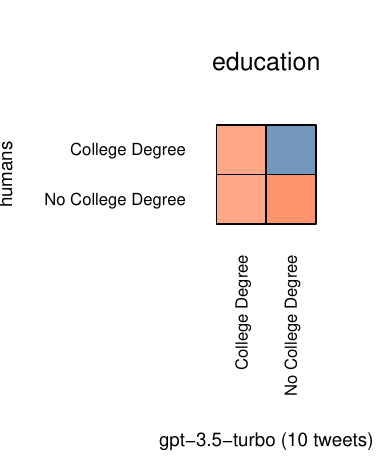}
         \caption{\\ \scriptsize{LLM $\rightarrow$ \emph{high-edu.} }}
         \label{net_human_gpt10_college}
     \end{subfigure}
     \hfill
     \begin{subfigure}[t]{0.29\textwidth}
         \centering
         \includegraphics[width=\textwidth]{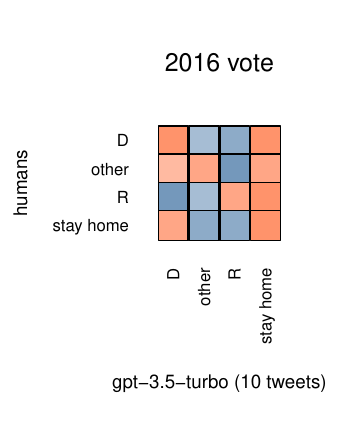}
         \caption{\\ \scriptsize{humans $\rightarrow$ Democrats \emph{stay home}\\ LLM $\rightarrow$ general \emph{stay home} }}
         \label{net_human_gpt10_vote2016}
     \end{subfigure} 
    \begin{tablenotes} 
    \item[]\footnotesize Note: Graphical representation of the relative frequency of posterior predictive incidence across annotations between humans and the \texttt{gpt-3.5-turbo} model prompted with $10$ tweets. See Figure \ref{perfect} for understanding scale and colour-coding. See Figure \ref{net_human_gpt10_state} for the variable `state'. Sub-captions hint at noticeable relative bias per rater.
        \end{tablenotes}
\end{figure}

\noindent We fit the model using \texttt{JAGS} \cite{plummer2003jags}, which offers the ability to sample from latent parameters without having to re-parametrise the model. Figure \ref{net_human_gpt10} presents the relative frequency of posterior predictive incidences for human versus gpt-3-5 turbo with 10 tweets (additional comparisons are provided in Figures \ref{net_human_gpt5} and \ref{net_gpt10_gpt5}). These comparisons generally display broad agreement between the annotators, as can be seen by the mostly orange tone of the diagonals of the posterior incidence matrices. Where there is disagreement, this is typically informative of relative `biases' of annotators - disagreement seems rarely `random', and more often `systematic' in a specific direction. Each panel of Figure \ref{net_human_gpt10} identifies the predominant disagreements for the variable in question.  A summary of the disagreements between models is tallied in Table \ref{(dis)agreement_interpretation}. Note that for the variable `state', the context does not play a role (see Figure \ref{location.prompt}). Hence any differences between \texttt{gpt-3.5-turbo} (10 tweets) and (5 tweets) are ultimately due to noise in the LLM output to the location prompt. \\

\pagebreak

\noindent From the annotations in Figure \ref{net_human_gpt10} and the summaries in Table \ref{(dis)agreement_interpretation} we see the large-context LLM tended to believe potential Democrats would stay home in $2020$ at higher rates than humans did. LLM tended to believe far more Twitter users would have stayed-home in $2016$, independent of their other characteristics, whilst humans tended to be more pessimistic about the chances that potential democratic voters would cast a ballot on election-day. Due to the relative importance of past-vote as a predictor of current vote, we can tentatively attribute at least part of the relatively strong perfromance of models fit to large-context AI-annotated data to the AI's ability to annotate samples such that the joint distribution of $2020$ and $2016$ vote would be more compatible with the true swing. At a macro-level, this can also be seen in Figures \ref{area_predictions_delta} and \ref{area_predictions_delta_538}, where the change in vote shares since the last election is substantially better predicted by the large-context AI. Large context LLMs further tended to believe Twitter users were generally somewhat older, richer, more highly-educated and more likely to have not voted in $2016$. It is difficult to say whether these biases contributed to the better performance. Linking Twitter user data to an auxiliary ground-truth dataset, such as Voter Registration Files \cite{cerina2020measuring}, could provide a strategy to better understand whether these disagreement represent higher-quality AI labels, or are simply noise which ends up having no meaningful effect on the predictions due to the relatively low magnitude of the demographic effects, net of past-vote. 


\pagebreak

\section{Discussion}\label{discussion}
We have presented \emph{Artificially Intelligent Opinion Polling}: a novel methodology to produce fully-automated high-quality pre-election polls from social-media data. We introduce the use of AI, in the form of Large Language Models, to extract pre-election polling features from the self-reported preferences and socio-demographics of Twitter users. We show LLMs tend to broadly agree with humans in their annotations of social-media users. We further propose a modification to the traditional Multilevel Regression and Post-stratification methodology to account for \emph{online selection}, a general selection framework applicable to many kinds of data where self-selection onto a given medium plays a role. We show this amendment should generate substantial gains in bias reduction, RMSE, pearson correlation and coverage via a simulation study. We further show that applying bias-corrected structured MrP to the AI-extracted social-media surveys can produce state-of-the-arts estimates of the vote in an application to the $2020$ US election. Our methodology could reduce the cost of pre-election public opinion polling, relative to random-digit-dial, by making it anywhere between $500$ and $2,500$ times cheaper (see Table \ref{sample_selection}). The methodology we outline here, and the results reported, suggest three broad areas for further research: increasing the automation of variable selection and modeling; improving on, and accounting for, the uncertainty in feature extraction from social media; and sampling multi-media content from more diverse social-media platforms. \\

\noindent \textbf{Automation.} We have outlined the prospect of a fully-automated polling machine. Data-collection for pre-election polls can already be fully automated, as the implementation of APIs to download social-media data and convert these into survey-formats is logistically straightforward, as well as being computationally and economically feasible for research. The major pain-points for the implementation of such a machine lie with the modeling framework, and in particular variable selection. In prior versions of this manuscript we experimented with using the horseshoe prior \cite{carvalho2009handling,piironen2017hyperprior} to select an optimal area-time level predictor. Unfortunately, though not outlandish, the predictions produced under this approach were substantially weaker than those produced by carefully selecting the predictor using topic-specific knowledge and expertise. Moreover, the use of the horseshoe negatively affected the coverage of the estimates. There are other approaches which have been suggested to automate variable selection within MrP \cite{bisbee2019barp,ornstein2019stacked,broniecki2022improved}, though each presents trade-offs, especially as it comes to the accounting of uncertainty, and some are not compatible with the linear-modeling specification necessary to thoughtfully account for online selection. In our view, this is a crucial area of research: whilst in the context of pre-election polls we benefit from knowing, at least partially, the functional form of the vote and its most relevant predictors, we do not benefit from the same knowledge in other areas of relevant opinion and behaviour. If one was able to successfully automate variable selection and mitigate selection bias, we could see a world in which real-time, population-representative monitoring of a variety of interesting variables - happiness, psychometrics, consumption, attitudes, pathogen spread, etc. - could happen effortlessly, in real time, and at a granular level of analysis.\\ 

\noindent \textbf{Feature Extraction.} In our application to the $2020$ US election we have concentrated exclusively on four types of self-reported variables: location, name, description and posts. All of these inputs to the LLM were text-based. Within these boundaries, we have attempted to test the power of context. We looked at how the quality of post-modeling predictive outputs would change under differing amounts of context. We also explored how the annotations would change, under different amounts of context. Our analysis suggests more context is better for predictive power. Future work should focus on a more formal investigation, controlling for the quality of the context, and exploring heterogeneity across opinion-domains. It is possible pre-processing tweets to extract only those with the highest proportion of `useful' context could improve the quality of the annotations.\\

\noindent Whatever the quality of the annotations might be, it is unlikely we will ever be able to extract individual-level features from social-media without adding some measure of noise and/or systematic bias into our sample. Future work should further consider enhancing the Hierarchical Bayesian model to account for such impurities. Simple extensions to the Bayesian logistic regression framework proposed here are available to account for measurement error \cite{muff2015bayesian} as well as contamination in the dependent variable \cite{cerina2020measuring}. Proper accounting of uncertainty at the annotation-level is liable to make the coverage of bias-corrected structured MrP estimates more robust. Ultimately we would benefit from having a probabilistic LLM which provides uncertainty estimates. In absence of this, future work should look at tuning the \texttt{temperature} hyper-parameter to produce desirable levels of uncertainty around the labels. Based on this parameter we could then sample multiple labels per user, essentially generating quasi-posterior samples from the LLM. Proper accounting for annotation uncertainty is an important area of further research.\\ 

\noindent Another concern specific to the study of vote-choice is what understanding the LLM has of the underlying social context. We use \texttt{gpt-3.5-turbo}, which was trained with data encompassing the $2020$ US election, and hence we could assume it was in some sense aware of the \emph{quality} of the candidates, and their specific relationships with various types of voters. Would the LLM produce judgments of vote-choice which are equally valid if its training data did not include information about a specific candidate? For example, if in the $2024$ US election we would see the Democratic or Republican party run some relatively unknown candidate, would the LLM be able to account for this? How about an insurgent third-party candidate with national appeal? If the LLM would limit itself to looking at the party of the candidate - which is what would remain stable in terms of relationships with voters from election to election - it would surely produce low-quality labels. This is where the work of Argyle et al. \cite{argyle2023out} on conditioning using specific context-cues may be most relevant. The pollster would need to provide extra context in the prompts to refine the LLM labels on new candidates, or to update it of specific changes in political dynamics since its last training day. How to best do this in the context of pre-election opinion polling remains an open area of research. \\

\noindent As a final comment on feature extraction, we note the future of this methodology clearly lies in Multimodal AI \cite{ngiam2011multimodal,radford2021learning}, which would enable the ability to perform feature-extraction from images, videos, recordings, or other media. A version of this feature extraction machine is already implementable: image-to-text \cite{vinyals2015show}, speech-to-text \cite{hannun2014deep} and video-to-text \cite{venugopalan2015sequence} models exist, and state-of-the-arts versions are typically accessible through various APIs\footnote{See the \texttt{Multimodal Models}, and associated APIs at \url{https://huggingface.co/docs/transformers}. OpenAI makes their speech to text model `\emph{Whisper}' available via their API - \url{https://openai.com/research/whisper}, and \texttt{gpt-4} accepts images as inputs and is able to use internally formulated descriptions of these images to answer specific prompts and complete classification tasks.}. Their outputs can be distilled and passed-on to a LLM for a more holistic feature extraction, which is not solely limited to text-inputs.\\

\noindent \textbf{Social Media Data.} In our US $2020$ application we focus on Twitter, which at the time of collection we considered to be the primary platform for varied online political discourse. It was also conveniently accessible via a free-to-use streaming API. Since then the social media landscape has been evolving. We cannot fail to mention a trend in making API basic-usage tiers more expensive. Both Twitter \footnote{\url{https://twitter.com/XDevelopers/status/1649191520250245121}} and \texttt{reddit} \footnote{\url{https://www.reddit.com/r/reddit/comments/145bram/addressing_the_community_about_changes_to_our_api/}} have reduced basic-usage API access and partially implemented subscription plans. It's too early to tell what impact this may have on our proposed modeling strategy, though we have shown a moderate amount of posts (in the order of 50,000) from a moderate number of users (in the order of 5,000) can be sufficient to make high-quality predictions. This amount of Tweets and users is still relatively feasible to collect over a few months for aspiring pollsters under a `hobbysit' Twitter plan, which comes at a cost of $\$100$ per month \footnote{\url{https://developer.twitter.com/en/products/twitter-api}}. A second important trend concerns the impact of polarisation on social-media usage. Though little has been done to systematically review this phenomenon, there is some evidence that the social-media space is fracturing according to partisanship. Facebook and Twitter have historically had a partisan and demographic bent \cite{mellon2017twitter} - though this may be changing. The well documented flight of users from Twitter to Mastodon \cite{zia2023flocking} has coincided with a general perception of Twitter become more right-aligned \cite{anderson2023after}. Platforms such as Truth-Social and Gab \cite{jasser2023welcome} appeal primarily to a subset of conservative and libertarian users. Interestingly, Mastodon and Gab's APIs remain free. This offers an unique opportunity for researchers, namely to fully embrace the sampling design advocated by King \& Zeng \cite{king2001logistic}, and sample partisans from fully siloed social-media platforms. Provided we can use the bias-correction to address the sampling protocol, we believe our approach to be robust to both changes in pricing and partisan siloing of social-media platforms - though ultimately these are empirical questions.\\




\subsection{Conclusion}

\noindent Advances in artificial intelligence will radically change how we conduct public opinion polling.  Our contribution is first to suggest how AI can be leveraged to transform digital traces into public opinion data.  Secondly, we propose a robust strategy for modeling public opinion preferences based on these transformed digital traces data.  The major modeling challenge we address is accounting for the non-representative nature of the online-generated digital trace sample.  Our work makes it clear that with the judicious use of AI and social-media data that builds on a robust inferential framework, we can significantly advance claims regarding the representativeness of digital trace data.  Hence, the research agenda for \emph{Artificially Intelligent Opinion Polling} is clear: to build automated pipelines, founded on flexible and interpretable models, that enable representative inference from easily obtainable, high-frequency unrepresentative samples. We hope others can build on this work. 

\pagebreak
\Urlmuskip=0mu plus 1mu\relax

\bibliography{main.bib}
\pagebreak

\renewcommand \thepart{}
\renewcommand \partname{}
\appendix
\pagenumbering{arabic}
\setcounter{page}{1}
\renewcommand*{\thepage}{A\arabic{page}}
\addcontentsline{toc}{section}{Appendix} 
\part{Appendix} 
\parttoc 
\thispagestyle{empty}

\clearpage
\pagenumbering{arabic}
\counterwithin{table}{section}
\counterwithin{figure}{section}
\begin{spacing}{1}

\newpage	

\newpage
\section{\texttt{Stan} listings}

\begin{lstlisting}[caption={\texttt{Stan} `Data' and `Transformed Data' Declaration Blocks. },captionpos=t,label={lst:stan_dist}]
data{

    int<lower = 1> N; // n. observations
    int Y[N]; // binary choice
    real offset; // king-zeng offset parameter

    int<lower = 1> P; // n. state- and day- level fxed effects
    matrix[N, P] X; // state- and day-  level covariate matrix

    int gender_id[N]; // gender level id
    int<lower=1> gender_N; // number of gender levels 

    int ethnicity_id[N]; // ethnicity level id
    int<lower=1> ethnicity_N; // number of ethnicity levels 

    int age_id[N]; // age level id
    int<lower=1> age_N; // number of age levels 

    int edu_id[N]; // education level id
    int<lower=1> edu_N; // number of education levels 

    int income_id[N]; // income level id
    int<lower=1> income_N; // number of income levels 

    int vote2016_id[N]; // 2016 vote level id
    int<lower=1> vote2016_N; // number of 2016 vote levels 

    int dte_id[N]; // days-to-election id
    int<lower=1> dte_N; // max number of days-to-election

    // SPATIAL-COMPONENT DATA

        int area_id[N]; // index of areas in the observed data
        int<lower=1> area_N; // no. of spatial units
        int<lower=1> k; // no. of separate inner-connected groups
        int group_size[k]; // observational units per group
        int group_idx[area_N]; // index of observations, ordered by group

        int<lower=1> N_edges; // number of adjacency instances
        int<lower=1, upper=area_N> node1[N_edges]; // node1[i] adjacent to node2[i]
        int<lower=1, upper=area_N> node2[N_edges]; // node1[i] < node2[i]
        int<lower=1, upper=k> comp_id[area_N]; // ids of groups by areas

        vector[k] inv_sqrt_scaling_factor ; // BYM2 scale factor, with singletons represented by 1

}

transformed data {

    int<lower=0,upper=1> has_phi=1; // turn phi on to include unstructured random effect in BYM2 spec.

}
\end{lstlisting}

\pagebreak

\begin{lstlisting}[caption={\texttt{Stan} `Parameters' Declaration Block.},captionpos=t,label={lst:stan_dist1}]
parameters{

    real alpha_star; // baseline rate of choice
    
    vector[P] beta_star; // fixed-effects
    
    vector[gender_N] gamma_gender; // gender unstructured random effects
    real<lower=0> gamma_gender_scale; // gender effects' scale

    vector[ethnicity_N] gamma_ethnicity; // ethnicity unstructured random effects
    real<lower=0> gamma_ethnicity_scale; // ethnicity effects' scale

    vector[age_N] gamma_age; // age autoregressive random effects
    real<lower=0>  gamma_age_scale; // age effects' scale

    vector[edu_N] gamma_edu; // education unstructured random effects
    real<lower=0>  gamma_edu_scale; // education effects' scale

    vector[income_N] gamma_income; // income autoregressive random effects
    real<lower=0>  gamma_income_scale; // income effects' scale

    vector[vote2016_N] gamma_vote2016; // 2016 vote unstructured random effects
    real<lower=0> gamma_vote2016_scale;// 2016 vote effects' scale

    vector[dte_N] delta; // days-to-election autoregressive random effects
    real<lower=0> delta_scale;// days-to-election effects' scale

    vector[area_N] phi; // unstructured area effect
    vector[area_N] psi; // spatial (ICAR) area effect
    real<lower=0,upper = 1> omega; // mixing parameter for structured/unstructured area effects
    real<lower=0> spatial_scale; // scale for structured area effects
  
}
\end{lstlisting}

\pagebreak

\begin{lstlisting}[caption={\texttt{Stan} `Transformed Parameters' Block.},captionpos=t,label={lst:stan_dist2}]
transformed parameters{
    
    vector[N] mu; // latent propensity for choice j
    
    // NON-CENTERED PARAMETRISATION 
    
        vector[gender_N] gamma_gender_star = gamma_gender * gamma_gender_scale;
  
        vector[ethnicity_N] gamma_ethnicity_star = gamma_ethnicity * gamma_ethnicity_scale;
        
        vector[age_N] gamma_age_star = gamma_age * gamma_age_scale;
  
        vector[edu_N] gamma_edu_star = gamma_edu * gamma_edu_scale;
  
        vector[income_N] gamma_income_star = gamma_income * gamma_income_scale;

        vector[vote2016_N] gamma_vote2016_star = gamma_vote2016 * gamma_vote2016_scale;

        vector[dte_N] delta_star = delta * delta_scale;

        vector[area_N] lambda_star = convolve_bym2(psi, phi, spatial_scale, area_N, k, group_size, group_idx, omega, inv_sqrt_scaling_factor);
        
    // LINEAR PREDICTOR 
  
        mu =
            offset +
            alpha_star +
            X*beta_star +
            gamma_gender_star[gender_id] +
            gamma_ethnicity_star[ethnicity_id] +
            gamma_age_star[age_id] +
            gamma_edu_star[edu_id] +
            gamma_income_star[income_id] +
            gamma_vote2016_star[vote2016_id] +
            lambda_star[area_id] +
            delta_star[dte_id];
            
}
\end{lstlisting}

\pagebreak 

\begin{lstlisting}[caption={\texttt{Stan} `Model' Declaration Block. },captionpos=t,label={lst:stan_dist3}]
model{
    
    // IID PRIORS

        alpha_star ~ std_normal();
    
        to_vector(beta_star) ~ std_normal() ;

    // UNSTRUCTURED RANDOM EFFECTS

        to_vector(gamma_gender) ~ std_normal();
        gamma_gender_scale ~ std_normal();

        to_vector(gamma_ethnicity) ~ std_normal();
        gamma_ethnicity_scale ~ std_normal();

        to_vector(gamma_edu) ~ std_normal();
        gamma_edu_scale ~ std_normal();

        to_vector(gamma_vote2016) ~ std_normal();
        gamma_vote2016_scale ~ std_normal();

    // STRUCTURED AUTOREGRESSIVE PRIORS 

        sum(gamma_income) ~ normal(0, 0.01 * income_N); // sum-to-0 constraint
        for(i in 2:income_N){ 
            gamma_income[i] ~ normal(gamma_income[i-1],1) 
        };
        gamma_income_scale ~ std_normal();

        sum(gamma_age) ~ normal(0, 0.01 * age_N); // sum-to-0 constraint
        for(i in 2:age_N){ 
            gamma_age[i] ~ normal(gamma_age[i-1],1) 
        };
        gamma_age_scale ~ std_normal();

        sum(delta) ~ normal(0, 0.01 * dte_N); // sum-to-0 constraint
        for(i in 2:dte_N){ 
            delta[i] ~ normal(delta[i-1],1) 
        };
        delta_scale ~ std_normal();

        psi ~ icar_normal(spatial_scale,node1, node2, k, group_size, group_idx, has_phi);
        phi ~ std_normal();
        omega ~ beta(0.5,0.5);
        spatial_scale ~ std_normal();
        
    // LIKELIHOOD

        Y ~ bernoulli_logit(mu) ;

}
\end{lstlisting}

\pagebreak

\section{Simulation Study}

\begin{figure}[!htb]
    \caption{Comparing quality of estimates of $\bm{\theta}_j$ under different scenarios.}
    \label{sims.comparison.theta}
    \includegraphics[width = \textwidth]{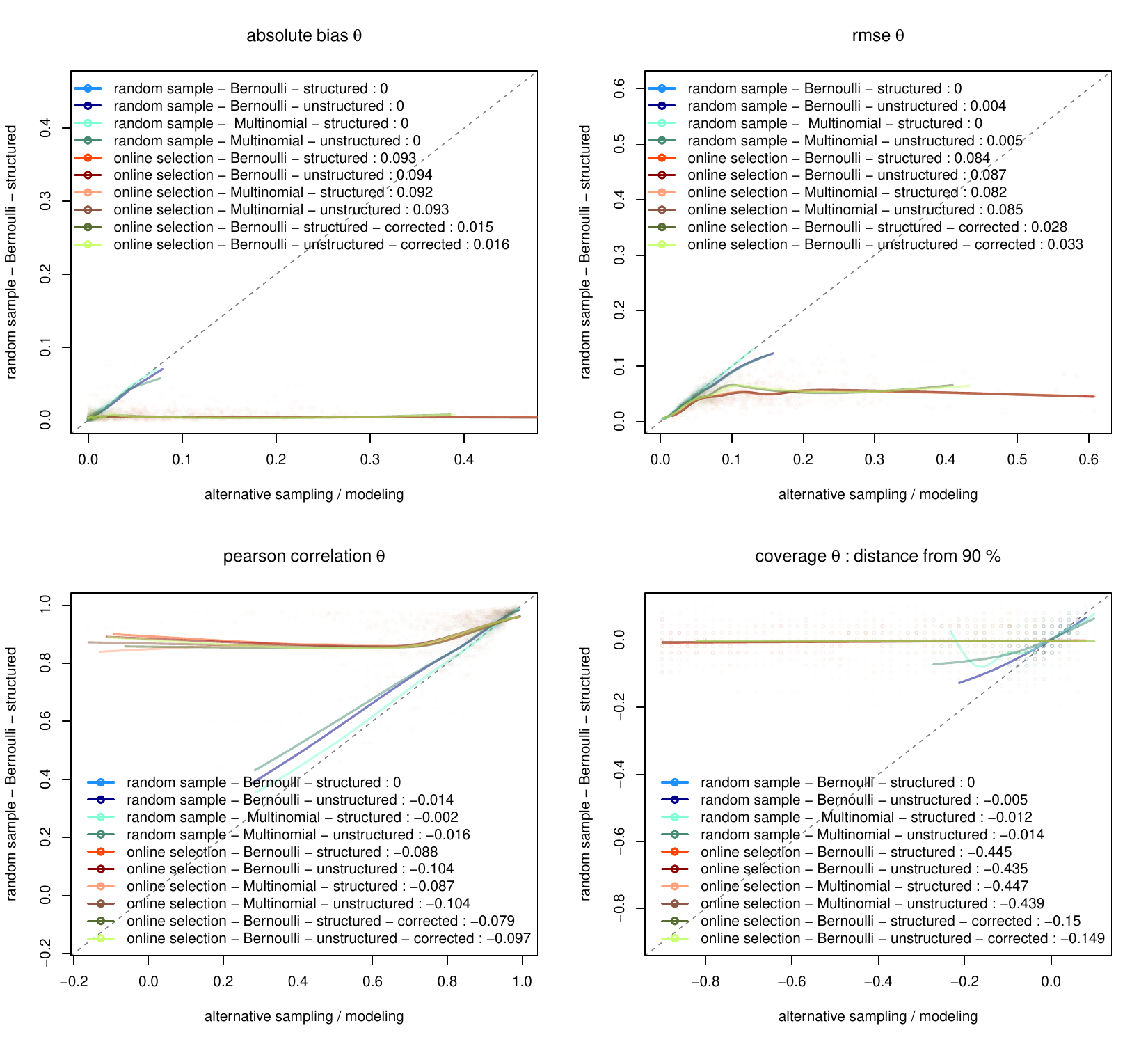}
    \begin{tablenotes}
    \item[] \footnotesize Notes: The y-axis represents the score of a structured-priors MrP model without bias-correction fit to a random sample from the population \texttt{(S.0)}; the x-axis represents performance of all other models, for a given simulated population. The legend reports the performance relative to \texttt{(S.0)}. Smooth curves are fit using the \texttt{mgcv} package \cite{wood2015package}.
    \end{tablenotes}
\end{figure}

\pagebreak 

\begin{figure}[!htb]
\caption{Comparing quality of estimates of $\bm{\pi}_j$ under different scenarios.}
  \label{sims.comparison.pi}
    \includegraphics[width = 0.825\textwidth]{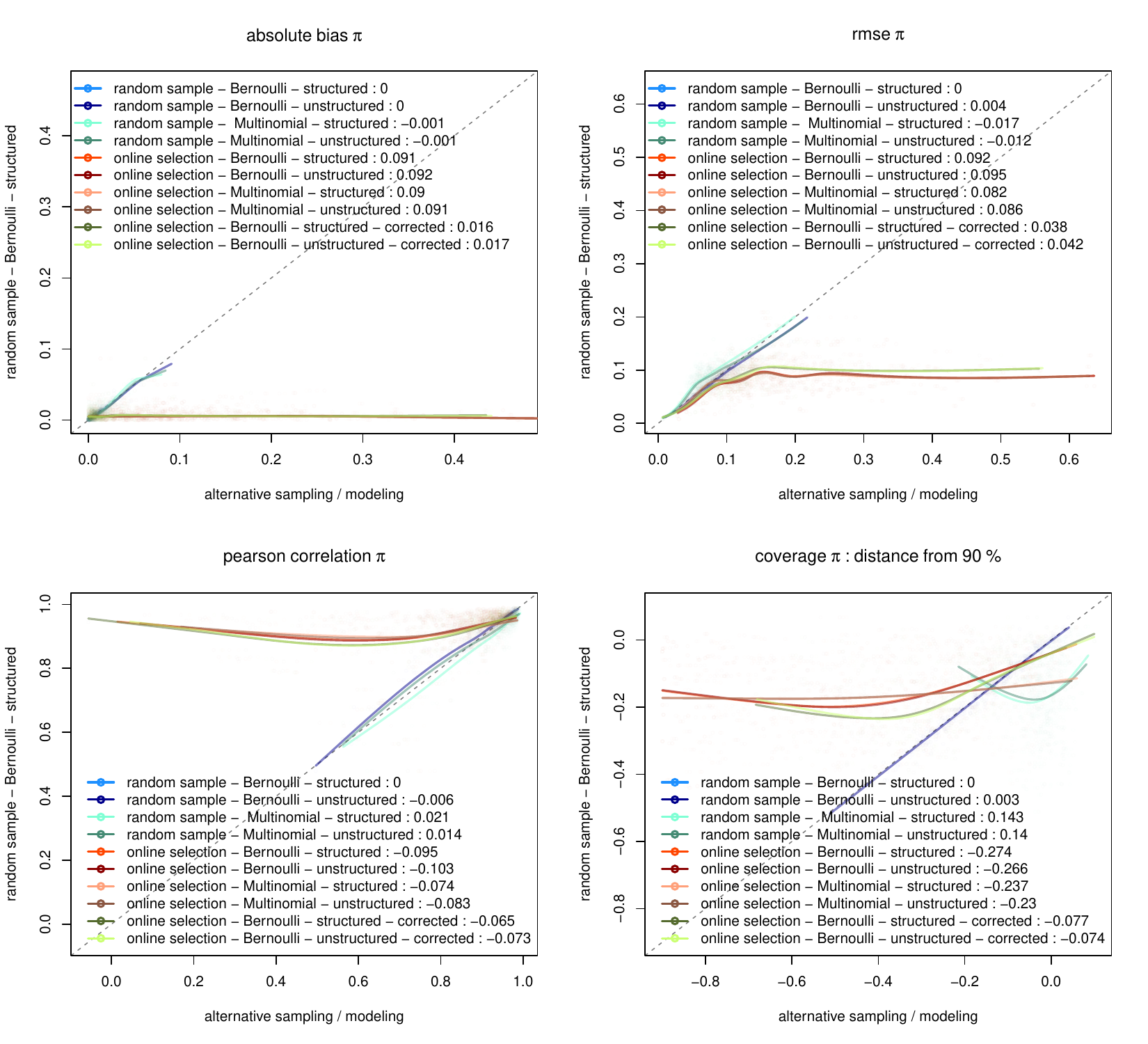}
    \begin{tablenotes}
    \item[] \footnotesize Notes: The y-axis represents the score of a structured-priors MrP model without bias-correction fit to a random sample from the population \texttt{(S.0)}; the x-axis represents performance of all other models, for a given simulated population. The legend reports the performance relative to \texttt{(S.0)}. Smooth curves are fit using the \texttt{mgcv} package \cite{wood2015package}.
    \end{tablenotes}
\end{figure}

\pagebreak 

\subsection{Examining Model Behaviour}
\begin{figure}[!htb]
    \caption{Effect of sample size $n$ on estimation performance for $\bm{\pi}_j$.}
    \label{sims.properties.pi.n}
    \includegraphics[width = \textwidth]{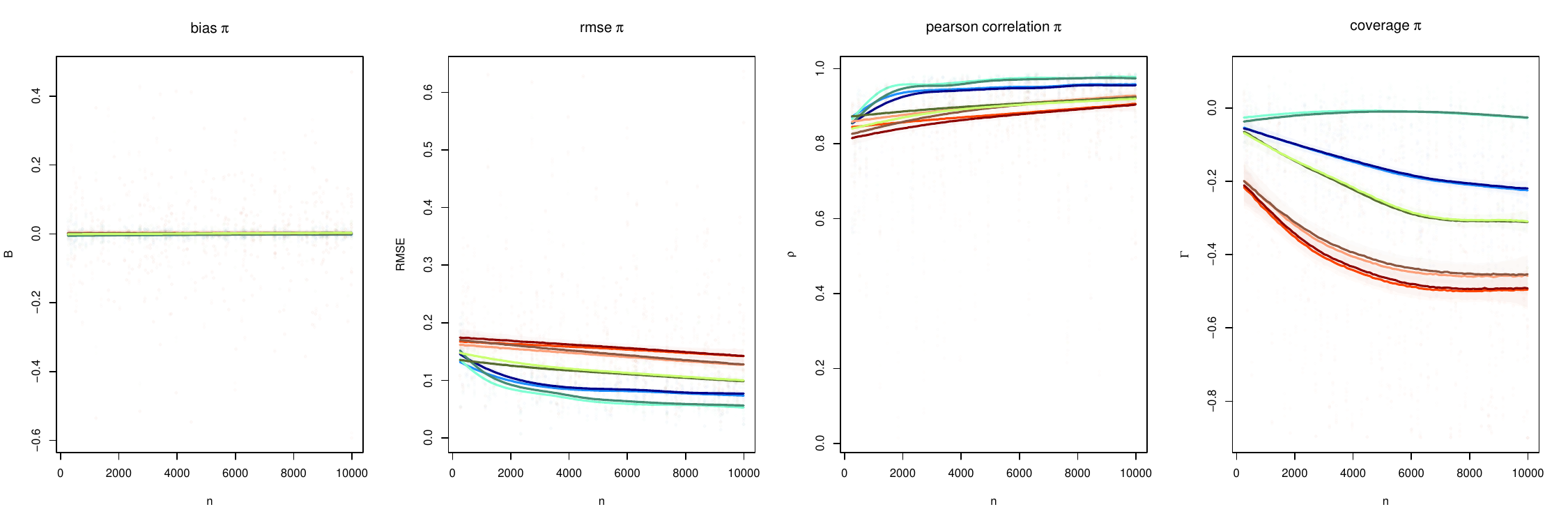}
    \begin{tablenotes}
    \item[] \footnotesize Notes: See Table \ref{simulation_scenarios} or Figure \ref{sims.comparison.theta} for colour coding.
    \end{tablenotes}
\end{figure}

The following emerges from a detailed look at the properties of each scenario relative to specific stimuli: 

\noindent \emph{i. \textbf{sample size}} $n$: Figures \ref{sims.properties.theta.n} and \ref{sims.properties.pi.n} demonstrate that more is better in terms of sample size for RMSE reduction and increases in point-estimate correlations. Regarding the stratified preferences $\bm{\theta}_j$, under random-sampling we see decreasing returns and a plateau around $n = 8,000$. For selected samples we never enter the plateau phase in the examined range. Coverage is negatively affected by increasing sample-size under online selection. We see logarithmic degradation of coverage under online selection. At the cell-level, estimates of $\bm{\pi}_j$ behave similarly, though we see one major difference - namely models endowed with Multinomial likelihood outperform Bernoulli models. Models with Bernoulli likelihood experience substantial coverage degradation at larger sample sizes, whilst Multinomial models remain robust at any sample-size;

\begin{figure}[!htb]
    \caption{Effect of population prevalence $\pi_j$ on estimation performance for $\bm{\pi}_j$.} \label{sims.properties.pi.prevalence.pi}
    \includegraphics[width = \textwidth]{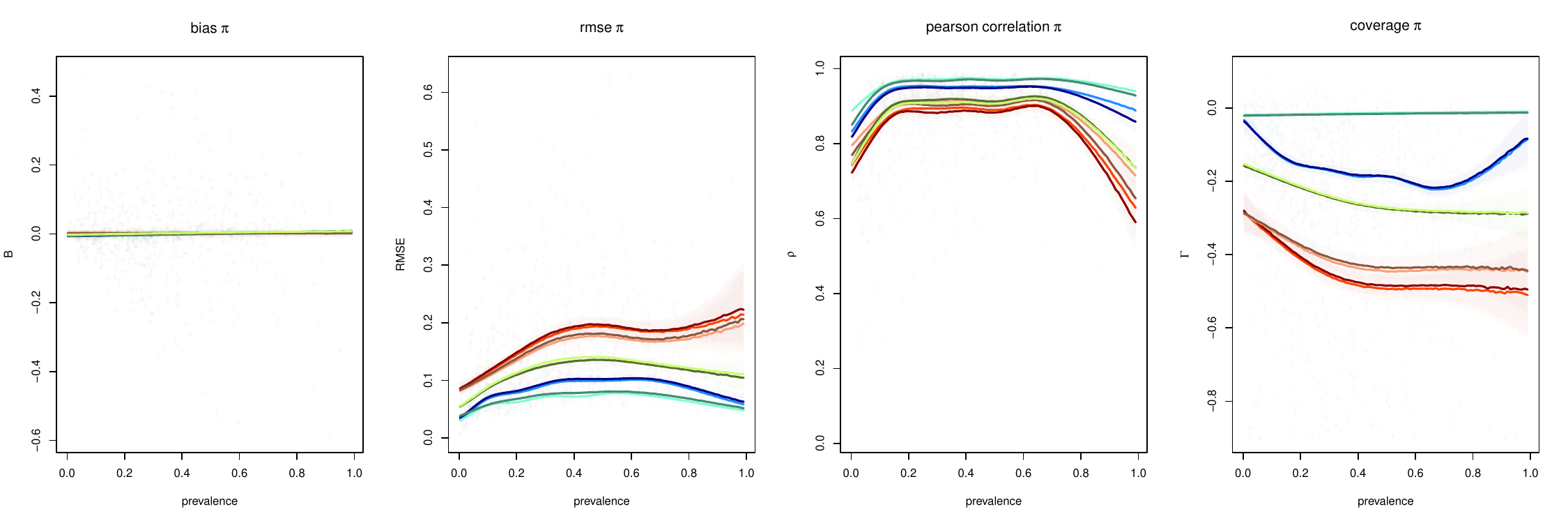}
    \begin{tablenotes} 
    \item[] \footnotesize Notes: See Table \ref{simulation_scenarios} or Figure \ref{sims.comparison.theta} for colour coding. \end{tablenotes}
\end{figure}

\emph{ii. \textbf{population prevalence} } $\pi_j$ : Figures \ref{sims.properties.theta.prevalence.pi} and \ref{sims.properties.pi.prevalence.pi} show that with respect to RMSE there is a general degradation moving away from \emph{rare} events - a property attributable to the larger scale of the prevalence. The DGP rarely produces prevalence close to $1$, so we cannot confidently asses properties for $\pi_j > 0.8$. However, behaviour from estimates of both $\bm{\theta}_j$ and $\bm{\pi}_j$ suggests a symmetric improvement in performance away from $\pi_j \approx 0.5$. Regarding correlation, here models show similar behaviour, with correlation peaking at $\pi_j \approx 0.2$, again showing hints of symmetry around $\pi = 0.5$ for selected samples. Coverage behaviour is similar to that which we see for sample-size, namely selected samples experience quasi-logarithmic decrease in coverage as prevalence increases, with bias-corrected models seemingly more robust to changes in prevalence. We further note that at the cell-level, Multinomial models outperform Bernoulli models, especially as it pertains to coverage - though we see no evidence of this at the stratified level;

\begin{figure}[!htb]
    \caption{Effect of online-selection penalty $\mu^\Upsilon_j$ on estimation performance for $\bm{\pi}_j$.} \label{sims.properties.pi.penalty}
    \includegraphics[width = \textwidth]{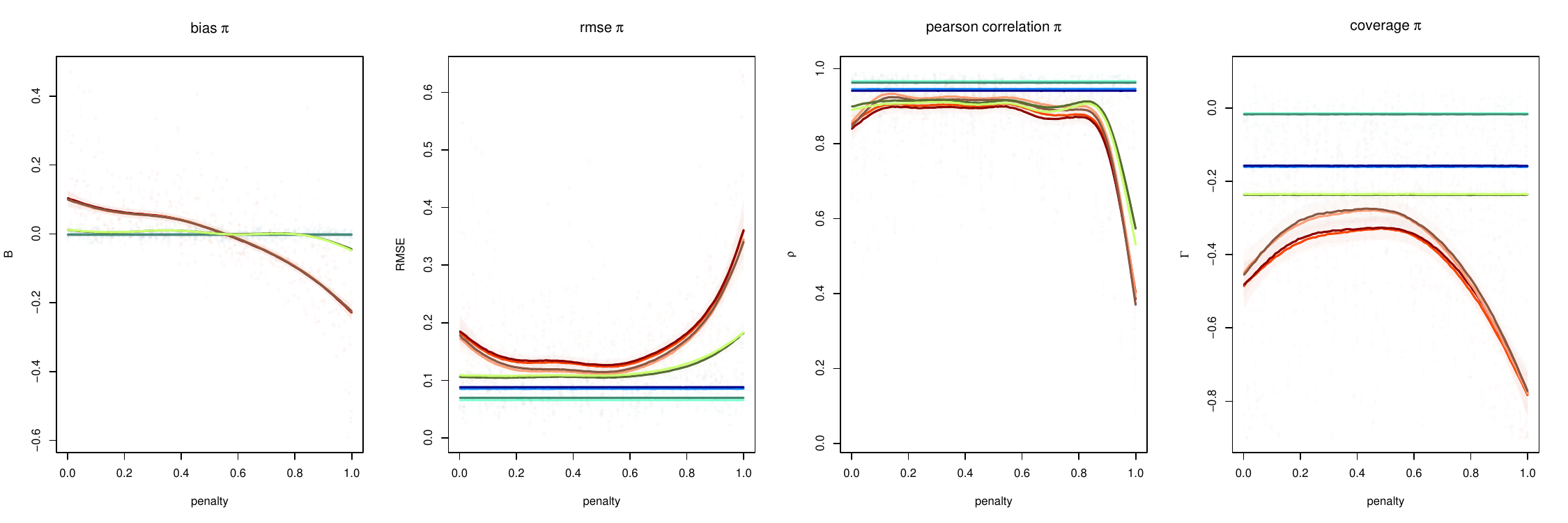}
    \begin{tablenotes}
    \item[] \footnotesize Notes: See Table \ref{simulation_scenarios} or Figure \ref{sims.comparison.theta} for colour coding. \end{tablenotes}
\end{figure}

\emph{iii. \textbf{central selection penalty} $\mu^\Upsilon_j$}: Figures \ref{sims.properties.theta.penalty} and \ref{sims.properties.pi.penalty} show the online selection penalty exponentially degrades estimates from uncorrected selected samples at $\mu^\Upsilon_j > 0.5$, across all metrics. Bias-corrected models appear broadly unbiased despite crippling penalty levels, only showing signs of degradation around $\mu^\Upsilon_j > 0.8$. RMSE starts degrading earlier, around at $\mu^\Upsilon_j > 0.6$, though the rate of degradation in RMSE is far slower for bias-corrected models than for uncorrected models.  Note further that, whilst the bias-corrected models are robust to bias introduced via relative over-sampling, the uncorrected models under online selection show an increase in positive bias proportional to the degree of over-sampling relative to the other choices. This can be seen in spikes of bias and RMSE when the $\mu^{\Upsilon}_j = 0$ and $\mu^{\Upsilon}_{j^{'}} > 0, \mbox{ } \forall {j^{'}} \neq j$. The impact of the penalty is especially severe on correlation estimates when $\mu^\Upsilon_j > 0.8$, at which point there is a massive drop in correlation between the estimates and the true values. Though coverage of bias-corrected models is not as good as random sampling, we see relatively robust coverage at any penalty level compared to otherwise abysmal levels (approaching $0$ as the penalty approaches $1$) under uncorrected selection. 

\begin{figure}[!htb]
    \caption{Effect of sample bias $(\hat{\pi}_j - \pi_j)$ on estimation performance for $\bm{\pi}_j$.} \label{sims.properties.pi.sample.prevalence.bias}
    \includegraphics[width = \textwidth]{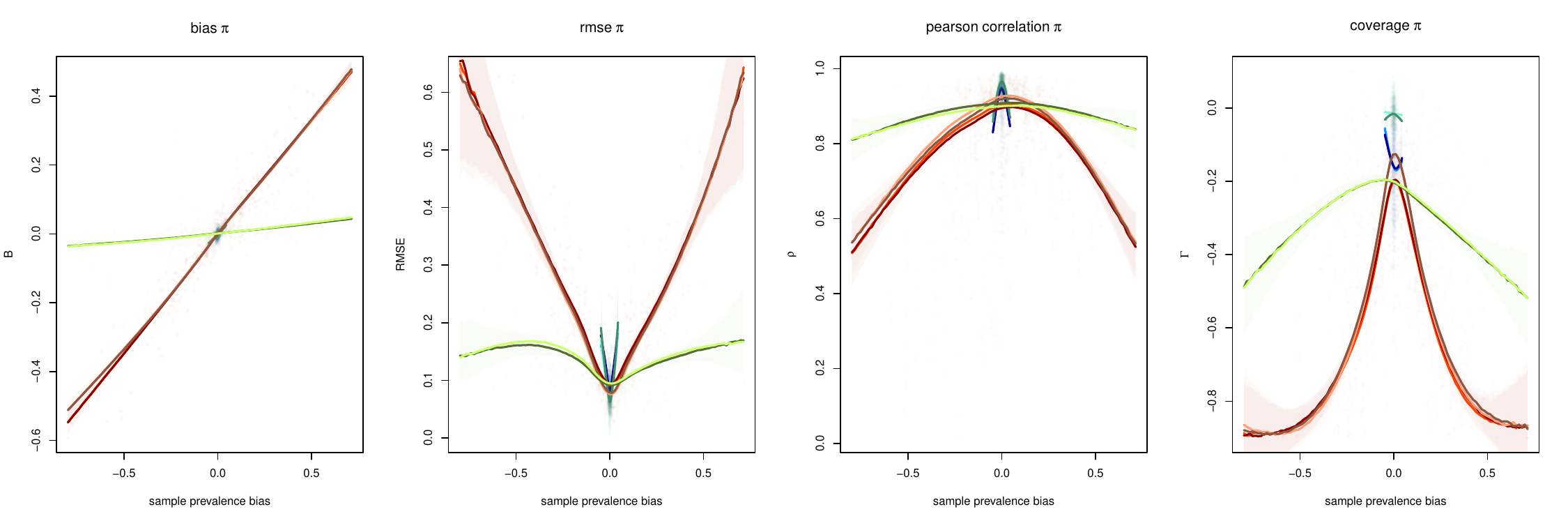}
    \begin{tablenotes}
    \item[] \footnotesize Notes: See Table \ref{simulation_scenarios} or Figure \ref{sims.comparison.theta} for colour coding.
    \end{tablenotes}
\end{figure}

\emph{iv. \emph{\textbf{sample prevalence bias}} $\hat{\pi}_j-\pi_j$}: this is a measure of the severity of the bias which is effectively faced by the sample. Note that random samples can also suffer - albeit more rarely than selected samples - from severely biased prevalence. Figures \ref{sims.properties.theta.sample.prevalence.bias} and \ref{sims.properties.pi.sample.prevalence.bias} suggest bias-correction is robust to bias in sample prevalence. Only under extraordinary sample prevalence bias $\mid \hat{\pi}_j-\pi_j \mid  > 0.1$ do we see a significant degradation in performance for bias-corrected models. The Figures clearly display the highly damaging impact of sample prevalence bias, which, if unaccounted for, translates almost linearly to increases in bias and RMSE, as well as exponential loss of coverage. The smooth curves imply some evidence that bias-correction further out-performs random-samples in the unlikely event of highly-unrepresentative random draws, though because these extreme draws are extremely rare, the intervals around these tail-events are large and we caution against over-interpreting these few limiting observations.


\pagebreak 

\pagebreak 

\pagebreak 

\section{Feature Extraction via LLMs}

\begin{figure}[!htb]
    \begin{framed}
\begin{footnotesize}
\texttt{demos\_string <- c(}\\

\texttt{`}ETHNICITY:$\backslash$nE1) White $\backslash$nE2) Black$\backslash$nE3) Hispanic$\backslash$nE4) Asian$\backslash$nE5) Other$\backslash$n\texttt{',}\\

\texttt{`}AGE:$\backslash$nA1) between 0 and 17 year old$\backslash$nA2) 18 to 24 years old$\backslash$nA3) 25 to 34 years old$\backslash$nA4) 35 to 44 years old$\backslash$nA5) 45 to 54 years old$\backslash$nA6) 55 to 64 years old$\backslash$nA7) 65 or older$\backslash$n\texttt{'},\\

\texttt{`}SEX:$\backslash$nS1) Male$\backslash$nS2) Female$\backslash$n\texttt{'},\\

\texttt{`}MARITAL STATUS:$\backslash$nM1) Married$\backslash$nM2) Not married$\backslash$n\texttt{'},\\

\texttt{`}HIGHEST EDUCATIONAL QUALIFICATION:$\backslash$nQ1) no formal education$\backslash$nQ2) completed high-school but did not go to college$\backslash$nQ3) obtained a Bachelor degree or higher$\backslash$n\texttt{'},\\

\texttt{`}HOUSEHOLD INCOME BRACKET:$\backslash$nH1) up to 25000 USD per year$\backslash$nH2) between 25000 and 50000 USD per year$\backslash$nH3) between 50000 and 75000 USD per year$\backslash$nH4) between 75000 and 100000 USD per year$\backslash$nH5) more than 100000 USD per year$\backslash$n\texttt{'},\\

\texttt{`}THIS INDIVIDUAL IS REGISTERED AS:$\backslash$nR2) a Democrat$\backslash$nR3) a Republican$\backslash$nR4) an Independent$\backslash$n\texttt{'},\\

\texttt{`}2016 US PRESIDENTIAL ELECTION VOTE:$\backslash$nL1) did not vote$\backslash$nL2) voted for Donald Trump, the Republican candidate$\backslash$nL3) voted for Hillary Clinton, the Democrat candidate$\backslash$nL4) voted for Gary Johnson, the Libertarian candidate$\backslash$nL5) voted for Jill Stein, the Green Party candidate$\backslash$n\texttt{'},\\

\texttt{`}2018 MIDTERM ELECTION VOTE:$\backslash$nT1) did not vote$\backslash$nT2) voted for the Republican Party$\backslash$nT3) voted for the Democratic Party$\backslash$nT4) voted for a third party$\backslash$n\texttt{'},\\

\texttt{`}2020 US PRESIDENTIAL ELECTION VOTE:$\backslash$nV1) did not vote$\backslash$nV2) voted for Donald Trump, the Republican candidate$\backslash$nV3) voted for Joe Biden, the Democrat candidate$\backslash$nV4) voted for Jo Jorgensen, the Libertarian candidate$\backslash$nV5) voted for Howie Hawkins, the Green Party candidate$\backslash$n\texttt{'}\\

\texttt{)}
\end{footnotesize}
    \end{framed}
    \caption{Format of the survey-like categories used for classification. Each socio-demographic and political category, along with its respective levels, is an element within the vector \texttt{demos\_string}. The elements of this vector are randomised and passed to the LLM within the prompt in Figure \ref{demo.prompt}. The \texttt{$\backslash$n} are necessary to appropriate ensure spacing in the prompt.}
    \label{demos}
\end{figure}

\pagebreak 

\section{Rater Agreement}

\begin{figure}[!htb]
\centering
\begin{subfigure}{\textwidth}
    \centering
    \includegraphics[width =\textwidth]{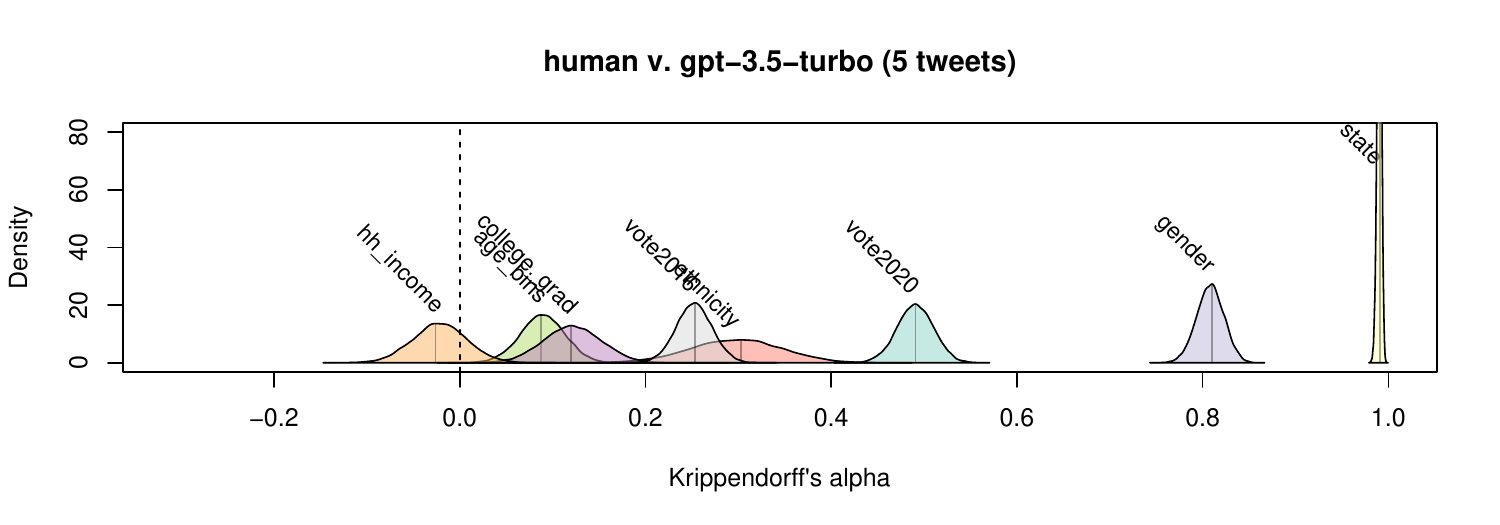}
    \caption{}
    \label{kip_human_gpt5}
\end{subfigure}

\hfill
\begin{subfigure}{\textwidth}
    \centering
    \includegraphics[width =\textwidth]{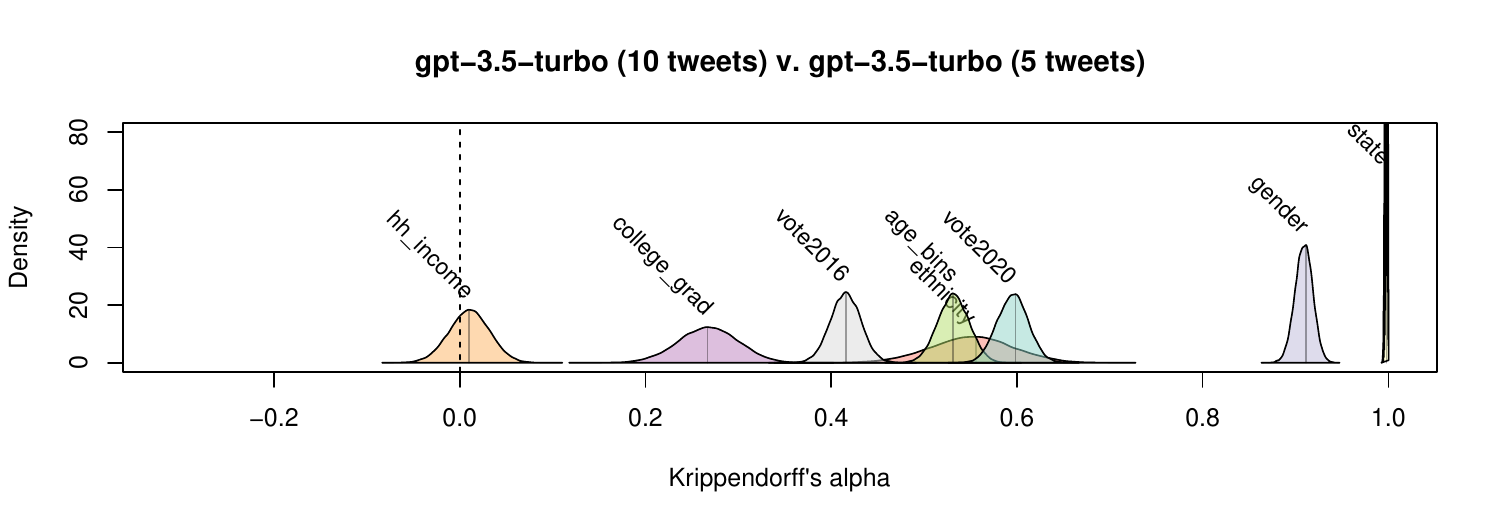}
    \caption{}
    \label{kip_gpt10_gpt5}
\end{subfigure}

\caption{Bootstrap distribution of Krippendorff's $\alpha$ for the annotated variables used in the model.}
\end{figure}

\pagebreak

\begin{figure}[!htb]
    \centering
    \includegraphics[width = \textwidth]{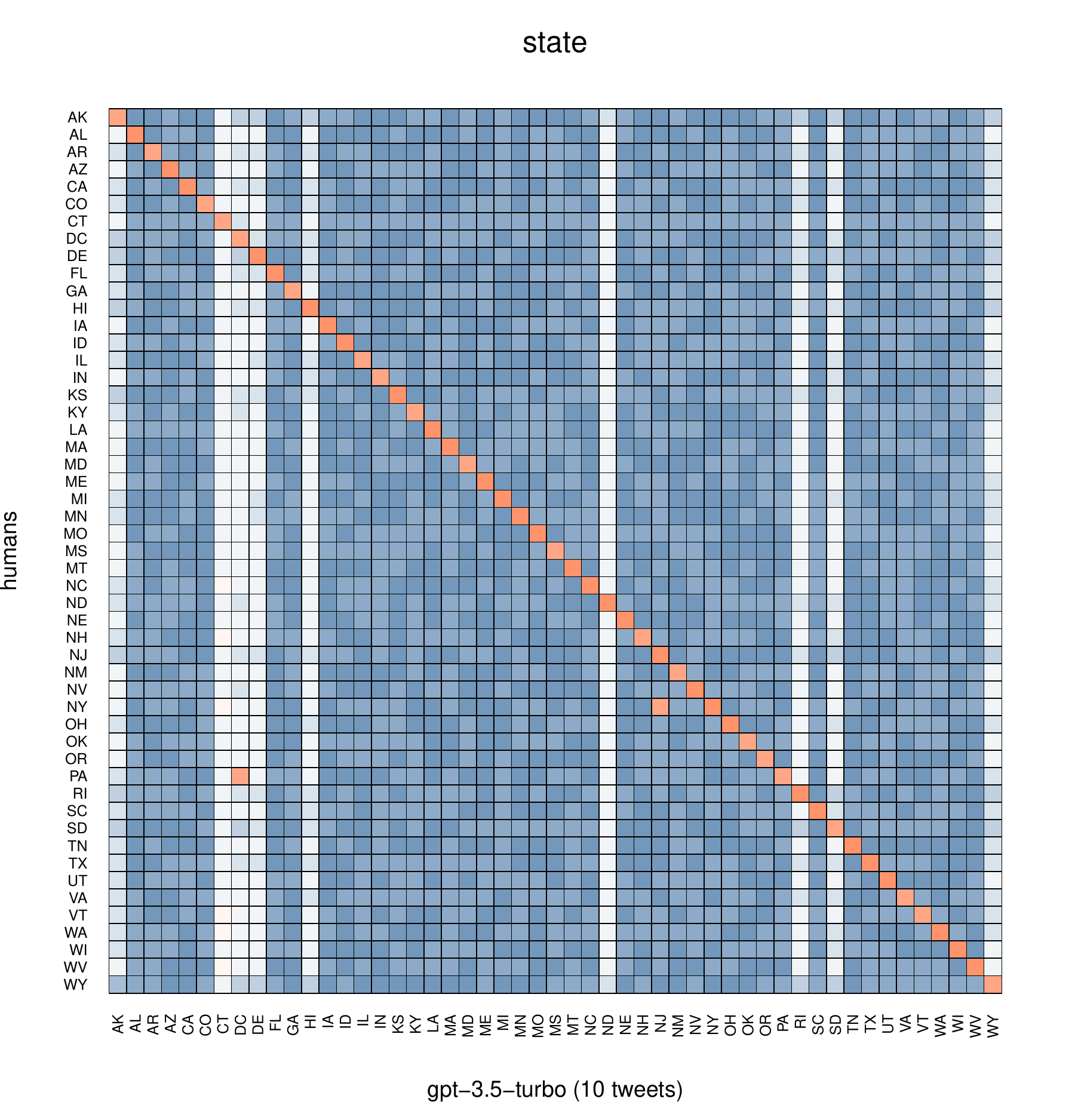}
    \caption{Graphical representation of the relative frequency of posterior predictive incidence across annotations between humans and the \texttt{gpt-3.5-turbo} model prompted with $10$ tweets, for the variable `state'.}
    \label{net_human_gpt10_state}
\end{figure}

\pagebreak

\begin{figure}[!htb]
\centering
     \begin{subfigure}{0.3\textwidth}
         \centering
         \includegraphics[width=\textwidth]{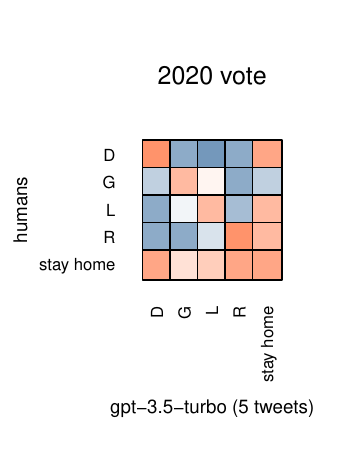}
         \caption{}
         \label{net_human_gpt5_vote2020}
     \end{subfigure}
     \hfill
     \begin{subfigure}{0.31\textwidth}
         \centering
         \includegraphics[width=\textwidth]{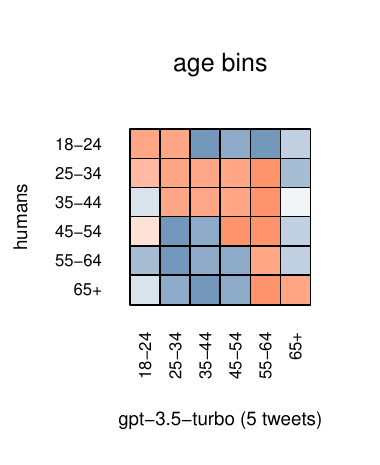}
         \caption{}
         \label{net_human_gpt5_age.bins}
     \end{subfigure}
     \hfill
     \begin{subfigure}{0.335\textwidth}
         \centering
         \includegraphics[trim={0 0.5cm 0 0},clip,width=\textwidth]{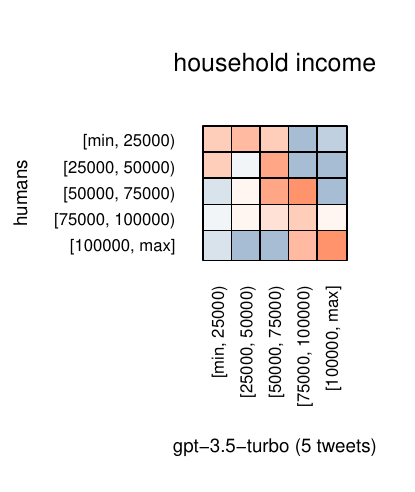}
         \caption{}
         \label{net_human_gpt5_hh.income}
     \end{subfigure}

     \hfill
     \begin{subfigure}{0.34\textwidth}
         \centering
         \includegraphics[width=\textwidth]{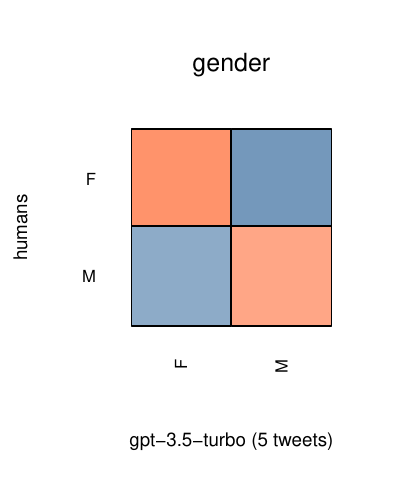}
         \caption{}
         \label{net_human_gpt5_gender}
     \end{subfigure}
     \hfill
     \begin{subfigure}{0.33\textwidth}
         \centering
         \includegraphics[width=\textwidth]{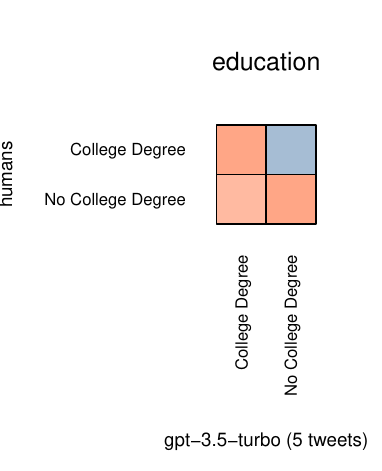}
         \caption{}
         \label{net_human_gpt5_college}
     \end{subfigure}
     \hfill
     \begin{subfigure}{0.29\textwidth}
         \centering
         \includegraphics[width=\textwidth]{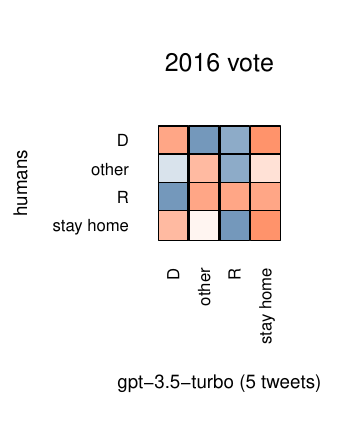}
         \caption{}
         \label{net_human_gpt5_vote2016}
     \end{subfigure}
     
        \caption{Graphical representation of the relative frequency of posterior predictive incidence across annotations between humans and the \texttt{got-3.5-turbo} model prompted with $5$ tweets. See Figure \ref{perfect} for understanding scale and colour-coding. See Figure \ref{net_human_gpt5_state} for the variable `state'.} 
        \label{net_human_gpt5}

\end{figure}

\pagebreak 

\begin{figure}[!htb]
    \centering
    \includegraphics[width = \textwidth]{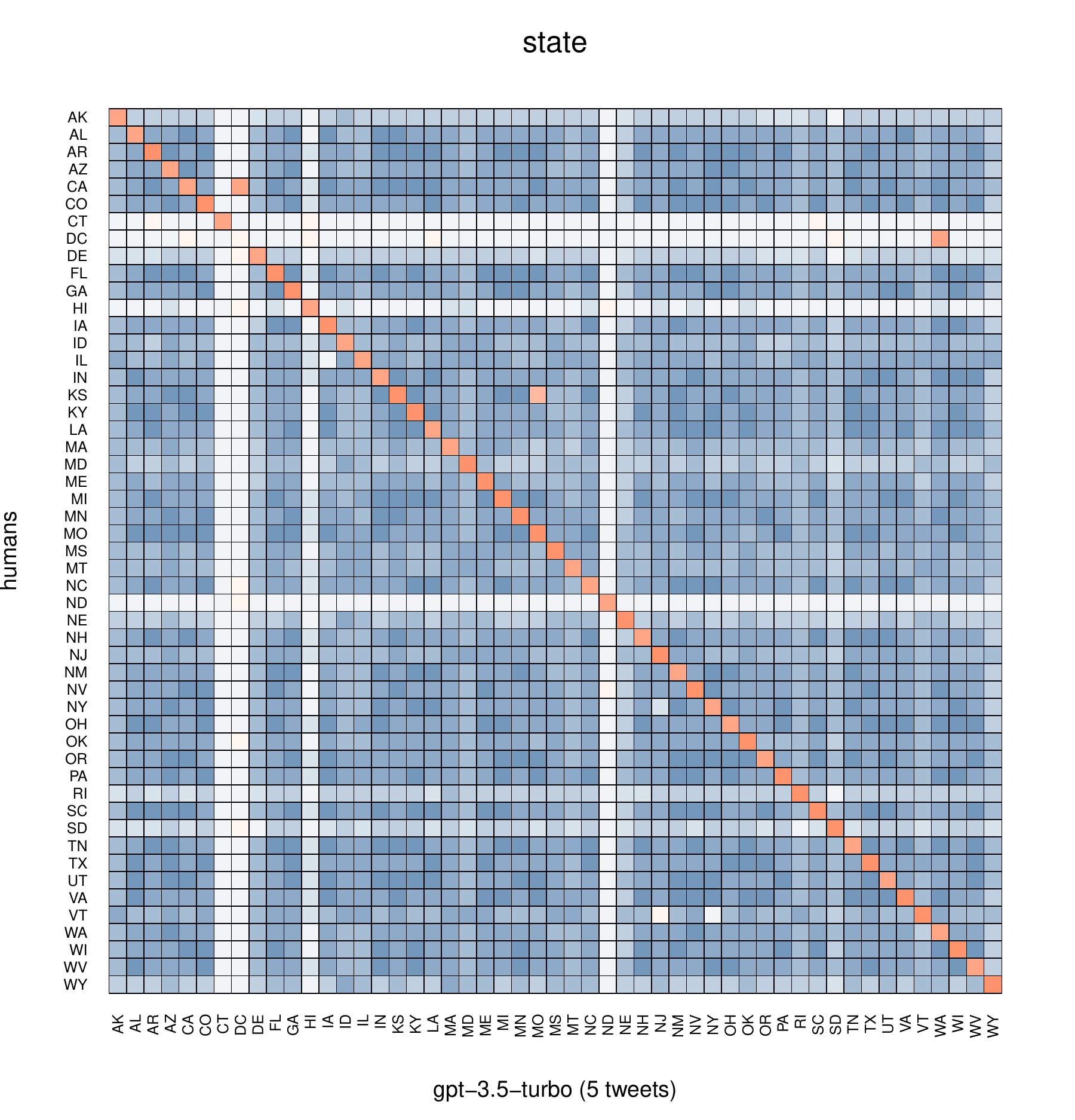}
    \caption{Graphical representation of the relative frequency of posterior predictive incidence across annotations between humans and the \texttt{got-3.5-turbo} model prompted with $5$ tweets, for the variable `state'.}
    \label{net_human_gpt5_state}
\end{figure}

\pagebreak

\begin{figure}[!htb]
\centering
     \begin{subfigure}{0.3\textwidth}
         \centering
         \includegraphics[width=\textwidth]{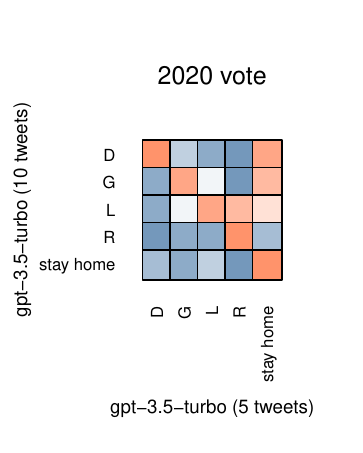}
         \caption{}
         \label{net_gpt10_gpt5_vote2020}
     \end{subfigure}
     \hfill
     \begin{subfigure}{0.31\textwidth}
         \centering
         \includegraphics[width=\textwidth]{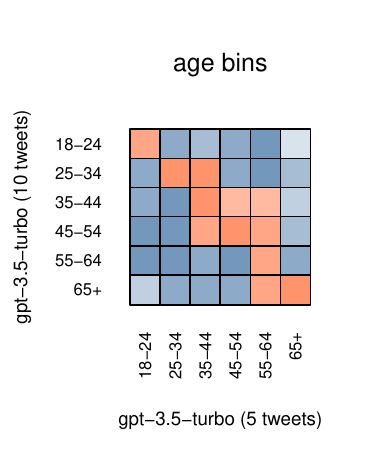}
         \caption{}
         \label{net_gpt10_gpt5_age.bins}
     \end{subfigure}
     \hfill
     \begin{subfigure}{0.335\textwidth}
         \centering
         \includegraphics[trim={0 0.5cm 0 0},clip,width=\textwidth]{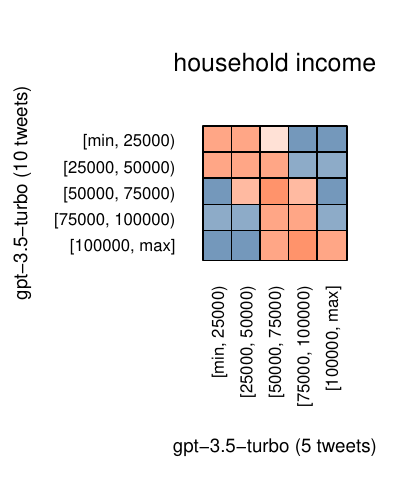}
         \caption{}
         \label{net_gpt20_gpt5_hh.income}
     \end{subfigure}

     \hfill
     \begin{subfigure}{0.34\textwidth}
         \centering
         \includegraphics[width=\textwidth]{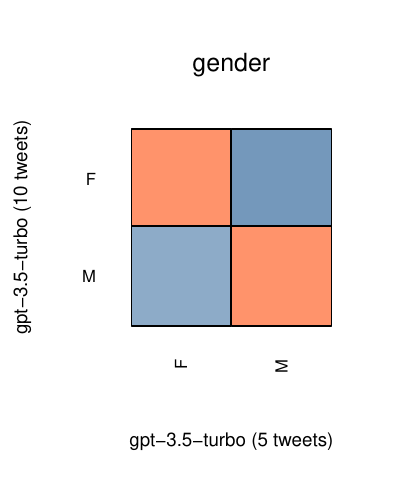}
         \caption{}
         \label{net_gpt10_gpt5_gender}
     \end{subfigure}
     \hfill
     \begin{subfigure}{0.33\textwidth}
         \centering
         \includegraphics[width=\textwidth]{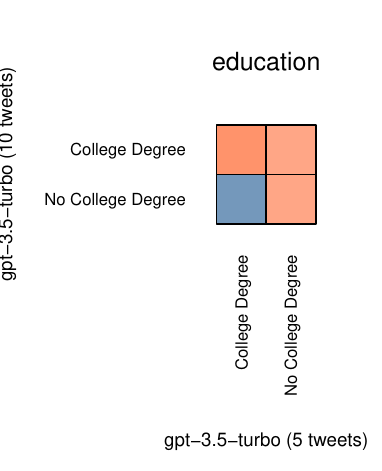}
         \caption{}
         \label{net_gpt10_gpt5_college}
     \end{subfigure}
     \hfill
     \begin{subfigure}{0.29\textwidth}
         \centering
         \includegraphics[width=\textwidth]{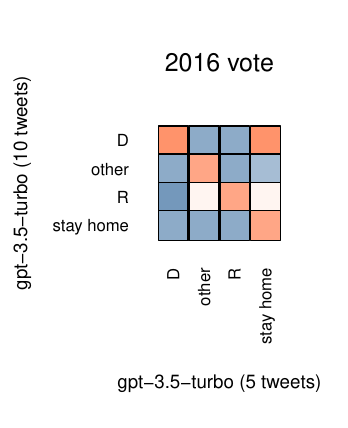}
         \caption{}
         \label{net_gpt10_gpt5_vote2016}
     \end{subfigure}
     
        \caption{Graphical representation of the relative frequency of posterior predictive incidence across annotations between the \texttt{got-3.5-turbo} model prompted with $10$ tweets and the same model prompted with $5$ tweets. See Figure \ref{perfect} for understanding scale and colour-coding. See Figure \ref{net_gpt10_gpt5_state} for the variable `state'.} 
        \label{net_gpt10_gpt5}
\end{figure}

\pagebreak

\begin{figure}[!htb]
    \centering
    \includegraphics[width = \textwidth]{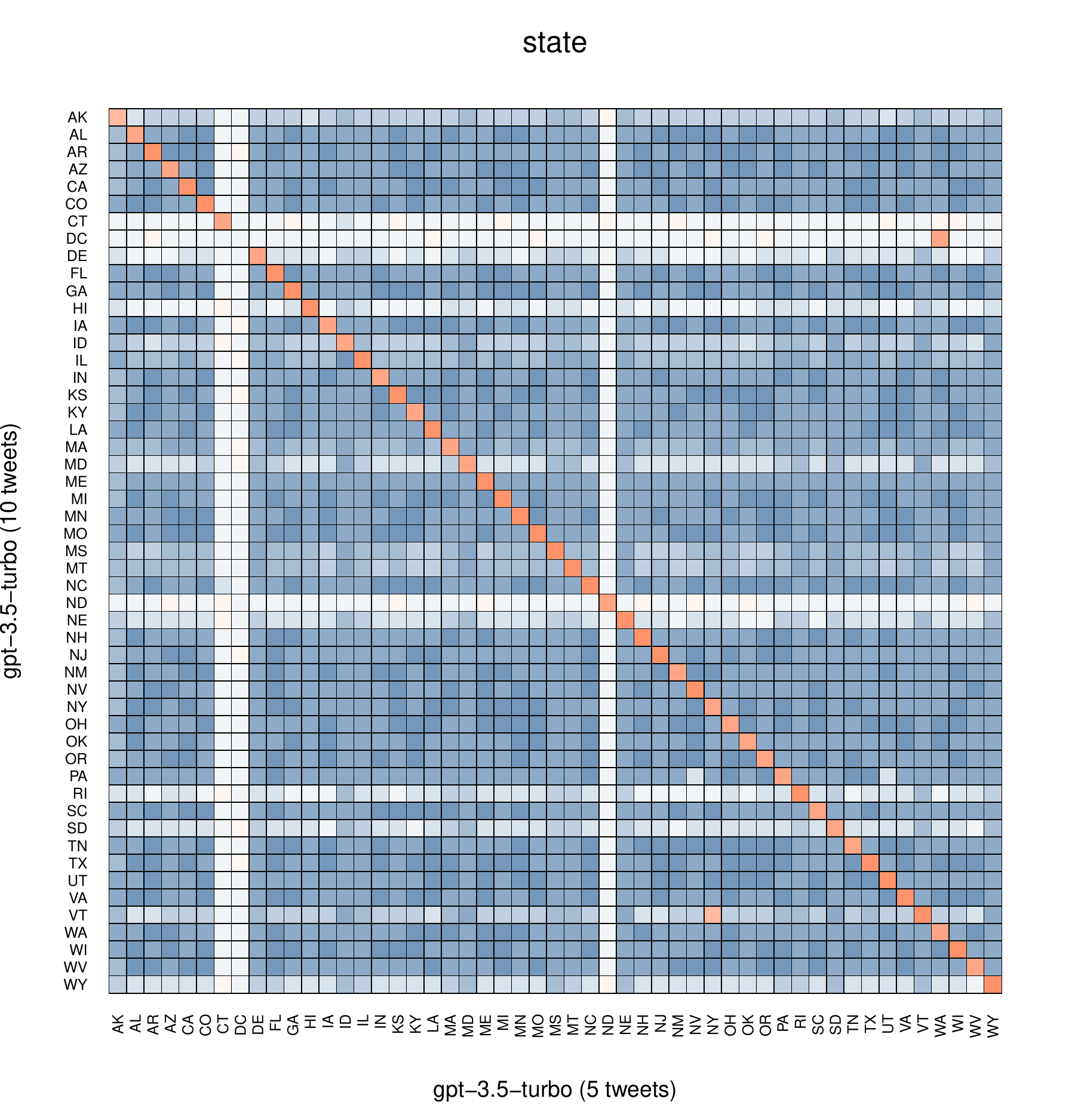}
    \caption{Graphical representation of the relative frequency of posterior predictive incidence across annotations between \texttt{gpt-3.5-turbo} model prompted with $10$ tweets and the same model prompted with $5$ tweets, for the variable `state'.}
    \label{net_gpt10_gpt5_state}
\end{figure}

\pagebreak

\begin{table}[!htb]
\scalebox{0.635}{
\begin{tabular}{r||l|l|l}
  & \begin{tabular}{@{}l@{}} humans v.\\\texttt{gpt-3.5-turbo} (10 tweets) \end{tabular}
  & \begin{tabular}{@{}l@{}} humans v.\\\texttt{gpt-3.5-turbo} (5 tweets) \end{tabular}
  & \begin{tabular}{@{}l@{}} \texttt{gpt-3.5-turbo}: \\(10 tweets) v. (5 tweets) \end{tabular} \\
  \hline
  \hline
  \emph{$2020$ vote} &
  \begin{tabular}{@{}l@{}} $\bullet$ Human $\rightarrow$ \emph{stay home} bias \\ $\bullet$ LLM $\rightarrow$ Democrats \emph{stay home} bias
  \end{tabular} &
    \begin{tabular}{@{}l@{}} $\bullet$ \emph{stay home} disagreement \\ $\bullet$ LLM $\rightarrow$ third-party mixing 
    \end{tabular} &   
    \begin{tabular}{@{}l@{}} $\bullet$ (5 tweets) $\rightarrow$ stay-home bias 
    \end{tabular}\\
  \hline
  \emph{age bins} &
  \begin{tabular}{@{}l@{}} 
  $\bullet$ \emph{middle-aged} confusion \\
  $\bullet$ LLM $\rightarrow$ some \emph{old-age} bias 
  \end{tabular} &
    \begin{tabular}{@{}l@{}} 
    $\bullet$ see humans v.\\ \texttt{gpt-3.5-turbo} (10 tweets)
  \end{tabular} &
    \begin{tabular}{@{}l@{}}
  $\bullet$ minor \emph{middle-age} noise
  \end{tabular}\\
  \hline
  \emph{hh income} &
  \begin{tabular}{@{}l@{}}
  $\bullet$ LLM $\rightarrow$ \emph{higher-income} bias 
  \end{tabular} & 
  \begin{tabular}{@{}l@{}}
  $\bullet$ noisy agreement
  \end{tabular} &
  \begin{tabular}{@{}l@{}}
  $\bullet$ noisy agreement
  \end{tabular} \\
  \hline
  \emph{gender} & & & \\
  \hline
  \emph{education} &
  \begin{tabular}{@{}l@{}}
  $\bullet$ LLM $\rightarrow$ \emph{high-edu.} bias 
  \end{tabular} & 
    \begin{tabular}{@{}l@{}}
    $\bullet$ see humans v.\\ \texttt{gpt-3.5-turbo} (10 tweets)
  \end{tabular}  &
      \begin{tabular}{@{}l@{}}
    $\bullet$ (5 tweets) $\rightarrow$ \emph{low-edu} bias
  \end{tabular} \\
  \hline
  \emph{$2016$ vote} &
  \begin{tabular}{@{}l@{}}
  $\bullet$ humans $\rightarrow$ Democrats \emph{stay home} bias\\
  $\bullet$ LLM $\rightarrow$ general \emph{stay home} 
  \end{tabular} & 
    \begin{tabular}{@{}l@{}}
      $\bullet$ see humans v.\\ \texttt{gpt-3.5-turbo} (10 tweets)
    \end{tabular}& 
    \begin{tabular}{@{}l@{}}
      $\bullet$ (5 tweets)  $\rightarrow$ Democrats \emph{stay home} bias
    \end{tabular} \\
    \hline
  \emph{state} &
  \begin{tabular}{@{}l@{}}
  $\bullet$ extremely minor disagreement\\
  $\bullet$ LLM = DC $\rightarrow$ humans = PA\\
  $\bullet$ LLM = NJ $\rightarrow$ humans = NY\\
  \end{tabular} & 
    \begin{tabular}{@{}l@{}}
  $\bullet$ extremely minor disagreement\\
  $\bullet$ LLM = DC $\rightarrow$ humans = CA\\
  $\bullet$ LLM = MO $\rightarrow$ humans = KS\\
  $\bullet$ LLM = WA $\rightarrow$ humans = DC\\
  \end{tabular} &
      \begin{tabular}{@{}l@{}}
 $\bullet$ extremely minor disagreement\\
   $\bullet$ (5 tweets) = NY $\rightarrow$ (10 tweets) = VT\\
   $\bullet$ (5 tweets) = WA $\rightarrow$ (10 tweets) = DC\\
     \end{tabular} \\
\hline
\end{tabular}
}
\caption{A qualitative summary of deviations from perfect agreement emergent from the (Dis)Agreement Network analysis. Any `bias' in this context should be interpreted as relative to the opposite rater.}
\label{(dis)agreement_interpretation}
\end{table}

\pagebreak

\newgeometry{left = 1.25cm,bottom = 0.5cm,top = 0.5cm}
\begin{landscape}

\begin{figure}
    \centering
    \includegraphics[width = 750pt]{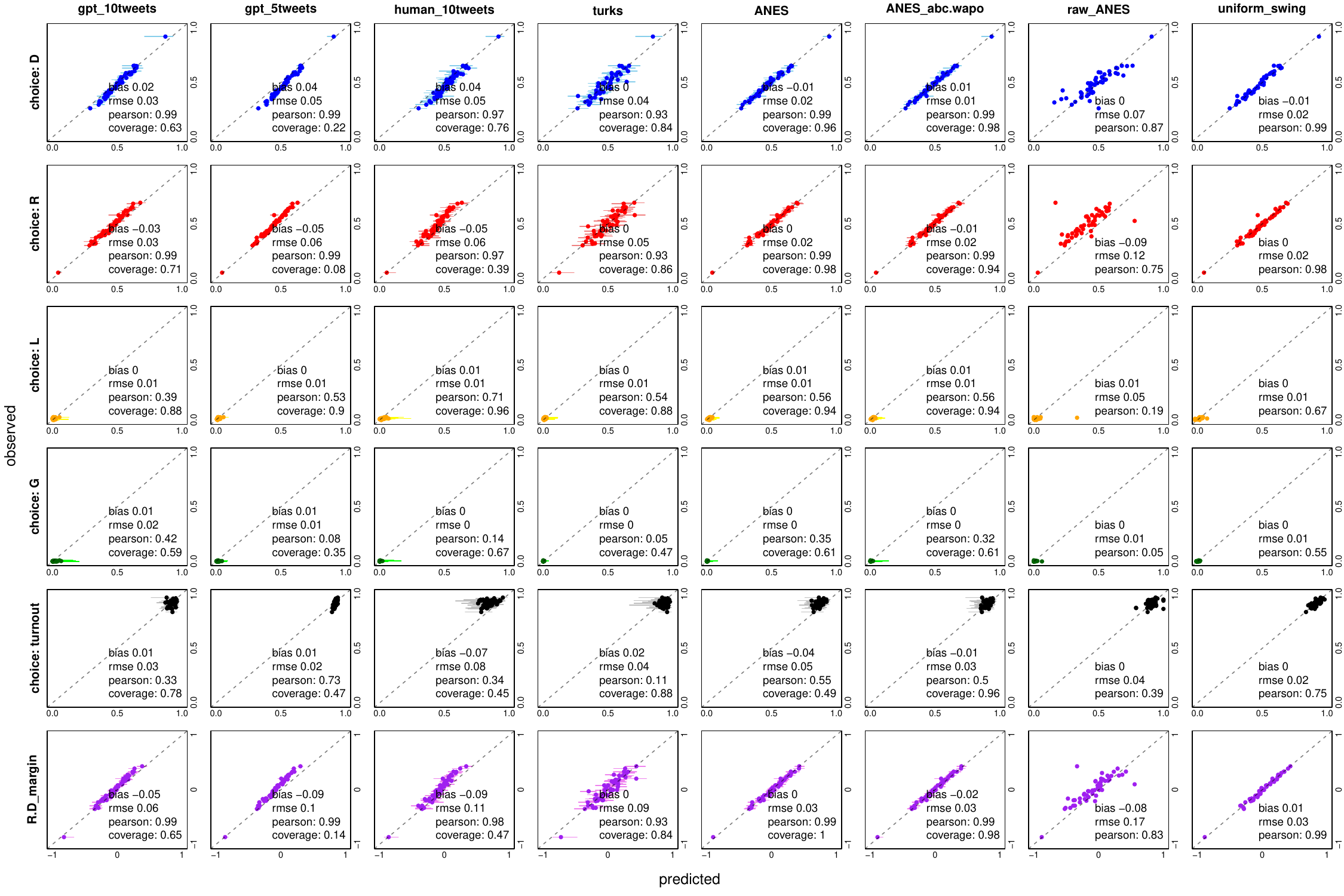}
    \caption{Predictive performance of the bias-corrected structured MrP under different training data. Each training-set is displayed in a column, whilst each choice in the $2020$ US election choice-set is displayed over rows. The Republican-Democrat margin is presented along with choice predictions.}
    \label{area_predictions}
\end{figure}

\end{landscape}
\restoregeometry

\pagebreak

\pagebreak 

\newgeometry{left = 3.5cm,right = 0.5cm, bottom = 0.5cm, top =1cm}
\begin{landscape}
\begin{figure}[!htb]
    \centering
    \includegraphics[width = 725pt]{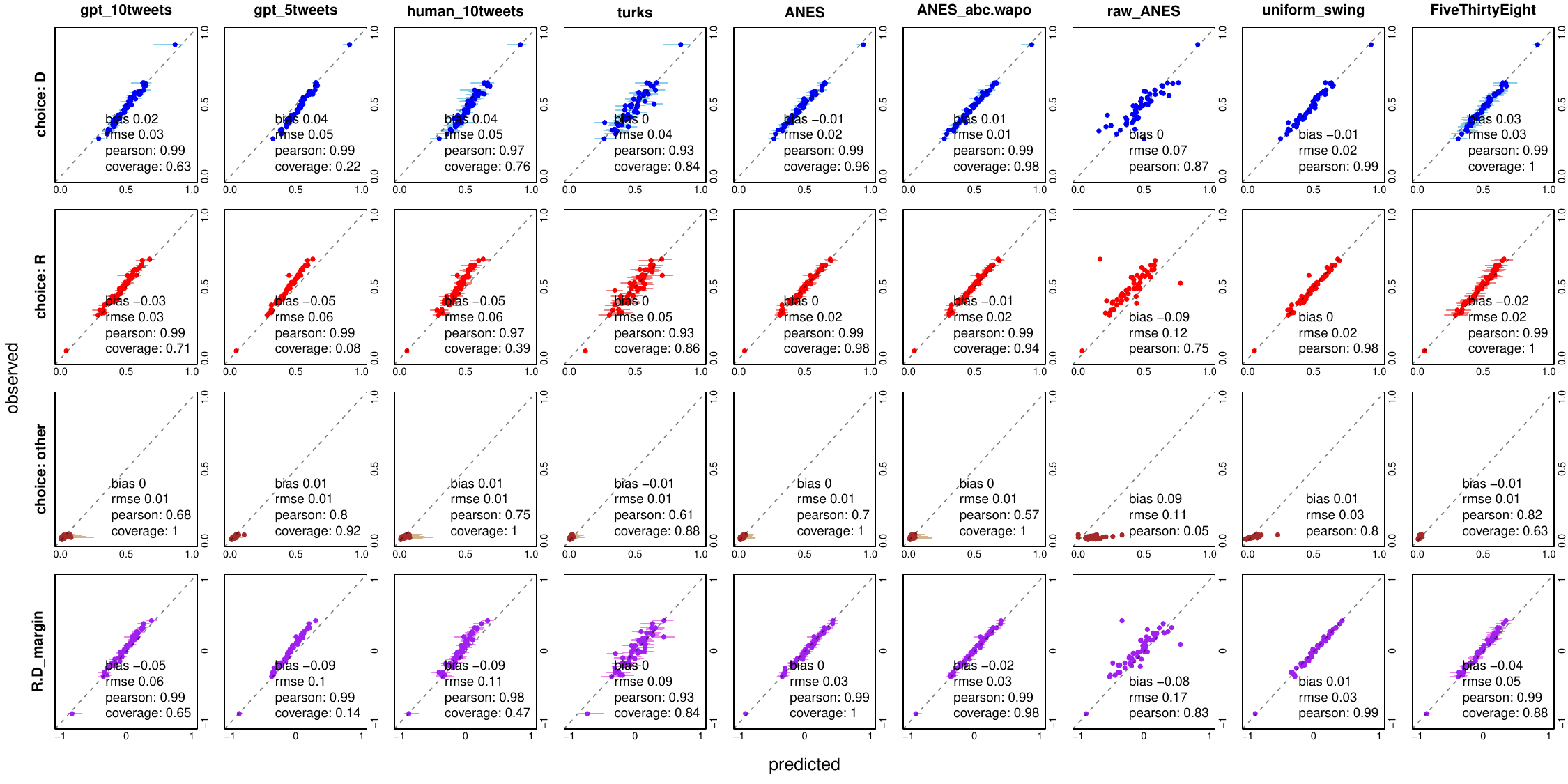}
\caption{Predictive performance of the bias-corrected structured MrP under different training data.  We aggregate Libertarians and Greens into the `other' category to allow comparison with \texttt{FiveThirtyEight}.}
    \label{area_predictions_538}
\end{figure}
\end{landscape}
\restoregeometry

\pagebreak 

\begin{landscape}
\newgeometry{left = 1.25cm,right = 0.5cm, bottom = 0.5cm, top =5cm}
\begin{figure}[!htb]
    \centering
    \includegraphics[width = 650pt]{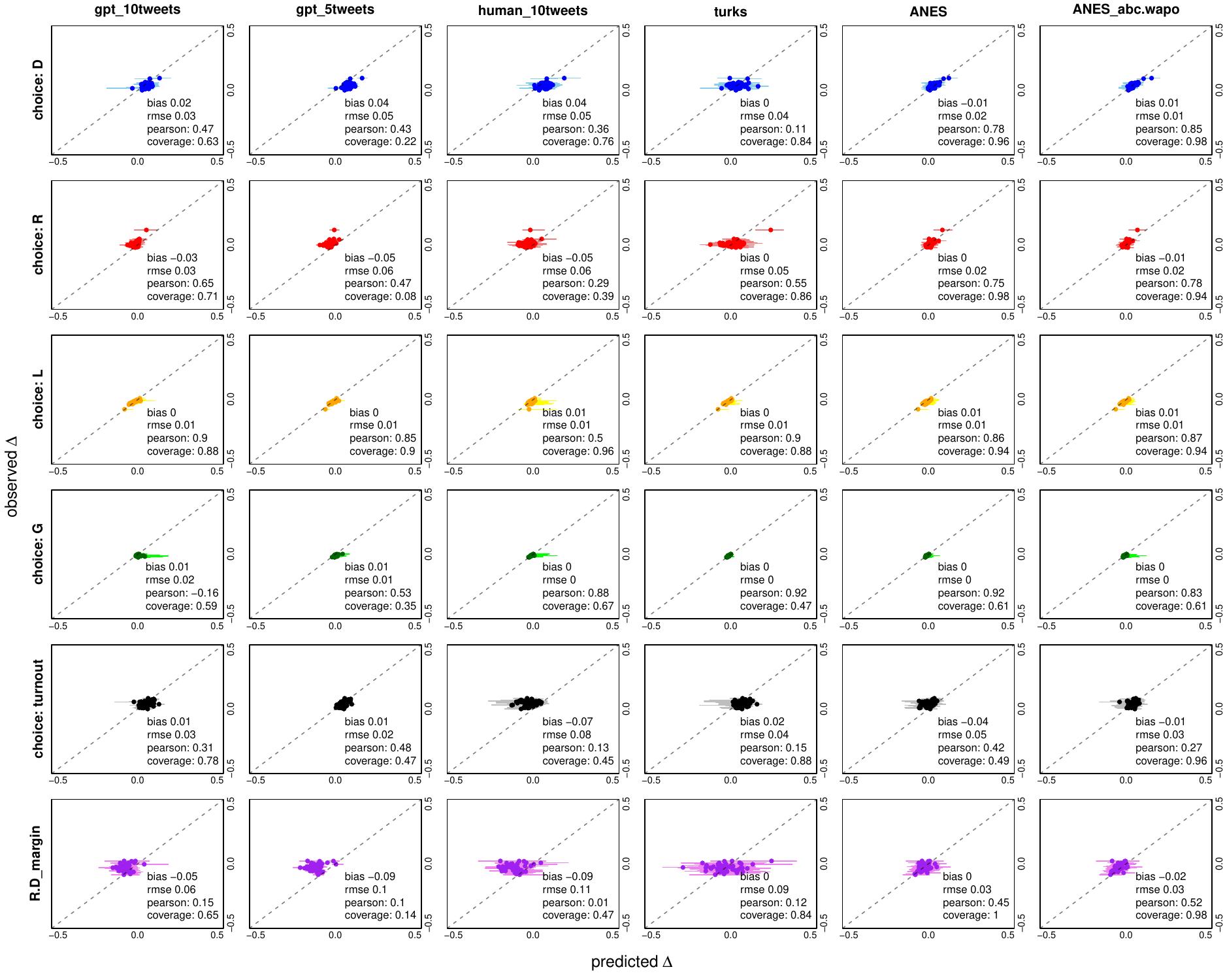}
    \caption{Ability to predict non-uniform change from last election.}
    \label{area_predictions_delta}
\end{figure}
\restoregeometry
\end{landscape}

\begin{landscape}
\newgeometry{left = 3.5cm,right = 0.5cm, bottom = 0.5cm, top =7.5cm}
\begin{figure}[!htb]
    \centering
    \includegraphics[width = 650pt]{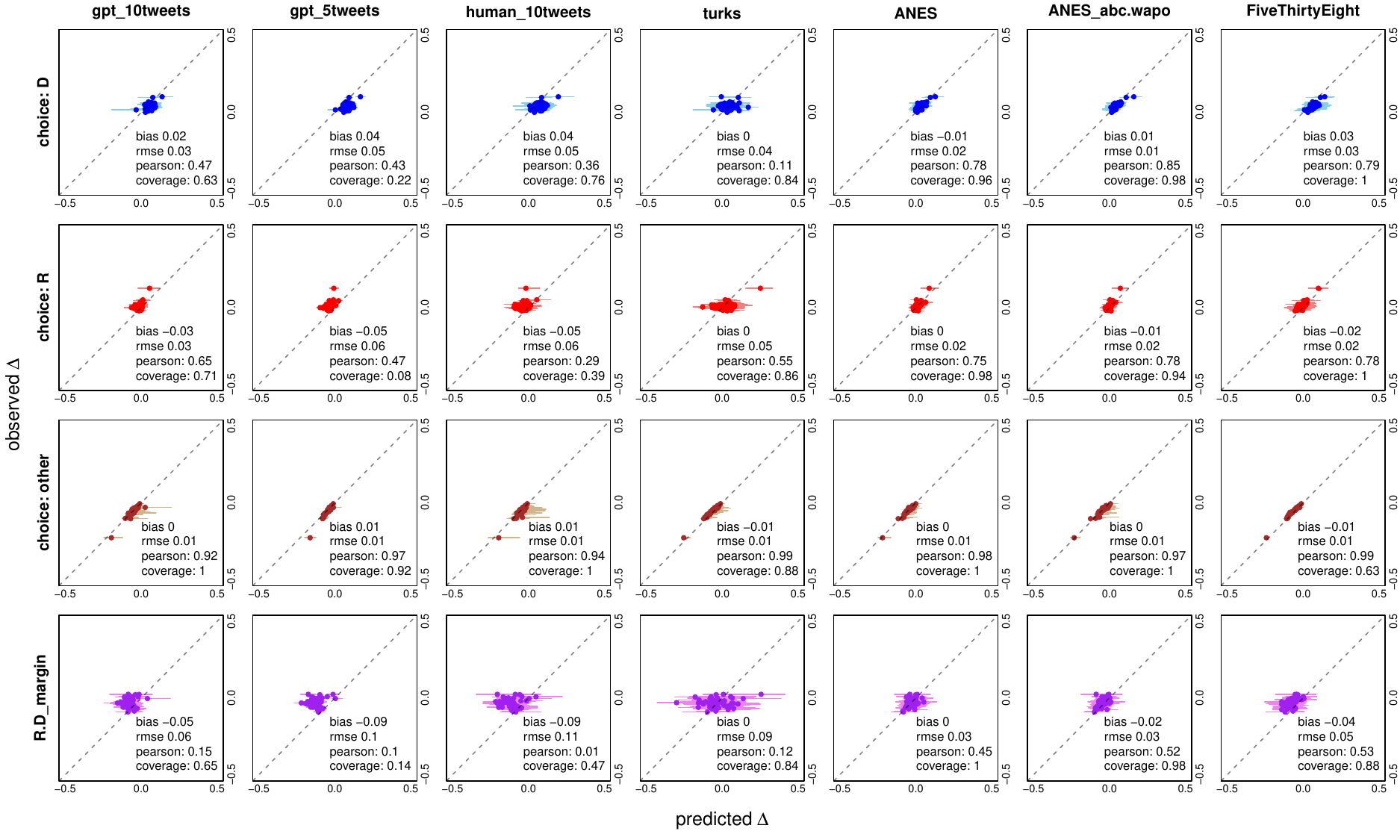}
\caption{Ability to predict non-uniform change from last election. We aggregate Libertarians and Greens into the `other' category to allow comparison with \texttt{FiveThirtyEight}.}
    \label{area_predictions_delta_538}
\end{figure}
\restoregeometry
\end{landscape}

\end{spacing}{}

\end{document}